\documentclass[longauth]{aa}  

\usepackage{natbib}
\bibpunct{(}{)}{;}{a}{}{,} 
\usepackage{hyperref}
\hypersetup{
    colorlinks=true,
    citecolor=blue,
    linkcolor=red,
    filecolor=magenta,      
    urlcolor=cyan,
    }

\usepackage{soul}
\usepackage[dvipsnames]{xcolor}
\usepackage{graphicx}
\usepackage{float}
\usepackage[caption = false]{subfig}
\usepackage[varg]{txfonts}
\usepackage{array}

\usepackage{placeins}

\defcitealias{bailer21}{BJ21}
\defcitealias{cowley65}{C65}
\defcitealias{gonzalez02}{GR02}

\usepackage[normalem]{ulem}

\begin{document}

   \title{Hidden massive eclipsing binaries in red supergiant systems}
   \subtitle{The hierarchical triple system KQ Puppis and other candidates\thanks{Based on observations made with the Very Large Telescope Interferometer (VLTI) at the Paranal Observatory under program 114.27NF.001 and 116.28P0.001 (PI: Jadlovsk\'y)}}

    \titlerunning{Hidden massive eclipsing binaries in red supergiant systems}
    \authorrunning{D. Jadlovský et al.}

   \author{D. Jadlovsk\'y\inst{1,2}
           \and
          L. Moln\'ar \inst{3,4,5}
         \and
          A. Ercolino\inst{6}
            \and
          M. Bernini-Peron\inst{7}  
            \and
          A. Mérand\inst{2}
            \and
          J. Krtička\inst{1}
            \and
          L. Wang\inst{9}
           \and
          R. Z. Ádám\inst{3,4}
          \and
          D. Baade\inst{2}
          \and
         A. Bayo\inst{2}
          \and
          G. González-Torà\inst{7}
          \and
          T. Granzer\inst{10}
         \and
          G. W. Henry\inst{11}
           \and
          J. Janík\inst{1}
                    \and
          J. Kolář\inst{1}
                    \and
         K. Kravchenko\inst{12}
                    \and
          N. Langer\inst{5}
          \and
           L. M. Oskinova\inst{13}
          \and
          D. Pauli\inst{14}
          \and
          V. Ramachandran\inst{7}
          \and
          A. C. Rubio\inst{15}
        \and
         A.A.C. Sander\inst{7,8}
        \and 
         K.G. Strassmeier\inst{10}
         \and
         M. Weber\inst{10}
        \and
        M. Wittkowski\inst{2}
        \and
       R. Brahm\inst{16,17}
       \and
       V. Schaffenroth\inst{18}
       \and
       L. Vanzi\inst{19}
       \and
       M. Skarka\inst{20}
    }
   \institute{Department of Theoretical Physics and Astrophysics, Faculty of Science, Masaryk University, Kotl\'a\v rsk\'a 2, 61137, Brno, Czech Republic  \\
            \email{jadlovsky@mail.muni.cz}
        \and
            European Southern Observatory (ESO), Karl-Schwarzschild Str. 2, 85748, Garching bei München, Germany
        \and
            Konkoly Observatory, HUN-REN Research Centre for Astronomy and Earth Sciences, MTA Centre of Excellence, Konkoly-Thege Mikl\'os \'ut 15-17, 1121, Budapest, Hungary
        \and
             MTA–HUN-REN CSFK Lendület "Momentum" Stellar Pulsation Research Group, Konkoly-Thege Mikl\'os \'ut 15-17, 1121, Budapest, Hungary
        \and
            E\"otv\"os Lor\'and University, Institute of Physics and Astronomy, 1117 P\'azm\'any P\'eter s\'et\'any 1/A, Budapest, Hungary
        \and
            Argelander-Institut für Astronomie, Auf dem Hügel 71, 53121 Bonn, Germany
        \and
            {Zentrum f{\"u}r Astronomie der Universität Heidelberg, Astronomisches Rechen-Institut, M{\"o}nchhofstr. 12-14, 69120 Heidelberg, Germany\label{inst:ARI}}
        \and  
            {Universität Heidelberg, Interdiszipli{\"a}res Zentrum f{\"u}r Wissenschaftliches Rechnen, 69120 Heidelberg, Germany\label{inst:IWR}}
                     \and
            Yunnan Observatories, Chinese Academy of Sciences (CAS), Kunming 650216, Yunnan, China
                         \and
              Leibniz-Institut für Astrophysik Potsdam (AIP), An der Sternwarte 16, 14482 Potsdam, Germany
                          \and
            Tennessee State University (retired), Nashville, TN 37209, USA
                        \and
            Max Planck Institute for Extraterrestrial Physics (MPE), Giessenbachstrasse 1, 85748, Garching bei München, Germany
            \and
            {Institute for Physics and Astronomy, University Potsdam, 14476 Potsdam, Germany\label{inst:UP}}
            \and
            {Institute of Astronomy, KU Leuven, Celestijnenlaan 200D, 3001 Leuven, Belgium\label{IvS}}
            \and
            Max-Planck-Institut für Astrophysik, Karl-Schwarzschild-Str. 1, 85748 Garching b. München, Germany
            \and
            Facultad de Ingenier\'ia y Ciencias, Universidad Adolfo Ib\'{a}\~{n}ez, Av. Diagonal las Torres 2640, Pe\~{n}alol\'{e}n, Santiago, Chile
            \and
            Millennium Institute for Astrophysics, Santiago, Chile
            \and
            Th\"{u}ringer Landessternwarte Tautenburg, Sternwarte 5, 07778 Tautenburg, Germany
            \and
            Department of Electrical Engineering and Center of Astro Engineering of Pontificia Universidad Catolica de Chile, Av. Vicu\~na Mackenna 4860, Santiago, Chile
            \and
            Astronomical Institute, Czech Academy of Sciences, Fri\v{c}ova 298, 25165 Ond\v{r}ejov, Czech Republic 
            }

   \date{ }

 
  \abstract
   {
   The majority of massive stars are part of binary systems and may interact with their companions during their evolution. This has important consequences for systems in which one star evolves into a red supergiant (RSG). However, not many RSGs have been found in binary systems, and only a few have constrained orbital parameters.}
   {We aim to better characterize and constrain the properties of some of the known RSGs in binaries. We were inspired to do so by recent observations where these RSGs transitioned to an earlier spectral type, likely following a major interaction event with their companions.}
   {We searched the available TESS photometry for eclipsing companions. We focused on the best candidate, the VV Cephei-type RSG binary KQ Pup, which is made up of an RSG, KQ Pup A, and a B-type companion, KQ Pup B, and has an orbital period of 26 years. For this system, we have enough data to constrain the system's properties. We used archival photometry and UV spectroscopy along with newly acquired optical spectra and interferometric data from VLTI-GRAVITY.} 
   {Using TESS light curves, we discovered eclipses with a period of $17.2596 \: \rm d$ associated with KQ Pup B, making it a binary. We refer to the two components as KQ Pup Ba and Bb. Meanwhile, the detection of the hydrogen Br$\gamma$ line with VLTI-GRAVITY enabled us to track the orbital motion of the Ba+Bb pair relative to A and thus determine the astrometric orbit of A+B. The dynamical masses agree with independent estimates from asteroseismology and evolutionary models. The results give a mass of $ \sim 10 \: \rm M_{\odot} $ for the RSG KQ Pup A and $ \sim 14 \: \rm M_{\odot} $ for the sum of the hot components Ba+Bb. The observed properties of the system are compatible with a coeval hierarchical triple star, where we constrained the mass of KQ Pup Bb to be $ \gtrsim 1.2 \: \rm M_{\odot} $. We determined an orbital parallax of $\pi = 1.24^{+0.05}_{-0.04}\, \rm mas $ based on the detailed orbital model of the system, which is the first such parallax measurement for an RSG, and the value is consistent with Gaia DR3.}  
   {KQ Pup represents a unique demonstration of mass transfer in a wide, eccentric RSG system. The variability of Balmer emission lines and the detection of Br$\gamma$ are a strong signature of accretion to the Ba+Bb pair near periastron. With the RSG filling its Roche lobe by only $\sim 70\%$ at periastron ($\phi_{\rm A+B} \sim 0$), the mass transfer is instead driven by accretion from its extended atmosphere via the wind Roche-lobe overflow. The accretion disk is formed already by $\phi_{\rm A+B} \sim 0.95$, while TESS light curves also begin to show a signature of occultations by the circumbinary disk, similar to young stellar objects. Through apastron ($\phi_{\rm A+B} \sim 0.5$), Balmer emissions weaken and indicate a decaying disk. 
   Overall, using TESS, we discovered that several previously assumed RSG binaries host eclipsing inner systems, corresponding to $\sim 10 \%$ of all known Galactic RSG binaries. This suggests that many of the other RSG binaries may also be hierarchical triples.
   } 

   
   \maketitle

\section{Introduction}
Red supergiants (RSGs) are a critical point of evolution for massive stars ($M > 8 \: \rm M_{\odot}$). During the RSG phase, stars experience increased mass-loss rates \citep{van_loon05}, which significantly contributes to the dust production in galaxies \citep{levesque17}. Red supergiants may explode as Type IIP/L supernovae (SN) or further evolve and explode as a different type of SN \citep{smartt09}. While the effect of the mass loss can significantly alter the final evolution of RSGs, there are many uncertainties in the mass-loss process. Recently, it has been shown that episodic mass-loss events may be the missing link to explain the mass-loss process, as shown, for example, through the dimming event of Betelgeuse \citep[e.g.,][]{montarg21, dupree22, humphreys22} and RW Cep \citep{anugu23}. The binary interaction can be one of the deciding factors \citep{ercolino24}, as the majority of massive stars have a companion that they will interact with at some point during their evolution \citep{sana12}. Nonetheless, there are only a few binary RSG systems that are well characterized \citep{patrick24a}.

Most of the few well-studied binary RSG systems belong to the VV Cephei-type binaries \citep{cowley69}, i.e., wide interacting systems with periods from years to decades, which consist of a cool K/M supergiant and an early B-type star. The optical spectra show broad emission wings of the Balmer series and many singly ionized emission lines, including forbidden emission lines. In the initial sample by \citet{cowley69}, there were 13 RSGs. Some of the VV Cephei binaries also belong to the rare type of $\zeta$ Aurigae-type binaries, i.e., an eclipsing binary (EB) system undergoing the chromospheric eclipse phenomenon \citep{ake15}. In VV Cephei and $\zeta$ Aurigae binaries, their companions essentially probe the subsonic part of the cool wind, which is otherwise difficult to constrain by observations. 
This diagnostic capability was exploited for determining the mass-loss relation of red giants and supergiants by \citet{reimers75}, \citet{dupree86}, and \citet{dupree87}. 

In recent years, the number of identified RSG binaries has increased considerably. For the most up-to-date list of Galactic RSG binaries, we consulted the list of Galactic RSGs and RSG candidates from \citet{healy24}, which includes about 79 binaries, out of which 44 are identified as RSGs in SIMBAD\footnote{\url{https://simbad.cds.unistra.fr/simbad/}} \citep{wenger00}. The situation has been further improved by newly detected populations of RSG+B binaries in the Local Group \citep{neugent18, neugent19, dorda21, patrick22, patrick25}, increasing the sample by a few hundred binaries. These results 
yielded the fraction of RSG binaries relative to single RSGs: $\sim 10-30 \%$ in Magellanic Clouds \citep{neugent20, dai25} and $\sim 15-40 \%$ in M31 and M33 \citep{neugent21}. Considering that the binary fraction of OB-type stars is between 80--100\% \citep{Chen-2024PrPNP}, these lower RSG fractions indicate that RSGs may lose companions (either through escape or mergers; \citealt{wheeler17}) or that they have gone undetected so far.

It is not clear what fraction of binary RSGs in these new samples show properties similar to those of the classical VV~Cephei binaries. \citet{patrick25} spectroscopically studied 16 RSG+B systems in the Small Magellanic Cloud and found that four of them also show signatures of being embedded in the wind of the RSG. They found that in several of their systems, the ages for the RSG star and hot companion disagree, suggesting a past mass transfer, while the rest of the sample can be explained by co-eval evolution. The recent temporary, dramatic transition of RSG WOH G64 in the Large Magellanic Cloud to an earlier spectral type, very likely due to an interaction with its companion \citep{munoz24, vanLoon26}, demonstrated that binary interaction can indeed have a strong impact on 
the final evolution of an RSG.  
Furthermore, other single RSGs, such as Betelgeuse, were recently hypothesized to have a low-mass companion \citep{goldberg24,macleod25}. Following several non-detections \citep{goldberg25,ogrady25}, it seems that the companion has finally been detected via speckle imaging 
\citep{howell25}, and there are signs of a possible chromospheric wake \citep{Dupree-2026}.

Studies of RSG systems are further complicated by their large stellar radii, which result in angular diameters that significantly exceed parallax angles. This coupled with the surface inhomogeneities caused by giant convective cells that affect the position of the photocenter of the star \citep{chiavassa11b, beguin24} and possible binary motion makes distance measurements complicated. Alternate distance inferences are therefore sought after. The seismic parallax method for RSGs, based on stellar pulsations, was pioneered by \citet{joyce20}, but orbital parallaxes, based on astrometric orbits, have mostly been limited to smaller stars \citep[see, e.g.,][]{Gallenne-2023}.

In this work, we report on the discovery of several possible hierarchical triple RSG systems identified with the Transiting Exoplanet Survey Satellite \citep[TESS;][]{ricker15}. We focus on the VV Cephei-type RSG binary KQ Pup \citep[e.g.,][]{rossi92}, for which we have sufficient data to constrain the system, and we also list other RSG candidates that will require further observations to confirm or rule out new eclipsing companions. KQ Pup is one of the most prominent members of the VV Cephei-type binaries, and it used to be one of the systems with the longest orbital periods known (26 yr, \citealt{gonzalez02}). It is also considered a southern analogue of the more famous VV Cep system \citep[e.g.,][]{bauer00}, although, unlike VV Cep, it is not known to undergo atmospheric eclipses. A unique UV spectrum of VV Cep during the atmospheric eclipse (hot component behind the RSG) was described by \citet{bauer07}.

The paper is structured as follows: In Sect. \ref{chapter:observations} we list the interferometric, spectroscopic, and photometric data used in this work. In Sect. \ref{chapter:candidates} we list candidates for hierarchical triple RSG systems based on TESS. In Sect. \ref{chapter:kq_pup_syst} we focus on the most promising candidate, KQ Pup, and derive the properties of the system, 
while in Sect. \ref{chapter:triple} we unveil the mass transfer mechanism and deduce the likely properties of the third component, KQ Pup Bb. Lastly, in Sect. \ref{chapter:discussion}, we discuss the classification of RSG binaries and the implications of the newly discovered systems to the evolution of massive stars, and in Sect. \ref{chapter:conclusions} we summarize our results.


\section{Observations and data reduction}
\label{chapter:observations}

\subsection{Photometry}
We used photometry from TESS \citep{ricker15}. We extracted TESS light curves from Full Frame Images \citep[TESS-SPOC;][]{caldwell20} for KQ Pup as well as other RSG targets (see Sect. \ref{chapter:candidates}) using the \textit{lightkurve} package \citep{lightkurve18}. We inspected observations in every sector and the available reduction pipeline. The chosen primary target, KQ~Pup, has been observed in four sectors so far, in Sectors 7, 34, 61, and 88. In all cases, 120\,s short-cadence postage-stamp data were collected, so we used these observations for our analysis. 

Additionally, for KQ Pup, we use time-domain photometry from Hipparcos \citep{van-Leeuwen-1997}, the All-Sky Automated Survey for Supernovae (ASAS-SN), using their new saturated stars pipeline \citep{Kochanek-2017,Winecki-2024}, and from the Kamogata/Kiso/Kyoto Wide-field Survey\footnote{\url{http://kws.cetus-net.org/~maehara/VSdata.py}} \citep[KWS,][]{KWS}. Hipparcos 
used its own specific, wide passband. ASAS-SN data were initially collected in the Johnson \textit{V}, and later in the Sloan \textit{g} passband. KWS is collected in the \textit{V} passband.
We also used differential photometry for KQ Pup taken with the T2 0.25\,m Automatic Photoelectric Telescope (APT) at Fairborn Observatory by one of the authors (G. W. Henry) in the $V$ filter.

\subsection{Interferometry}
We obtained spectro-interferometric observations of KQ Pup using the VLTI-GRAVITY instrument \citep{GRAVITY17} at Paranal Observatory. The observations are part of the observing programme 114.27NF.001 and 116.28P0.001  (PI: Jadlovsk\'y) focused on the analysis of the mass-loss process of a large sample of southern RSGs. The majority of observations were taken using a small configuration of Auxiliary Telescopes (ATs), with baselines between $ \sim 10 - 35 \: \rm m $. 
The instrument operates in the near-infrared $K$-band ($1.98 - 2.40 \: \rm \mu m $), and we used the high spectral resolution mode ($R \sim 4000$). For each observation, we used two calibrators with the CAL-SCI-CAL sequence, thereby improving the calibration quality. The observing logs and properties of the calibrators are listed in Appendix \ref{chapter:vlti}. Considering the high brightness of KQ~Pup ($K \sim 0.11 \: \rm mag$), we used the split polarization instrumental mode.

We reduced the raw data of KQ Pup and its calibrators using the \textit{ESO Reflex} workflow for VLTI-GRAVITY in its version 1.9.0.\footnote{\url{https://www.eso.org/sci/software/pipelines/gravity/gravity-pipe-recipes.html}} Interferometric visibility of the calibrators was used to calibrate the visibility of KQ Pup. The final interferometric calibrated dataset includes the visibilities ($|V|$) and differential phases ($\rm DPHI$) for six baselines as well as closure phases ($\rm T3PHI$) for four telescope triangles.

\subsection{Spectroscopy}
\label{chapter:spectra_method}
To obtain new high-cadence spectral time series of KQ Pup, we used the Stellar Activity (STELLA) echelle spectrograph (SES) mounted on the fully robotic 1.2\,m telescope at the Izanã Observatory in Tenerife, which is operated by the Leibniz Institute for Astrophysics Potsdam \citep[AIP,][]{stella,stella_1}. The spectra cover the $390{-}880 \: \rm nm$ wavelength range with $R\sim 55\,000 $. Our time series consists of 66 spectra taken between April 2024 and February 2026. Spectra were reduced using the IRAF{-}based SESDR 4.0 pipeline \citep{stella_3,stella_2}. The determination of radial velocities (RVs) is described in Sect. \ref{chapter:rv}.

We also took new optical spectra for KQ Pup as well as for other candidates (see Table \ref{table:table_candidates}) using PLATOSpec\footnote{\url{https://stel.asu.cas.cz/plato/spectrograph.html}} \citep{kabath26} mounted on the ESO 1.52m telescope at the La Silla Observatory, which covers the wavelength range from 380~nm to 680~nm with $R \sim 70\,000$. By default, data from PLATOSpec are processed using the CERES+ pipeline \citep{2017PASP..129c4002B}, but for the stars we measured, where emission lines are observed, standard IRAF routines were used. Additionally, we analyzed the archival ESPaDOnS \citep{ESPaDOnS03} spectra downloaded from the Polarbase database\footnote{\url{https://www.polarbase.ovgso.fr/}} and archival FEROS spectra downloaded via SPLAT.\footnote{\url{https://www.g-vo.org/pmwiki/About/SPLAT}}

Lastly, we employed archival spectra of KQ Pup taken with the International Ultraviolet Explorer (IUE) satellite during 1978-1995.
We downloaded IUE spectra from the MAST archive\footnote{\url{https://mast.stsci.edu/portal/Mashup/Clients/Mast/Portal.html}}.
The total dataset of high-resolution UV spectra for KQ Pup consists of 23 spectra in the short wavelength region (SWP; $1150-1980 \: \rm \AA$) and 26 in the long wavelength region (LWP/R, $1850-3350 \: \rm \AA$). However, some spectra associated with KQ Pup are at a distance of $\sim 7$ arcsec and $\sim 48$ arcsec, which we excluded from our analysis, as well as a few low signal-to-noise spectra.

\section{RSG systems with eclipsing companions in TESS}
\label{chapter:candidates}
Single companions to RSGs are very difficult to detect from photometric variations alone, unless they themselves are very luminous, and produce measurable eclipses, which is known to be the case only for three RSGs, i.e., AZ~Cas, VV~Cep, and 32 Cyg \citep[e.g.,][]{patrick24a}, similar to $\zeta$~Aurigae-type systems \citep{ake15}. Recently, long secondary periods in RSGs have been linked to close-by low-mass companions that may cause photometric variations through interaction and modulation of the circumstellar medium (CSM), but this has only been investigated in detail for Betelgeuse so far \citep{goldberg24,macleod25,howell25,Dupree-2026}. 
However, if the companion itself is a binary system, it can produce their own eclipses, albeit heavily diluted by the luminosity contribution of the RSG in the system. The orbital periods of such inner binary companions are expected to be significantly shorter than their orbits around the outer RSG. 

As such, high-precision, space-based photometric missions, such as TESS \citep{ricker15}, could reveal their presence through the detection of low-amplitude eclipses in the system. TESS observes the sky in Sectors: a $24^\circ \times 96^\circ$ area with four cameras for 27 days. 
For many stars, data coverage is typically limited to one or a few sectors per observing season, which is too short to follow the slow pulsations and oscillations of RSGs. However, TESS is very well-suited to detect faster variations even at very low amplitudes.

\subsection{List of candidates}

We cross-checked the list of 651 Galactic RSGs (and RSG candidates) by \citet{healy24} with the TESS catalogue, and searched for eclipses in RSG systems. At first, we identified 13 candidates. To remove false candidates, we verified that there are no blends with nearby EBs with a similar period within 5 arcmin, and we also verified the validity of the association of the light curve with the candidate systems by plotting the nearby \textit{Gaia} sources onto the Target Pixel Files (TPF) with the \texttt{lightkurve} package  \citep{lightkurve18}. We identified seven of the initial candidates as likely blends with near EBs. Thus, we are left with a sample of six promising candidates, see Table \ref{table:table_candidates}. TESS data for the promising candidates are shown in Fig. \ref{fig:tess_other_candidates}. We also list the discarded candidates in Appendix \ref{appendix:tess}.

All promising candidates show closely consecutive eclipses, on the order of several days.
Considering the short orbital periods such eclipses indicate, they cannot occur for the RSG component. The shortest orbital periods of known companions of RSGs have periods of at least several years \citep[e.g.,][]{patrick24a}. From Kepler's third law, a period of the order of several weeks would result in an orbit of several $\sim 10 \: \rm R_{\odot}$, i.e., inside the RSG. This further validates our sample of promising candidates, as all are known or suspected RSG+B binaries. 
This scenario is more likely, as there are many short-period companions of B-type stars known in the literature. For example, several papers report on eclipses and variability detected with TESS, such as for B stars \citep[e.g.,][]{sharma22}, as well as for Be stars \citep[e.g.,][]{labadie22}, with the shortest periods of the order of days.
Therefore, the detected eclipses can likely be associated with the companions of RSGs. We test this hypothesis on the KQ Pup system, and discuss the eclipses of the other candidates briefly in Appendix \ref{appendix:tess}. We also took new PLATOSpec spectra for the promising candidates to verify whether they display binary spectral properties, see Fig.~\ref{fig:platospec}.

We note that four targets from our sample were part of the TESS Eclipsing Binary stars catalogue \citep{prsa22}, which consists of several thousand candidates identified with an automatic algorithm. KQ Pup was also part of the list, but with incorrect ephemeris and period. For the other three stars, HD 300933, V340 Sge, and V349 Car, there is a period in their catalog that is similar. Furthermore, two of our candidates, AZ Cas and V1092 Cen, are also included in \citet{dorn20}, who studied the stochastic variability in selected evolved supergiants with TESS, but did not discuss any eclipses.

\begin{table*}[htbp]
        \centering
        \caption{List of promising hierarchical triple RSG candidates from TESS.} 
        \label{table:table_candidates} 
        \setlength{\extrarowheight}{2pt}
        \begin{tabular}{ccccccc}

\multicolumn{7}{c}{Confirmed TESS candidates} \\ 
\hline \hline 
\text{Star} & \text{Spectral type} & \text{Period [d]} & \text{\ion{H}{i} em.} & \text{[\ion{Fe}{ii}] em.} & \text{Multi-core Na I}  & \text{Notes}  \\ 
\hline
KQ Pup & M2Iab+B0V & 17.26 & \checkmark & \checkmark & \checkmark & \text{VV Cep type} \\  
\hline 
\\ 
\multicolumn{7}{c}{Promising TESS candidates} \\ 
\hline \hline 
\text{Star} & \text{Spectral type}  &  \text{Period [d]} & \text{\ion{H}{i} em.}  & \text{[\ion{Fe}{ii}] em.} & \text{Multi-core Na I}    &  \text{Notes}  \\ 
\hline
AZ Cas & K5Iab+B  & 1.25 & \checkmark & \checkmark  & \checkmark  & \text{VV Cep type, irregular ecl.} \\ 
HD 300933 & M3Iab+B2V & 2.91 & X & X & \checkmark & \text{binary RSG} \\ 
HD 303344 & M2Ib+B & 2.59 & \checkmark  & X  & \checkmark & \text{binary RSG} \\ 
FR Sct & M0+B & 3.53 & \checkmark & \checkmark & \checkmark & \text{VV Cep type} \\ 
V340 Sge & K4Ib(+A)\tablefootmark{*} & 1.63 & X & weak & broad & \text{single RSG?} \\  
\hline

        \end{tabular}
\tablefoot{The spectral binary signatures are listed based on our newly taken PLATOSpec spectra (Fig. \ref{fig:platospec}). The discarded candidates are listed in Table \ref{table:table_candidates_discarded}. 
\tablefoottext{*}{V340 Sge has a wide companion at 30 arcsec \citep{ccdm}, but its light variations cannot be separated with TESS. The eclipses appear to be present in the full PSF of the RSG.}}
\end{table*}

\subsection{Eclipses in KQ Pup and the discovery of KQ Pup Bb}
\begin{figure}
    \includegraphics[width=0.5\textwidth, keepaspectratio]{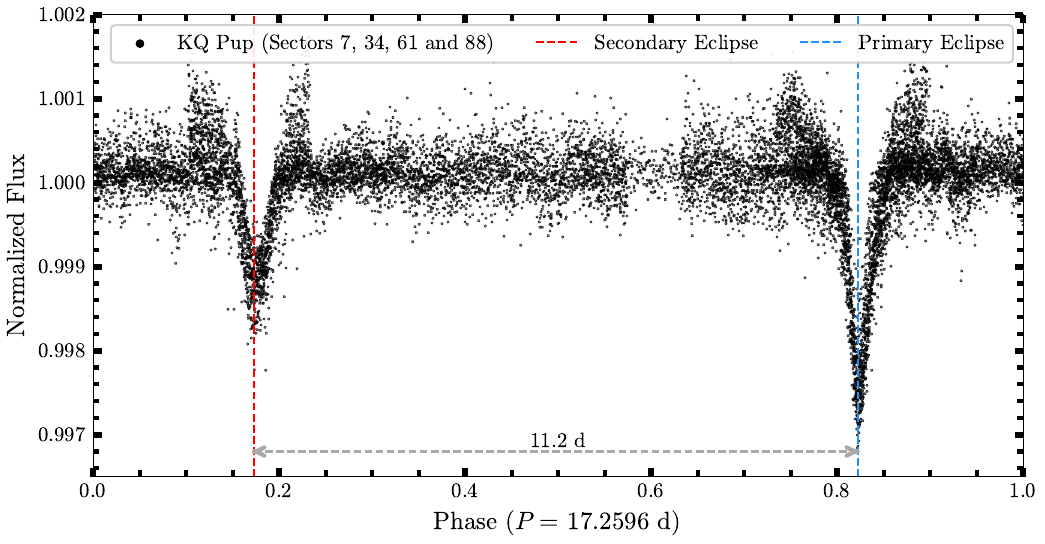}
    \caption{
    Cleaned and normalized TESS-SPOC data of KQ Pup (Sectors 7, 34, 61 and 88) phased with the period of eclipses $P_{\rm Ba+Bb}=17.2596 \: \rm d$, which corresponds to the Ba+Bb orbit and shows an eccentric orbit. Brightenings are artifacts from detrending some sectors. Each sector is shown separately in Fig. \ref{fig:kq_pup_tess_raw}. }
    \label{fig:kq_pup_tess}
\end{figure}

We chose the TESS data for the KQ Pup system for further investigation.   
This system exhibits clear primary and secondary eclipses in all four sectors covered with TESS between 2019--2025, with a period of about $P_{\rm Ba+Bb}=17.2596 \: \rm d$ and the secondary eclipse offset from the midpoint of the orbital period, which indicates that the related orbit is eccentric. In Fig.~\ref{fig:kq_pup_tess_raw}, we show the full TESS-SPOC pipeline data for KQ Pup, and in Fig.~\ref{fig:kq_pup_tess}, we show cleaned, phased, and normalized data. The TESS light curves were detrended using Chebyshev polynomials. Each sector was divided into smaller parts to better cover the trend influence and fitted by a polynomial of degree $n$ between 5 and 10. 

After normalizing the light curve and folding it with the orbital period, the eclipses can be seen clearly in Fig.~\ref{fig:kq_pup_tess}. Both eclipses are distinctly V--shaped, indicating that the eclipses are not central but partial or even grazing. The depths are also slightly different, which for a circular orbit would indicate small differences in stellar radii $R$ and/or effective temperature $T_{\rm eff}$ values. However, since the phase offset between the main and secondary minima indicates an eccentric orbit, the impact parameters ($b$) of the eclipses may also differ, if the orbit is not perfectly planar $(i<90^\circ)$ and/or if the semi-major axis is not perpendicular to parallel to the line of sight. In that case, breaking the degeneracies between $R$, $T_{\rm eff}$, and $b$ would require RV observations of the orbit, as well. We also note that the normalized light curves in Fig.~\ref{fig:kq_pup_tess} show brightenings, which are likely artifacts caused by residual variations the detrending did not remove completely, in order to preserve the shapes of the eclipses.

While Sectors 7 and 34 are dominated by the eclipses, all sectors show significant variations beyond them, and the amplitude of these additional variations increased significantly by Sectors 61 and 88. 
Namely, the last two sectors show additional semi-regular dips of about $\sim 10 \: \rm d$. These variations are reminiscent of dipper star light curves, where similar irregular or semi-regular variations are explained by occultations by a circumstellar or a circumbinary disk \citep[see, e.g.,][]{gillen14,Cody-2014,Roggero-2021}. Such variability is common in the UX Orionis class of young stellar objects, for example, for the AA Tau star \citep{bouvier13, mcginnis15}. In these systems, high-density inhomogeneities in the warped protostellar/planetary disk cause such longer light variability, for example, because of vertical structures due to accretion streams. In Sector 61, there is also a sudden increase in brightness following an eclipse, which could be related to accretion bursts to one of the components. To confirm the existence of an eclipsing pair and whether it hosts a disk, we focus on the KQ Pup system for the rest of the paper. After better constraining the properties of the KQ Pup system in Sects. \ref{chapter:asteroseismology}--\ref{sect:spectro-interf}, modeling of the orbit and eclipses of Ba+Bb is performed in Sect. \ref{chapter:BaBb}, while the properties of the disk are discussed in Sect. \ref{chapter:disk}.

\section{KQ Puppis system}
\label{chapter:kq_pup_syst}

\subsection{Literature parameters}
\label{chapter:kq_pup_literature}
The KQ Pup system was assumed to consist of an RSG component, KQ Pup A (M2Iab, HD 60414), and a B-type star, KQ Pup B (B0V, HD 60415), as classified by \citet{rossi92}. Based on the discovery of the third component in the KQ Pup system, we adjust the system configuration to A + (Ba+Bb), with A being the RSG. When the Ba+Bb components are considered jointly in the following analysis (e.g., center of mass), we refer to them as B or the hot component of the system. 
\citet{cowley65} analyzed RV measurements of KQ Pup from 1918-1964 and derived an eccentric orbit, with a period of $P_{\rm A+B} = 9752 \pm 85 \: \rm d$, eccentricity $e=0.46 \pm 0.03$ and argument of periastron $w=202.6 \pm 2.1^\circ$ (conjuctions close to periastron). 
In a related study, \citet{rossi92} expanded on these results and estimated a mass ratio $q_{\rm AB}= M_{\rm B}/{M_{\rm A}}  =0.8-1.2$ and a semi-major axis $a \sin(i)=13.6 \: \rm au $. For the hot companion, KQ~Pup~B, they found a spectral type of B0 with $T_{\rm eff} = 30\,000 \: \rm K$. In the most recent study on the KQ Pup system, \citet{gonzalez02} derived a revised period of $P_{\rm A+B} = 9500 \pm 50 \: \rm d$ ($\sim 26 \: \rm yr$). 

In these studies, a distance of $1400\pm200 \: \rm pc$ was assumed for the KQ Pup system (parallax $\pi \sim 0.71 \: \rm mas$). However, revised Hipparcos data give a parallax $\pi = 2.12 \: \rm mas$ \citep{hipparcos07}, while Gaia gives $\pi = 1.36 \pm 0.14 \: \rm mas$ \citep{gaia21}. In this work, we adopt a geometric Gaia distance of $\sim 779 \: (693 - 920)  \: \rm pc $ from \citet[][hereafter \citetalias{bailer21}]{bailer21}, and refine the distance of KQ~Pup based on that.

\subsection{Long-term photometric variability}
\label{chapter:asteroseismology}
\begin{figure*}[htbp]
    \includegraphics[width=0.33\textwidth, keepaspectratio]{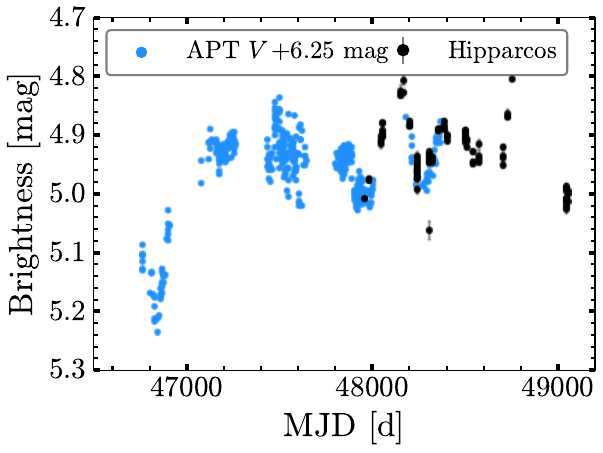}
    \includegraphics[width=0.66\textwidth, keepaspectratio]{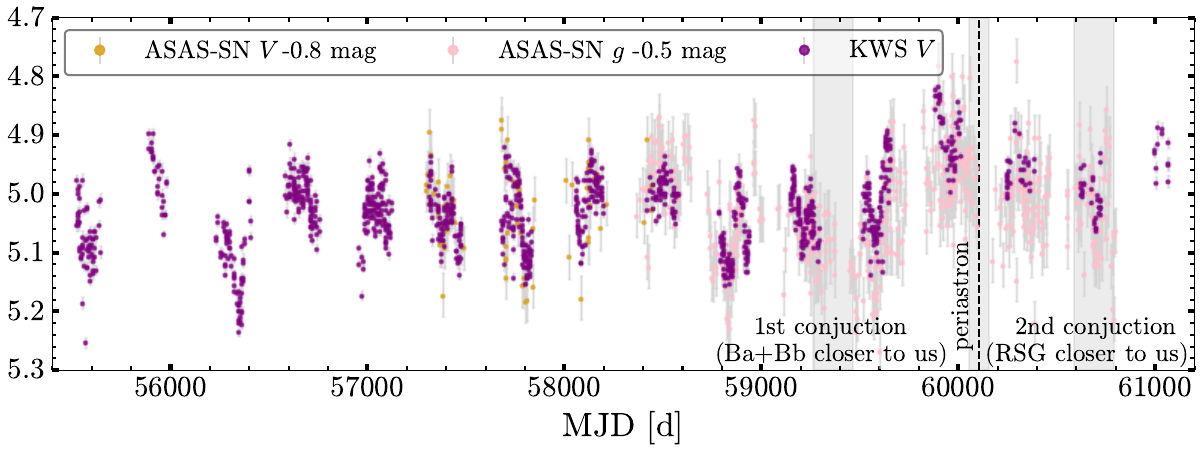}
    \caption{
    Photometric light curve of KQ Pup based on several archival sources: APT, Hipparcos, KWS, and ASAS-SN. Other data have been shifted to match Hipparcos and KWS. We also mark the approximate times of conjunctions and the periastron based on our new ephemeris from Sect. \ref{chapter:global_fit}. 
    }

    \label{fig:kq_pup_photo}
\end{figure*}

In Fig. \ref{fig:kq_pup_photo}, we show the photometric variability of KQ Pup during the last 40 years. 
The photometric variations are primarily caused by the variability of the RSG. 
The system is not known to undergo eclipses by the RSG, although it shows other variations due to the 26-yr orbit. \citet{gonzalez02} reported a significant decrease of far-UV flux near periastron. Meanwhile, we notice a clear periastron brightening in the optical region, i.e., by $V\sim 0.1 \: \rm mag$ during the last periastron (2023-2024). Periastron brightening events were reported for other interacting eccentric systems as well, including for the VV Cephei binary AZ Cas, and hypothesized to be due to accretion events or tidal heating \citep{koen24}. 

The beginning of the APT light curve in 1986-1987 shows a $\sim$0.3~mag dimming event that clearly exceeds the amplitude of the oscillations. A similar dimming is also repeated a full orbital period later, in 2013, as shown by KWS data. These events are at least 5 years away from any conjunction or potential eclipse, so geometric effects from the companion(s) can be ruled out. The dimming resembles the great dimming event of Betelgeuse \citep[e.g.,][]{montarg21}, which was caused by partial obscuration by a dense dust cloud; however, in this case, it could instead be caused by the binary interaction, such as a tidally induced outflow or trailing wake \citep[e.g.,][]{landri25, Dupree-2026}, or by the material surrounding the system.

A frequency analysis of the light curves revealed various low-amplitude periodicities ranging between 85--180\,d (0.06--0.14\,$\mu$Hz in frequencies), and possibly extending to even longer periods. However, aliasing from the annual gaps present in the data makes the existence of longer periods questionable, which hinders the accurate determination of the mode content. The variations of the star appear to be more similar to solar-like oscillations seen in luminous RGB stars and M giants, composed of multiple short-lifetime modes, rather than to coherent pulsations \citep{Banyai-2013,Yu-2020,Xiong-2025}. Incoherent oscillations in RSG stars have been detected before \citep{kiss06, joyce20}. However, mode identification for asteroseismic analysis can be difficult, with the aliasing present in the data.

We therefore decided to treat the star as an oscillating giant and attempted to measure global seismic parameters of the star. The power of asteroseismology lies in the fact that we can extrapolate our helioseismic knowledge and scale the solar values of various physical parameters, including $L$, $M$, $R$, $T_{\rm eff}$, as well as global asteroseismic parameters, to other stars \citep{kjeldsen-1995,Huber-2011}. To have a rough estimate on the possible seismic mass range of KQ~Pup~A (RSG), we first determined the frequency of maximum oscillation amplitude, or $\nu_{\rm max}$ parameter. As a first estimate, we can use simply the midpoint of the clearly identified periodicities in frequency space, which yields an approximate value of $\nu_{\rm max} = 0.12\pm0.02\,\mu{\rm Hz}$ (or $97^{+20}_{-14}$\,d in period). Unfortunately, at this point, we could not determine $\Delta \nu$, the frequency spacing between successive radial overtones, clearly.

Multiple scaling relations have been defined in the literature, but since we only have an estimate for $\nu_{\rm max}$, we chose the following scaling relation:
\begin{equation}
    \frac{M}{M_\odot} = \left(\frac{f_{\nu}\,\nu_{\rm max} }{\nu_{\rm max,\odot} }\right) \left(\frac{L}{L_\odot}\right)  \left(\frac{T_{\rm eff}}{T_{\rm eff,\odot}}\right)^{-7/2},
\end{equation}
\noindent where the solar values are $\nu_{\rm max,\odot} = 3090\pm30\,\mu{\rm Hz}$ and $T_{\rm eff,\odot}=5772\,{\rm K}$, following \citet{Huber-2011}. The $f_{\nu}$ parameter is a correction factor accounting for the structural differences between the target and the Sun. Here we set it to $f_{\nu}=1.05$, based on the results of \citet{Ash-2025}, who found that this type of scaling relation underestimates the masses at the lowest $\nu_{\rm max}$ values. However, this analysis was based on low-mass RGB stars: the correction factors for RSG stars have not been tested yet. We use $T_{\rm eff}=3660\pm170\,{\rm K}$ \citep{healy24} and two different values for the luminosity: $\log{L/{\rm L_\odot}}=4.55$ and $4.77$ based on \citet{healy24} and \citet{rossi92}, respectively.

This preliminary estimate gives us two seismic mass ranges for KQ Pup A: $M_{\text{low}L} = 7.1\pm1.7\,{\rm M_\odot}$ and $M_{\text{hi}L} = 11.8\pm2.8\,{\rm M_\odot}$ for the low and high luminosity values. These large uncertainties can be lowered with a more accurate determination of $\nu_{\rm max}$ and with stronger constraints on the luminosity of the RSG. A detailed asteroseismic analysis of KQ~Pup A will be the topic of a separate paper.

\subsection{Spectral variability}
\label{chapter:spectra}

A multi-wavelength spectral analysis is possible due to an abundance of archival IUE spectra covering most of the 26-year orbital period (1978-1995). Following the most recent periastron passage around January 2024 (based on ephemeris from \citealt{gonzalez02}), we have also obtained new high-resolution spectra in the optical with STELLA and PLATOSpec, and in the near-IR with VLTI-GRAVITY. As shown by \citet{rossi98}, but also recently by \citet{neugent18}, for RSG+B binaries, the spectral energy distribution at wavelengths longer than $\sim 4000 \: \rm \AA$ is typically dominated by the RSG. Nonetheless, for systems where the companion is embedded in the dense, cool shocked wind of the RSG, as in the case of VV Cep or systems studied by \citet{patrick25}, the photospheric lines of the hot companion may not be directly observed in the spectrum even below $\sim 4000 \: \rm \AA$. 

At longer wavelengths, for VV Cephei binaries, the contribution from the hot component can still be seen as double-peaked emission profiles in the Balmer series \citep{cowley69}, which is shown in Fig. \ref{fig:kq_pup_spectra_opt}. For KQ Pup, the ratios between the violet- and red-shifted emission peaks (V/R ratio) of the Balmer lines depend on the 26-year orbital phase \citep{cowley65, rossi92, rossi98}. The emission components become the weakest around apastron ($\phi_{\rm A+B} \sim 0.5$), and nearly fully disappear except for H$\alpha$. Meanwhile, the emission components become the strongest near periastron, showing strongly violet-shifted emission for all transitions until the Balmer break, appearing as inverse P Cygni profiles. 
The highest Balmer transitions are typically not detected in the spectra of single RSGs. We compare the profiles to new RSG+B binaries by \citet{neugent19}. Unlike in their case, here, the Balmer lines are narrow and do not show wide rotationally broadened absorption profiles typical of hot stars. That shows that the hot companion, KQ Pup B, is embedded in the dense CSM of the RSG, and therefore the photospheric lines of KQ Pup B are not directly seen in the spectra, similarly to VV Cep B \citep{bauer00}. The narrow absorption component of Balmer lines is similar to the typical RSG absorption forming in the extended CSM. 
Optical spectra also include forbidden emission lines (primarily [\ion{Fe}{ii}]). Other prominent photospheric lines of hot stars, \ion{He}{i} lines, are not detected in the optical spectrum, but below $\sim 4000 \: \rm \AA$, weak profiles may be associated with the hot component, most notably the rotationally broadened \ion{He}{i} $3819 \: \rm \AA$ line, as also noted by \citet{rossi92}. We also note that the \ion{Na}{i} doublet shows a~strong secondary component in all observations, with the sharp central component not moving and the wider typical RSG component showing shifts related to the orbital phase $\phi_{\rm A+B}$, see Fig. \ref{fig:kq_pup_spectra_nai}.

At shorter wavelengths, the hot companion dominates, and therefore UV spectra can be used to study more of its properties, especially related to its wind \citep{altamore82, altamore92, rossi92, muratorio92, bauer00, gonzalez03}. As analyzed in detail in the aforementioned studies on KQ Pup, there are several line systems present, superimposed on each other. First, in the far UV, there are broad absorptions of ionized resonance lines (e.g., \ion{C}{ii}, \ion{N}{v}), as well as other lines typical of the wind of hot stars. There are also absorptions of excited lines of doubly ionized \ion{Fe}{iii}. Second, in the near UV, there are many P Cygni profiles of singly ionized elements present (primarily \ion{Fe}{ii}), where the absorption part likely corresponds to the RSG wind as seen against the hot companion, and the emission part is likely produced in the cool wind of the RSG ionized by the radiation of the hot companion. In the UV, the appearance of the spectrum is strongly variable with the orbital phase $\phi_{\rm A+B}$, as shown in Fig. \ref{fig:kq_pup_spectra_uv}. Emission lines of permitted transitions, including those in P Cygni profiles, move to shorter wavelengths between the orbital phase $\phi_{\rm A+B} \sim 0.3-0.95$, while the absorption cores of \ion{Fe}{ii} P Cygni profiles move to longer wavelengths (resulting in inverse P Cygni appearance near the periastron). In the far UV, broad lines corresponding to \ion{Fe}{iii} near $\sim 1900 \: \rm \AA$ nearly disappear around apastron, while \ion{Al}{iii} becomes narrower. Meanwhile, prominent double-peaked UV emission lines (such as \ion{Mg}{ii} and \ion{Fe}{ii} multiplets at $\sim 2800 \: \rm \AA$ and $\sim 1787 \: \rm \AA$, respectively) also show a clear relation to the orbital phase. 

In addition, UV spectra of KQ Pup and VV Cep exhibit a shell spectrum (Fig. \ref{fig:kq_pup_spectra_uv_fit_full}), similar to those observed in the episodic Be shell star Pleione (28 Tau) and the B[e] star FS CMa \citep{bauer00}. Be stars are classified as shell stars when their spectra display, in addition to broad photospheric lines, narrow absorption features that are inconsistent with the stellar spectral type. These shell lines originate in the circumstellar disk and become visible when the Be star is viewed edge-on, such that the stellar photosphere is viewed through the equatorial disk. \citep[e.g.,][]{porter03}. 
For KQ Pup, the shell absorption spectrum (most notably $\ion{Fe}{ii}$ between $\sim 1550 -1750 \: \rm \AA$) originates in a region close to the hot component, as demonstrated by the disappearance of the shell absorptions during the eclipse of VV Cep \citep{bauer07}. As opposed to other narrow circumstellar absorptions forming in the RSG wind, the shell lines do not show such deep absorptions and may show broader variable profiles. The shell spectrum of KQ Pup becomes broader and deeper as the system approaches periastron (see Fig. \ref{fig:kq_pup_spectra_uv}). There are also some deep CSM or shell absorptions appearing only near periastron (see Fig. \ref{fig:kq_pup_spectra_uv_2}). Meanwhile, the broad resonance lines (e.g., \ion{Al}{iii}, \ion{Si}{iv}) assumed to correspond to the hot component did not disappear during the eclipse of VV Cep \citep{bauer07}, while a weak far-UV continuum also remained visible, suggesting these features are likely formed due to the scattered light from the hot component in the extended CSM. Another major insight learnt from the eclipse of VV Cep is that when the B companion is eclipsed by the RSG, all emission in higher Balmer lines disappeared, while H$\alpha$ emission significantly weakened \citep{pollmann18}. This confirmed that Balmer emissions form close to the companion.

Lastly, in Fig. \ref{fig:kq_pup_spectra_ir}, we show the near-infrared $K$-band spectrum obtained with VLTI-GRAVITY. Similar to the optical region, the entire spectrum is dominated by the RSG, showing typical RSG features, such as strong CO molecular bands and neutral metals. The only line clearly associated with the hot companion is the hydrogen Brackett gamma (Br$\gamma$) line at $2.167 \: \rm \mu m $. Br$\gamma$ was also detected for the likely binary WOH G64 by \citet{munoz24}, showing a similar profile as H$\alpha$.

\subsection{Model atmosphere fitting}
\label{chapter:powr}

\begin{figure*}[htbp]
    {\includegraphics[width=1\textwidth, keepaspectratio]{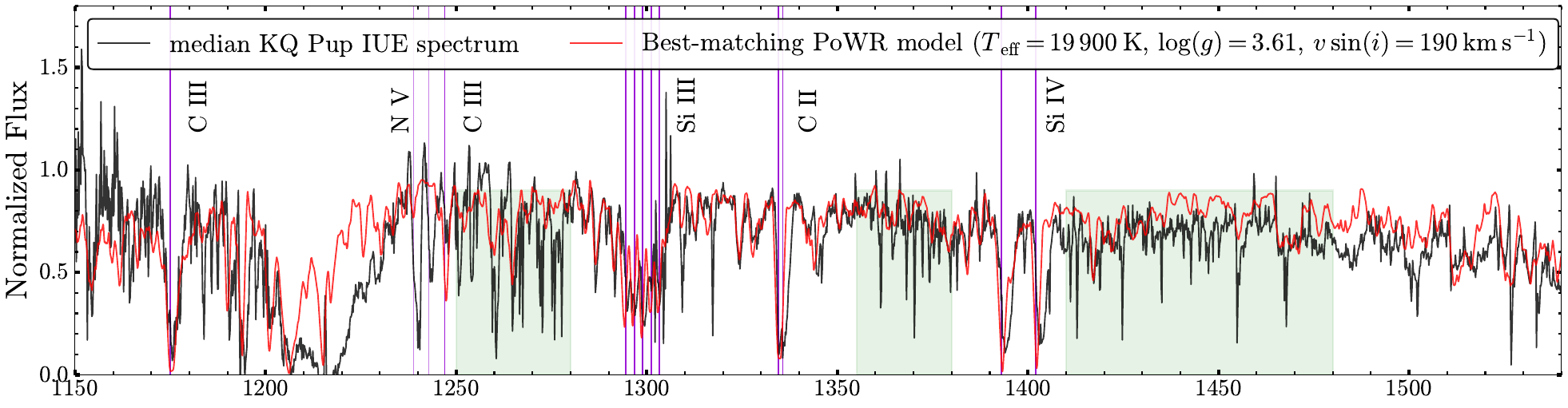}}
    {\includegraphics[width=1\textwidth, keepaspectratio]{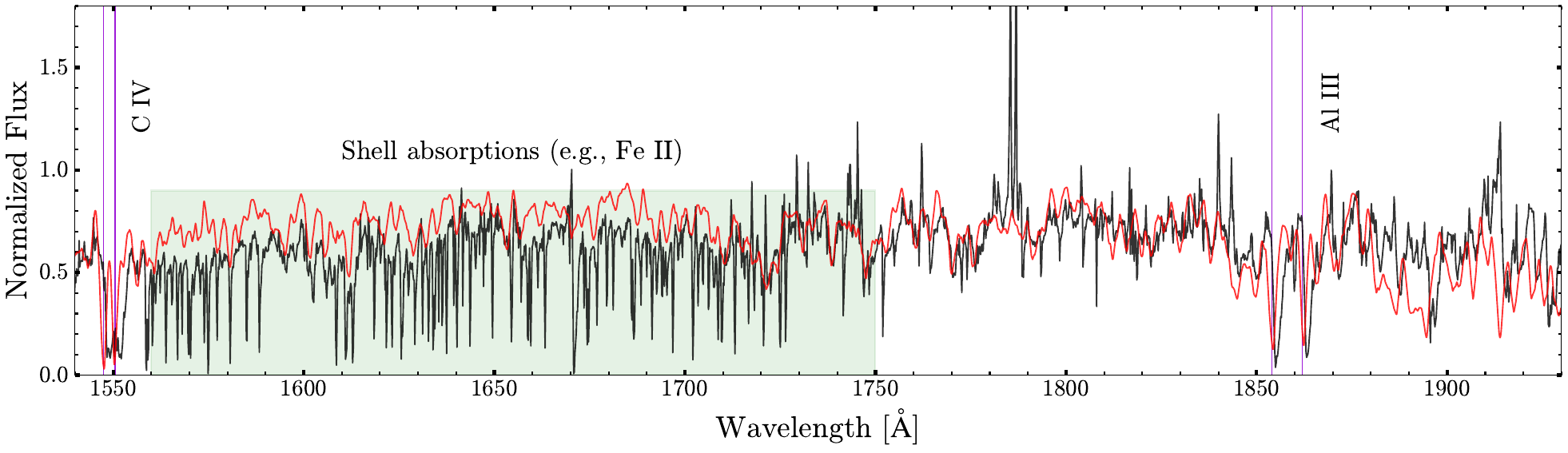}}
    \caption{Best fit of the PoWR model atmosphere to median IUE data of KQ Pup, showing the full SWP range, for $T_\mathrm{eff} = 19.9$\,kK. The main features of hot stars are shown (purple lines). The high-rotation model does not reproduce the narrow lines (green background), which come from shell absorption near the hot component, while some could also be CSM absorptions in the RSG wind. Detailed plots focused on specific lines are available in Fig. \ref{fig:kq_pup_spectra_uv_fit}.}

    \label{fig:kq_pup_spectra_uv_fit_full}
\end{figure*}

\begin{figure}[t]
    {\includegraphics[width=0.5\textwidth, keepaspectratio]{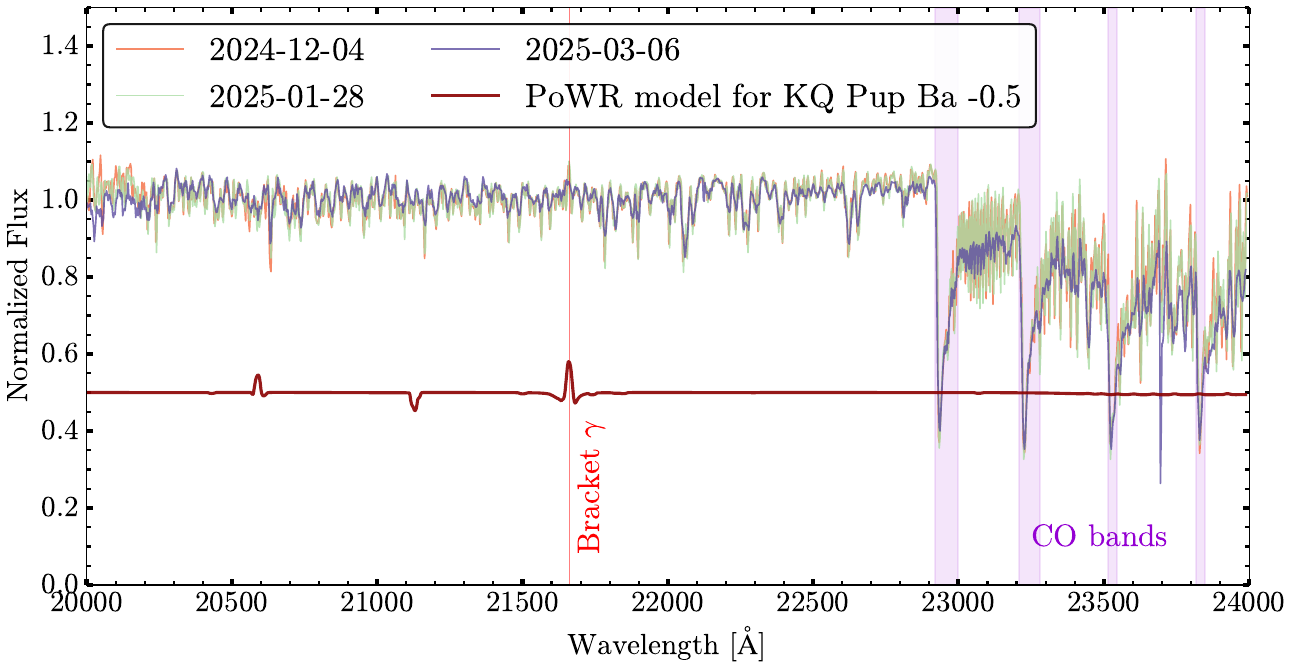}}
    \caption{   
    Near-IR spectra of KQ Pup from VLTI GRAVITY and our PoWR model for KQ Pup Ba. The spectra are shown relative to their continuum and have been normalized (the PoWR model has been vertically shifted). The observed spectra show prominent spectral features related to RSGs, such as the CO bands at  $\sim 2.29-2.4 \: \rm \mu m $. Based on the interferometric properties, the only feature clearly corresponding to the hot companions is the hydrogen Br$\gamma$ line at $ 2.167 \: \rm \mu m $. The only other two prominent lines of the hot stars in the region, \ion{He}{i} at $ 2.059 \: \rm \mu m $ and $ 2.113 \: \rm \mu m $, are not detected. The model for Ba assumed a single star; therefore, it does not reproduce the observed profiles, while the continuum contribution of Ba in the near-IR is negligible compared to that of the RSG. 
    }
    \label{fig:kq_pup_spectra_ir}
\end{figure}

To estimate the stellar parameters of the hot component, we compared the IUE spectra to synthetic spectra from atmosphere models computed with the Potsdam Wolf-Rayet (PoWR) model atmosphere code \citep{graefener02, sander15}. The PoWR model solves the radiative transfer simultaneously with the rate equations (i.e., in 
non-local thermodynamic equilibrium) in the comoving frame of a spherically symmetric, stationary expanding atmosphere (i.e., an outflowing wind).

As a starting point, we tried to reproduce the IUE spectra using a single stellar component, KQ Pup Ba. We selected a B-star model from the Galactic OB-Vd3 grid \citep{Hainich+2019, pauli26}, namely $T_* = 20.0$\,kK and $\log g = 3.6$. The choice of $T_*$ was motivated by the UV spectral signatures (
see the next paragraph). To improve the UV fit, we then calculate a tailored model, which is compared to the median IUE SWP spectra.
As the optical spectrum of KQ\,Pup is dominated by the RSG (KQ\,Pup\,A) and no clear photospheric lines of hot stars were found, we do not have clear diagnostics for the surface gravity ($\log g$) for KQ Pup Ba, which prevents the determination of the stellar mass. The chosen grid model has $\log g$ typical of B dwarfs of similar temperatures \citep[e.g.,][]{daflon03, ramachandran19}. 
By fitting the UV pseudo-continuum and weak rotationally broadened \ion{He}{i} lines below $4000 \: \rm \AA$ (e.g., \ion{He}{i} 3819$\AA$), we find a projected rotation velocity $v \sin i = 190 \pm 70$\,km\,s$^{-1}$ (see Fig. \ref{fig:kq_pup_spectra_uv_fit}).
To obtain an upper limit on the luminosity $L$ of KQ Pup Ba, we fit the synthetic model SED of the B-star to the flux-calibrated IUE spectra, considering a distance of 780\,pc from Gaia \citepalias{bailer21}, see Fig.\,\ref{fig:KQPUP_UV_SED}.
We estimate $\log L/\mathrm{L_\odot} \sim 3.95$ and a reddening of $E(B-V) = 0.30$ with $R_\mathrm{V} = 3.3$ assuming the extinction curve from \cite{Fitzpatrick+2019}. The reddening was obtained based on the 2175\,$\mathrm{\AA}$ absorption bump. The contribution of the RSG to the UV continuum is negligible compared to the hot companion \citep[e.g.,][]{bauer07}.

We noticed the presence of \ion{C}{ii} 1335, \ion{Si}{iv} 1400, \ion{C}{iv} 1550, and \ion{Al}{iii} 1855 in the UV spectra (see Figs. \ref{fig:kq_pup_spectra_uv_fit_full} and \ref{fig:kq_pup_spectra_uv_fit}). The simultaneous presence of these lines hints at $T_{\rm eff}$ of $\sim$$20$\,kK.
However, these diagnostic lines do not show up as typical P Cygni profiles, which are observational signatures of spherically symmetric outflows (see Fig.\,\ref{fig:kq_pup_spectra_uv_fit}). Instead, the profiles are broad absorptions (extending up to $\sim$200\,km\,s$^{-1}$), seemingly centered around the respective transition wavelengths and asymmetrically redshifted, which may suggest a different wind geometry. In the optical, we also notice a similar behavior in the Balmer lines with inverted P Cygni profiles (see FEROS spectra from 1997 in Fig. \ref{fig:kq_pup_spectra_opt}).
Such behavior can be explained by the geometry of the system: i.e., the wind of the B star interacts with the dense RSG wind before it develops to its terminal velocity. Winds in such systems of RSG + hot companion were discussed in, for example,  \citet{dupree87}.
The lack of clear P Cygni profiles throughout the UV also limits a precise determination of mass-loss rates ($\dot{M}$). The grid models by default assume that $\dot{M}$ follows the \cite{vink+2001} mass-loss recipe divided by 3, which yields $\log \dot M/(\rm M_{\odot} \: yr^{-1}) = -8.5$ in this case. 
For reproducibility, we list all the input stellar and wind parameters in Table\,\ref{table:table_results}. We note that in our model, Br$\gamma$ and other lines appear in emission (Fig. \ref{fig:kq_pup_spectra_ir}), which is a known effect of mass loss for some hot stars \citep[e.g.,][]{lenorzer04, najarro11}. Such lines could be observed if Ba were a single star. Instead, the observed emissions are dominated by the interaction in the system.

In Fig\,\ref{fig:kq_pup_spectra_uv_fit}, we also notice the presence of a strong \ion{N}{v} 1240 doublet, which is incompatible with the derived $T_{\rm eff}$, as this line would require a higher $T_{\rm eff}$.
In fact, this profile can only appear in early B and O stars in the presence of X-rays or extreme UV radiation as an extra source of ionization \citep[e.g.,][]{oskinova16, puebla16}.
As a~presumably mid-type B-star, a typical amount of X-ray luminosity cannot produce such a strong profile \citep[see][]{bernini-peron23,bernini-peron24}.
To account for the X-rays in our modeling, we included measurements for KQ Pup from the SRG/eROSITA survey ($\log L_\mathrm{X}/L \sim -6.5$ from \citealt{merloni24}). This helped as an extra source of ionization to strengthen \ion{C}{iv} 1550, which cannot be produced without X-rays in such a temperature.

We were able to fit features corresponding to the wind of hot stars in the IUE spectra with a single hot component, reproducing the majority of the spectral features (Fig. \ref{fig:kq_pup_spectra_uv_fit_full}), apart from \ion{N}{v} 1240. This could imply stronger X-ray or UV radiation. It could as well be potentially explained by a higher $T_{\rm eff}$ than $20\,000$ for the Bb component, if it were a very hot stripped star \citep[e.g.,][]{ramachadran24, mullerhorn26}. However, we have not detected any other typical signatures of hot stripped subdwarfs, i.e., \ion{He}{ii} lines, nor strong \ion{He}{i} lines. Therefore, in the absence of any other spectral features that would allow us to properly characterize the Bb component, we worked under the assumption that component Bb does not significantly contribute to the spectrum. We tested this hypothesis further in the following sections and took into account the possible contribution of Bb to the total $L$ based on the determined mass ranges (see Sect. \ref{chapter:evol}).

\subsection{Radial velocities}
\label{chapter:rv}

In this subsection, we focus on identifying line systems corresponding to KQ Pup A and B, which would allow us to determine RVs for both components and therefore also constrain the orbital parameters and mass ratio. The optical region of the spectrum is dominated by the RSG. Therefore, we determine the RV shifts of KQ Pup A using order-by-order cross-correlation of the optical spectra from STELLA (2024-2026) with a MARCS model\footnote{We downloaded a template with $ T_{ \rm eff} = 3600 \: \rm K$, $ \log{g} = -0.5$, $ v_{\rm microturb} = 2 \: \rm km \, s ^{-1}$ from \url{https://marcs.oreme.org}} \citep{gustaf08} close in parameters to KQ Pup A, using the full wavelength range, but excluding Balmer regions.
For KQ Pup B, we study its approximate RV using the lower Balmer emission lines, H$\alpha$ and H$\beta$, which have the strongest double-peaked emission features and the best signal-to-noise ratio compared to higher Balmer emission lines in the STELLA data. For VV Cep, the Balmer emission was used to determine the orbital velocity of the hot component as well \citep{hutch71, wright77}. When viewed (nearly) edge-on, a double-peaked emission profile is a classic signature of a star with a rotating disk \citep[e.g.,][]{rivinius24}, such as for Be shell stars. 
However, here, the Balmer profile is contaminated by the deep central absorption forming in the extended CSM of the RSG. Therefore, only the emission component can be clearly associated with the hot component, while part of the profile is missing due to the RSG absorption. To estimate the RVs of the emission centroid, we used the mirroring method \citep[e.g.,][]{arcos18}, i.e., comparing direct and flipped Balmer emission profiles. 
This method is also commonly used for Be stars \citep[e.g.,][]{wolf21, harmanec25}. For the determined RVs, we assumed that the Balmer emission lines form in the disk and therefore approximately track the center-of-mass of the Ba+Bb pair rather than specific components.

\begin{figure*}[htbp]
    \subfloat[]{\includegraphics[width=0.485\textwidth, keepaspectratio]{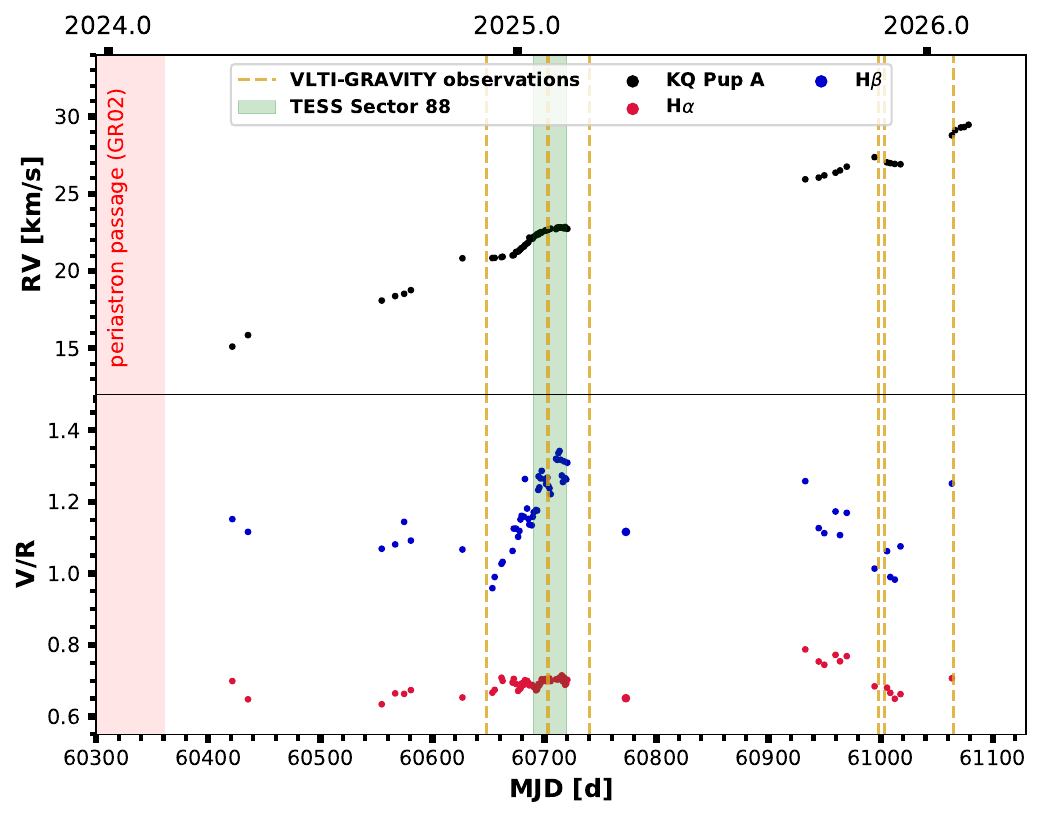} \label{fig:kq_pup_spectra_rv_1} }
    \subfloat[]{\includegraphics[width=0.5\textwidth, keepaspectratio]{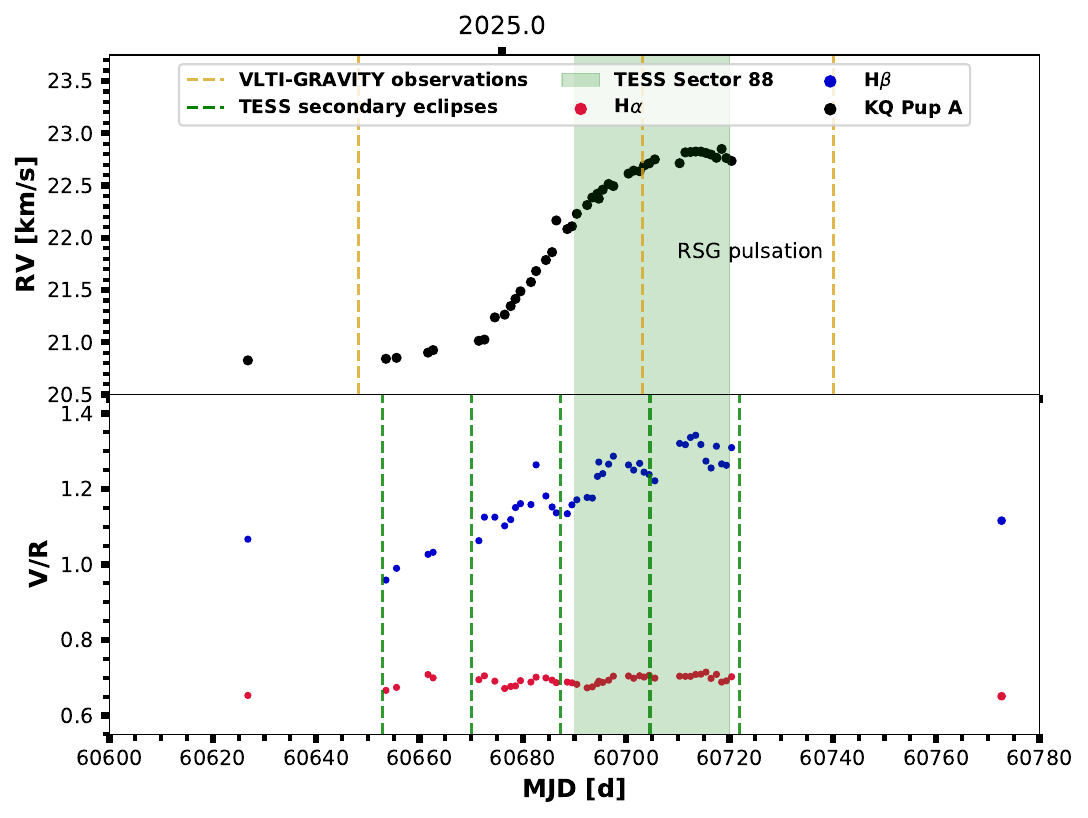} \label{fig:kq_pup_spectra_rv_2}  }
    \caption{
    {\em Left panel:}
    Radial velocities of KQ Pup A and V/R variations of the Balmer emission lines during the periastron epoch covered by STELLA (2024-2026). The vertical lines show the dates of the VLTI-GRAVITY observations. The shaded areas show the time of periastron passage (red) and TESS sector 88 (green).
    {\em Right panel:}
    Same as the left panel but zoomed in to the beginning of 2025. There are $\sim 17 \: \rm d$ period variations in H$\beta$, likely due to the Ba+Bb orbit, while strong longer trends are also present. 
    The PLATOspec spectrum taken in April 2025 (near MJD $ \sim 60\,780 \: \rm d$) is also included.}

    \label{fig:kq_pup_spectra_rv}
\end{figure*}

In Fig. \ref{fig:kq_pup_spectra_rv}, we show the obtained high-cadence STELLA RV curve. We can see that the velocity of KQ Pup A is increasing, following the periastron around January 2024 \citep{gonzalez02}. The overall velocity of KQ Pup B (Balmer emissions) is decreasing, verifying that the emissions approximately track the hot companions. However, during the post-periastron epoch covered by STELLA, the Balmer emission lines show variable asymmetric profiles and large velocity shifts. We found that these strong trends and asymmetric profiles are likely caused by the geometrical effects due to the proximity to the RSG; therefore, the determined RVs at this epoch may not be suitable for determining the orbital velocity of Ba+Bb. We further discuss the complicated variability of the Balmer lines near periastron in Sect. \ref{chapter:disk}. Therefore, from the available optical spectra, we use only the ESPaDOnS spectra for determining the orbital velocity of the Ba+Bb pair, as the spectra are taken at apastron, $\phi_{\rm A+B}\sim0.5$ (November-December 2010), and show symmetric H$\alpha$ profiles ($V/R \approx 1$), free from any interaction with the RSG.

Next, we also determined RVs from the IUE spectra (1978-1995). We checked the wavelength calibration of the SWP observations by performing an interstellar medium (ISM) line alignment and removed these lines from the spectra. A co-added average spectrum was constructed as the template spectrum, and we then cross-correlated each observed spectrum with the template to obtain the cross-correlation functions (CCFs), using the full wavelength range, except for the spectrum edges.
The RVs were then determined from the peak position of the calculated CCFs. The SWP region is dominated by the broad wind and narrow shell absorption lines, both associated with the CSM of hot companions. Therefore, the determined RVs should track KQ Pup Ba+Bb. In the IUE LWP/R region, which is dominated by strong emission lines forming in the ionized wind of the cool RSG, the determined RV appears to track KQ Pup A. Therefore, the emission lines likely form near the RSG. This corresponds to the shift of absorption and emission components described in Sect. \ref{chapter:spectra}, and is in agreement with the expected orbital shift of both components between $\phi_{\rm A+B} \sim 0.3-0.95$ \citep{cowley65}. The RVs determined from IUE spectra are used for the orbital fit (see Sect. \ref{chapter:global_fit}).

\begin{figure*}[htbp]
    \includegraphics[width=1\textwidth, keepaspectratio]{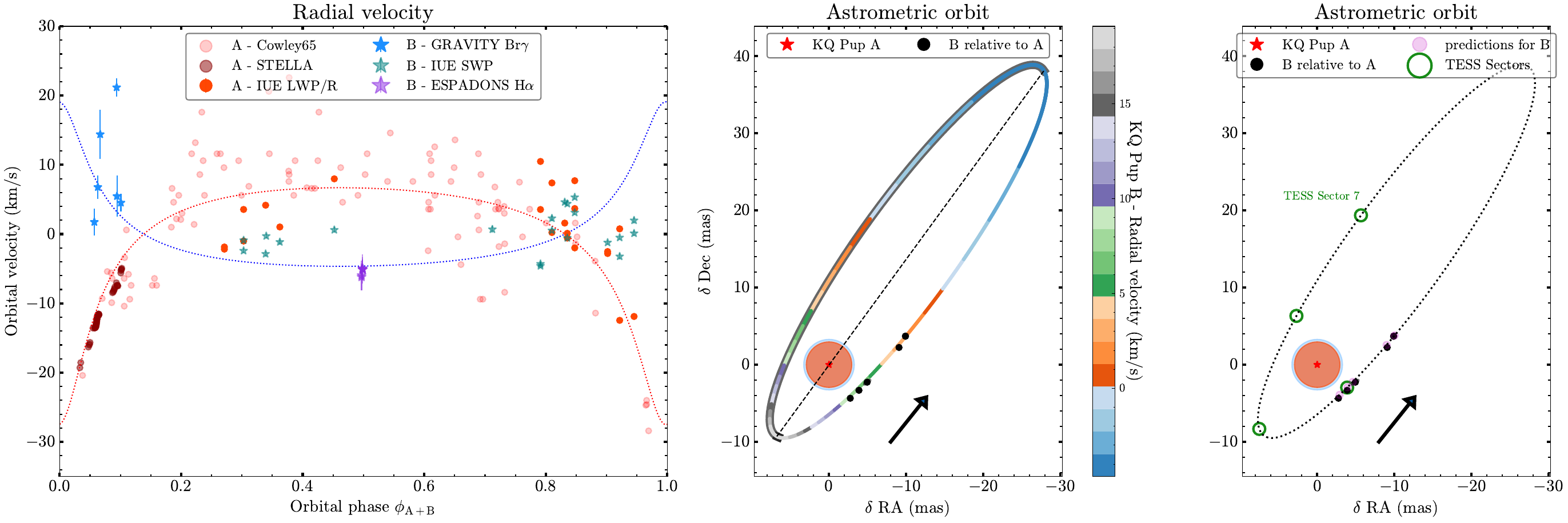} \label{fig:pmoired_0}
    \caption{Orbital solution for the KQ Pup system based on relative astrometry with VLTI-GRAVITY and the RVs.
    {\em Left panel:}
    Phased RVs, including the archival data.
    {\em Middle panel:}
    Fitted orbit of KQ Pup B using the Br$\gamma$ feature. The colors are based on the RV for KQ Pup B in the first plot, while the shaded part of the orbit shows when Ba+Bb are closer to us than A. 
    {\em Right panel:}
    Same as the middle panel but showing only the measured positions of B relative to A, including the direction of motion, as well as approximate locations of the TESS sectors. The size of KQ Pup A in the plots corresponds to its average measured angular diameter of $\theta_{\rm A}\sim 5.9 \: \rm mas$ (see Appendix \ref{chapter:vlti}).
    }
    \label{fig:pmoired}
\end{figure*}

We also further investigated the available spectra in search of spectral features of KQ Pup Bb. However, as discussed in previous sections, the hot companions in the KQ Pup system are embedded in the dense wind of the RSG. Furthermore, as is evidenced by the strong shell spectrum in the UV and also the possible occultations by the disk in TESS, the hot companions are also embedded in a dense, edge-on disk (see discussion on the disk properties in Sect. \ref{chapter:disk}). Therefore, apart from a weak signature of rotationally broadened \ion{He}{I} at $3819 \: \rm \AA$, we have not detected any photospheric lines of the hot components. The RVs determined from the shell spectrum (IUE SWP) would also not be expected to be sensitive to RVs of the separate Ba and Bb components. Indeed,
the RVs determined from IUE (Fig. \ref{fig:pmoired}) do not show large departures from the expected orbital velocity \citep{cowley65}, up to $ \pm 10 \: \rm km s^{-1}$. Likewise, no secondary peak was found in the CCFs (see Fig. \ref{fig:ccfs}), even when different regions in the IUE spectra were cross-correlated separately. Therefore, at the moment, it appears not possible to separate and further characterize KQ Pup Ba and Bb based on RV shifts, as their photospheric lines are absent and only lines forming in their disk and CSM are observed.

\subsection{Astrometric orbit}
\label{chapter:astrometry}

\begin{figure*}[htbp]
    \includegraphics[width=1\textwidth, keepaspectratio]{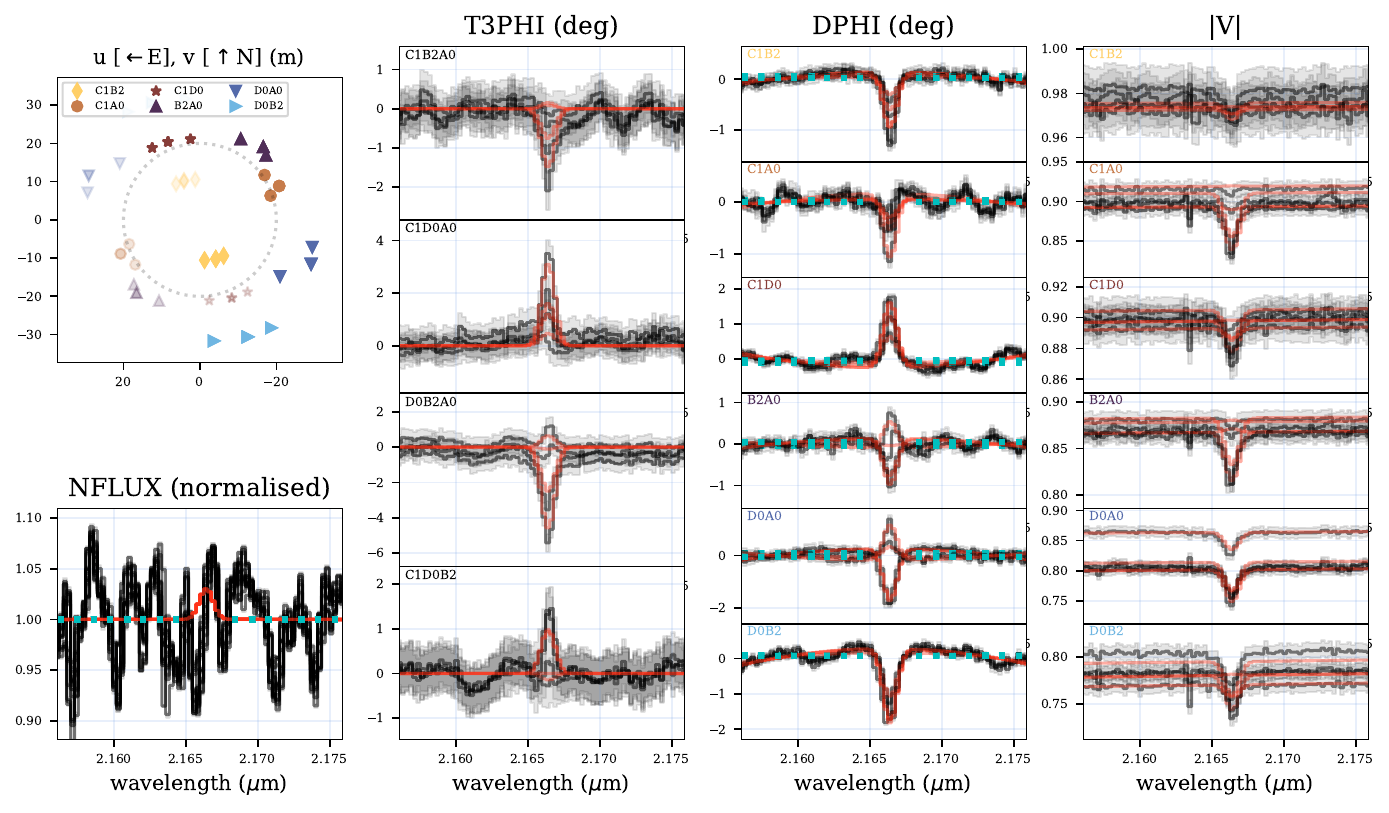}
    \caption{PMOIRED fit of the hydrogen Br$\gamma$ line for the five VLTI epochs (SMALL configuration). The red line shows the best fit in each epoch. The $uv$ coverage is shown in the upper-left corner, and the flux is in the bottom-left corner. In the remaining plots, the three main interferometric observables are shown: closure phase (T3PHI), differential phase (DPHI), and absolute visibility ($|V|$). The full spectral range is shown in Fig. \ref{fig:pmoired_obs_full}.}
    \label{fig:pmoired_obs}
\end{figure*}

We used the VLTI-GRAVITY dataset in order to determine the astrometric positions of the hot component KQ Pup B using the Python code Parametric Modeling of Optical InteRferomEtric Data\footnote{\url{https://github.com/amerand/PMOIRED}} \citep[PMOIRED,][]{merand22}. PMOIRED allows fitting interferometric data with multi-component geometric models, using least-squares minimization to obtain the best fit. 
In the near-IR interferometric VLTI dataset, the flux contribution of the hot component is too small to cause significant observable binary modulation; thus, the continuum would not be sufficient to determine the binary positions. Fortunately, relative astrometry was enabled by the detection of the hydrogen Br$\gamma$ line in our VLTI-GRAVITY data, which is the first time that this line could be used for an RSG binary system. This feature is very prominent in DPHI, suggesting a photocenter shift (see Fig. \ref{fig:pmoired_obs}). The line is not so well visible in the flux, likely due to the spectral resolution and contamination by other lines from the RSG.

For Be stars with circumstellar disks, typical double-peaked hydrogen spectral lines are usually S-shaped in DPHI and V-shaped in VIS \citep[e.g.,][]{meilland12}, with point sources or uniform disks used to represent the stellar components \citep[e.g.,][]{frost22, klement25}. However, in our interferometric data, the Br$\gamma$ line has the shape of a single-component unresolved emission at all baselines. For a similar system, $\epsilon$ Aur, its transiting disk was imaged during its eclipse using interferometry \citep{kloppenborg10, kloppenborg15}.  Therefore, we assume that the Br$\gamma$ feature traces the unresolved disk and/or hot components Ba and Bb, which we cannot distinguish between, given the spatial frequencies probed by our observations.

We proceeded to fit the Br$\gamma$ line in $|V|$, $\rm DPHI$ and $\rm T3PHI$ in PMOIRED using an unresolved point source model for KQ Pup B, while the rest of the spectrum is fitted with a uniform disk representing KQ Pup A. Based on the $a \sin(i)$ value from \citet{rossi92} and the distance of 779 pc from \citetalias{bailer21}, the semi-major axis of the system is $\sim 16 \: \rm mas$, while KQ Pup B should be closer to KQ Pup A than that, following the recent periastron. We were able to recover the position of KQ Pup B relative to KQ Pup A via the Br$\gamma$ line in all VLTI snapshots. The best and most stable position solution gives a separation of $\sim 5-10 \: \rm mas$ at different epochs. There are other local minima of the fit at larger separations, which would result in unrealistic masses for the system. We show an example of the data and its fit in Fig. \ref{fig:pmoired_obs}. We list the results of the fitting in Table \ref{table:table_vlti_results}, including the determined values of the angular diameter of KQ Pup A and the spectral parameters of Br$\gamma$. 

\subsection{Global A+B orbital fit and orbital parallax}
\label{chapter:global_fit}
We fit the recovered positions of KQ Pup B with an astrometric orbit in PMOIRED. We do the fitting in two ways. First, we fit only the VLTI-GRAVITY data (VLTI epochs), including the RVs from the spectro-interferometric data. For the fitting procedure, we fixed the well-known parameters by \citet{cowley65, gonzalez02} such as - $P_{\rm A+B}$, $K_{\rm A}$, $\gamma$, and the time of periastron, while for other parameters, we use their results as starting values. Second, we leave all parameters free and augment the fit with the archival and new RV data from other instruments (VLTI global+RV), allowing us to independently determine all parameters, including $P_{\rm A+B}$ and the time of periastron. Specifically, for KQ Pup A, we used RVs from STELLA, \citet{cowley65}, and IUE LW spectra, while for KQ Pup B, we used RVs from Br$\gamma$ (VLTI), IUE SWP spectra, and H$\alpha$ from ESPaDOnS taken at apastron. The list of RVs used for the orbital fit is listed in Table \ref{table:table_rvs}. To ensure that the orbital solution is not affected by intrinsic variability in our targets, we increase the RV errors for the fitting based on the observed variability in Fig. \ref{fig:kq_pup_spectra_rv_1}, i.e., we set the velocity error for KQ Pup A to $2.5 \: \rm km\,s^{-1}$ (RSG oscillations) and to $5 \: \rm km\,s^{-1}$ for KQ Pup B (variability in the disk and wind). Errors were determined using the bootstrapping method in PMOIRED.

In Table \ref{table:table_results}, we list the resulting orbital parameters for both fitting procedures, with the latter including the newly determined ephemeris based on $\sim$108 years of RV data. The new $P_{\rm A+B}=9534.2_{\pm6.7} \: \rm d$ lies within the uncertainties of \citet{gonzalez02}, although the time of periastron occurs about half a year earlier. With our updated ephemeris, the new phased RV measurements give the best agreement with the archival RV measurements (e.g., near-periastron RVs from 1918), compared to when using the literature ephemeris. Therefore, we conclude that our orbital solution gives the best agreement between all available datasets.
In Fig. \ref{fig:pmoired}, we show our best orbital fit of the recovered B positions (Br$\gamma$) and RVs. In both variations of the fitting, we obtain similar results, demonstrating that a few VLTI snapshots can give a reasonable result, although including the RVs from other sources covering the rest of the orbit significantly improves the fit. Our obtained parameters are not far from \citet{rossi92}, but in this work, we obtained for the first time the inclination $i$, the longitude of the ascending node $\Omega$, and most importantly, the projected semimajor axis $a \sin(i)$. We also found a high mass ratio of $q_{\rm AB} \sim 1.44$. Such a high mass ratio supports the binarity of KQ Pup B, as a single B-type companion cannot be both more massive and less evolved than the RSG primary (for a coeval evolution). We explore possible evolutionary scenarios in Sect. \ref{chapter:triple}. 

\begin{figure*}[t]
    \subfloat[]{\includegraphics[width=0.36\textwidth, keepaspectratio]{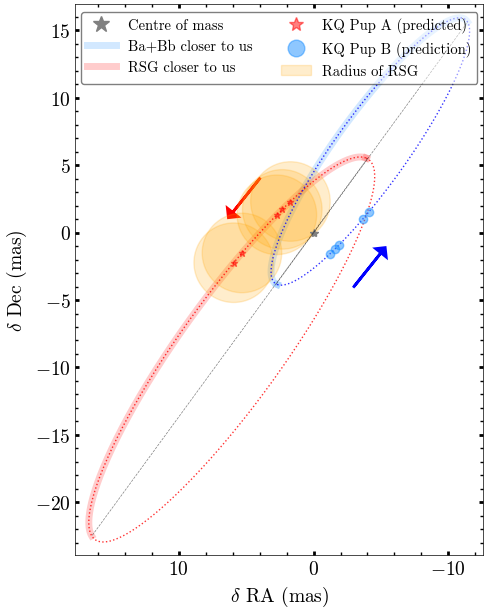} \label{fig:pmoired_2a} }
    \subfloat[]{\includegraphics[width=0.2775\textwidth, keepaspectratio]{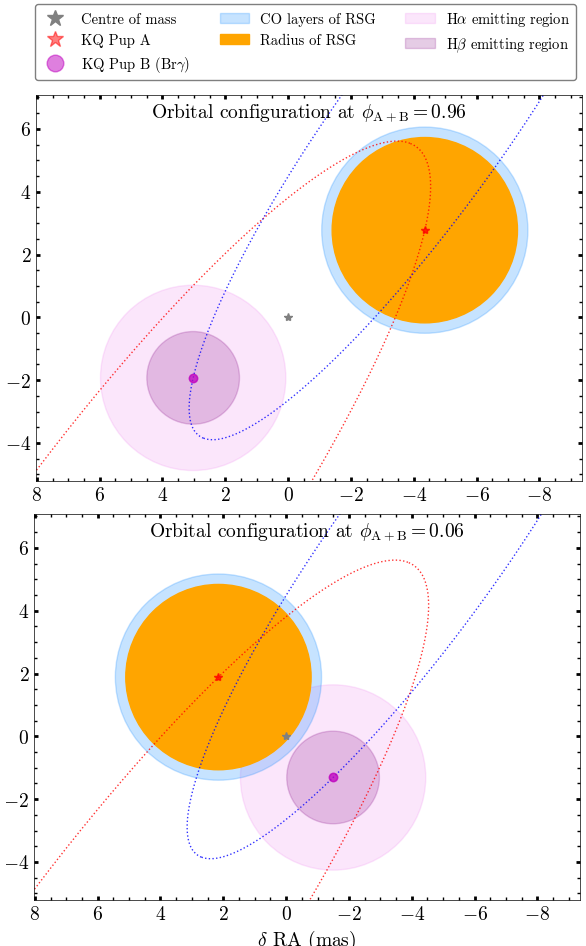} \label{fig:pmoired_2b}}
    \subfloat[]{\includegraphics[width=0.31\textwidth, keepaspectratio]{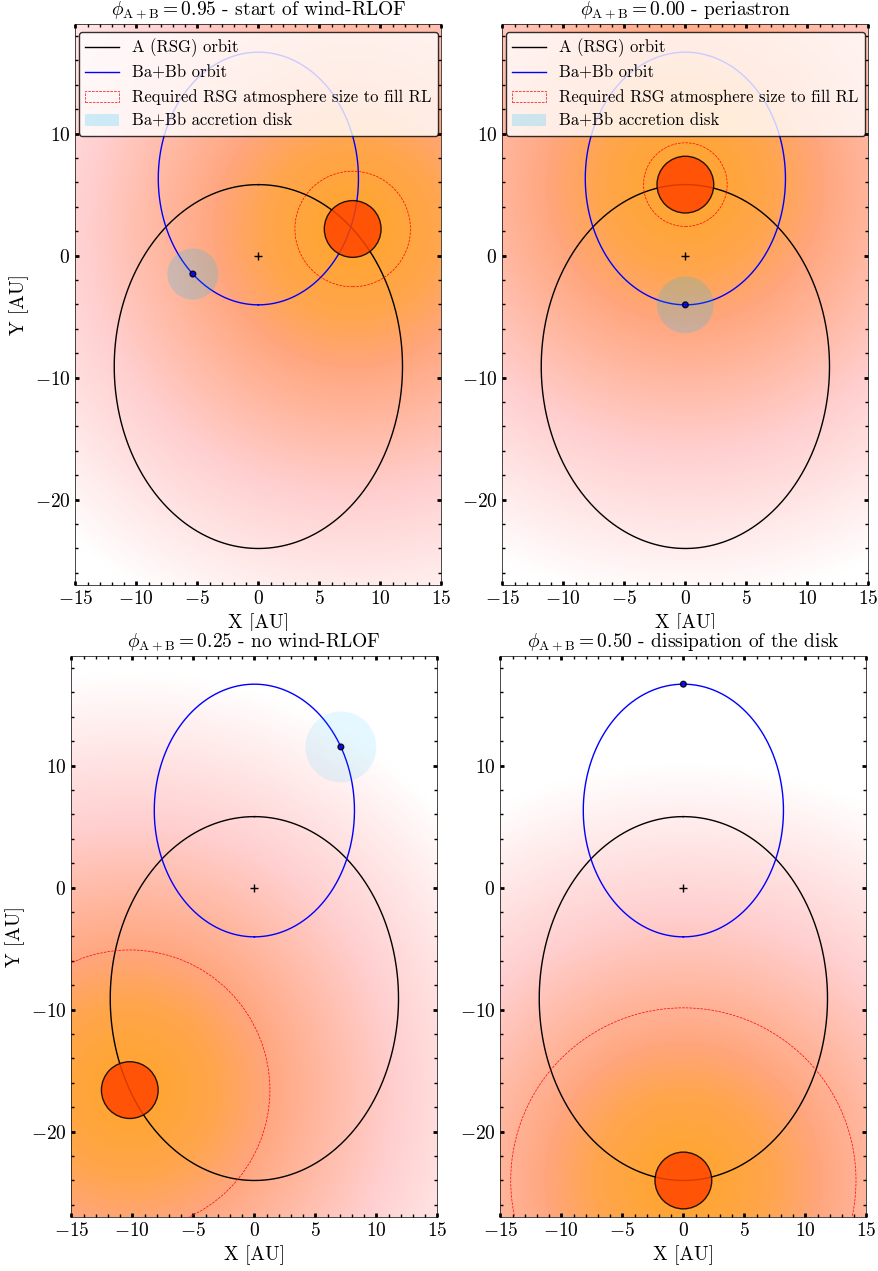} \label{fig:pmoired_2c}}
    \caption{Reconstructed orbit for the KQ Pup system.
    {\em Panel a):} 
    Predicted orbital motion of KQ Pup A and B relative to their center of mass during the time of our VLTI observations. See Appendix \ref{chapter:vlti} for the center-of-mass determination. 
    {\em Panel b):}
    Orbital configurations at selected $\phi_{\rm A+B}$, demonstrating the outer parts of the disk are eclipsed by the RSG after the periastron. The sizes of the disks are estimated and shown pole-on for better visualization.
    {\em Panel c):}
    Sketch of the system showing the physical distances between the stars and the wind-RLOF scenario. The blurred background represents the likely extended atmosphere of the RSG.
    }
    \label{fig:pmoired_2}
\end{figure*}

The astrometric solution itself would be sufficient to determine the semi-major axis $a_{\rm A+B}$, and thus the total mass of the system, if the distance is well-known. Nonetheless, combining this with the RV of both components, we can also find the orbital parallax of the KQ Pup system. From a global fit to all data, we find $\pi = 1.24^{+0.05}_{-0.04} \rm \: mas$, which is consistent to within the uncertainties with the parallax of $1.36 \pm 0.14 \: \rm mas$ obtained by \textit{Gaia} \citep{gaia21}, supporting the validity of our orbital fit. 
The binarity of the system was not detected directly by the mission. However, the RUWE (Renormalized Unit Weight Error), a goodness-of-fit metric of the astrometric fit of KQ Pup, is $\sim1.61$, which is above the $\approx1.4$ threshold of what is considered a good fit \citep{Castro-Girardi-2024}. While multiple effects can cause a large RUWE value, it is often caused by applying a single-star solution to an unresolved binary, and as such, the current \textit{Gaia} parallax is potentially affected. Significant discrepancies between Gaia parallaxes and orbital parallaxes from VLTI-GRAVITY were found even for stars with RUWE values of $1.0-1.4$ \citep{Gallenne-2023}. The periastron passage of the A+B pair was covered by the extended mission of \textit{Gaia}, and thus binarity could still be detected by the mission in later data releases (upcoming DR4 and later DR5). The reported proper motion of the system could also be affected by the A+B orbit. Indeed, Hipparcos \citep[$\phi_{\rm A+B} \sim 0.7-0.8$,][]{hipparcos07} gives a $\sim +1 \: \rm mas/yr$ larger proper motion in DEC than \textit{Gaia} DR3 \citep[$\phi_{\rm A+B} \sim 0.6-0.7$,][]{gaia21}, which could also support our orbital solution (see the RSG motion in Fig. \ref{fig:pmoired_2a}). Likewise, based on proper motion anomalies from Hipparcos and \textit{Gaia}, \citet{kervella22} identified possible companions for many systems, including for KQ Pup.

High-precision astrometry revealed the Ba+Bb pair as it leaves the periastron, moving by about $\sim 10 \: \rm mas$ toward the NW direction between December 2024 and January 2026. In relative astrometry, the absolute position of the primary star is not measured. However, using the mass ratio, we can also reconstruct the orbit of both stars relative to their center of mass, as shown in Fig. \ref{fig:pmoired_2a}. This shows that at the time of the VLTI observations, KQ Pup A (RSG) would be at a similar distance from the center of mass as KQ Pup Ba+Bb, but on the opposite side, moving toward the SE. 
Considering the determined average angular diameter of KQ Pup A of $\sim 5.89 \: \rm mas$, its stellar radius is $R_{\rm A} = 509^{+19}_{-17} \: \rm R_\odot $ (based on the determined $\pi$), so there would be a separation of only of several RSG radii between the components. Based on the RV difference up to $\sim 50 \: \rm km s^{-1}$ between the A and B components near periastron, Ba+Bb are moving supersonically through the wind of the RSG near periastron and likely also for a large part of the orbit (considering the typical sound velocity of $v_{\rm sound} = 5-15 \: \rm km s^{-1}$ in the wind, see Appendix \ref{appendix:mesa:atmosphere}).

\begin{table}[htbp]
        \centering
        \caption{Best-fit orbital parameters of KQ Pup A and Ba+Bb based on VLTI astrometry and archival RVs along with the properties derived from asteroseismology, evolutionary models and model atmospheres.} 
        \label{table:table_results} 
        \setlength{\extrarowheight}{2pt}
        \begin{tabular}{c|cc}
\multicolumn{3}{c}{\textbf{Astrometric solution for the KQ Pup system}} \\
\hline \hline 
\text{ } & \text{VLTI epochs} & \text{VLTI global} 
\\
\text{ }  & \text{only } & \text{+ RV}  
\\
\hline
$P \: \rm [d]$ & 9500 \citepalias{gonzalez02} & $9534.20_{\pm 6.70}$   \\ 
$\rm MJD_{\rm periastron} \: \rm [d]$ & 60310 \citepalias{gonzalez02}  & 60104.20$_{\pm 24}$  \\ 
$\gamma \: \rm [km \: s^{-1}] $ & 34.4 \citepalias{cowley65}   & $34.56_{\pm 0.22}$   \\ 
$K_{\rm A} \: \rm [km \: s^{-1}] $ & 17.1  \citepalias{cowley65} & - \\ 
\hline
$e$ &  $0.72_{\pm 0.03}$  & $0.61^{+0.01}_{- 0.01}$  \\ 
$\omega \: \rm [^\circ] $ & $178.03_{\pm 5.46}$  &  $184.3^{+2.0}_{- 2.3}$ \\ 
$\Omega \: \rm [^\circ] $ & $328.11_{\pm 3.10}$  &  $323.82^{+1.33}_{-0.66}$\\ 
$i \: \rm [^\circ] $ & $67.45_{\pm 2.18}$  &  $73.91^{+1.14}_{- 0.88}$\\ 
$a \sin(i) \: \rm [mas] $ &  $27.19_{\pm 1.39}$  &  $30.14^{+0.67}_{-0.97}$\\ 
$q$ &  -\tablefootmark{*} & $ 1.44^{+0.06}_{\pm 0.05}$ \\  
$  \pi \: \rm [mas]$ & -\tablefootmark{**} & $1.242^{+ 0.044}_{- 0.044}$ \\  
\hline
$a_{\rm A+B} \: \rm [au] $ &  $23.06^{+4.24}_{-2.88}$ & $25.26^{+1.15}_{-1.10}$\\ 
$M_{\rm A} \: \rm [M_{\odot}] $ & $7.38^{+4.77}_{-2.43}$ & $9.67^{+1.39}_{-1.22}$ \\ 
$M_{\rm Ba+Bb} \: \rm [M_{\odot}] $ & $10.63^{+6.90}_{-3.50}$ &  $ 13.95^{+1.99}_{-1.75}$ \\ 

        \hline
        \end{tabular}
        \tablefoot{ 
        \tablefoottext{*}{Not fitted; 
        the value from the global fit is used.}  
        \tablefoottext{**}{Not fitted; the \citetalias{bailer21} distance is used to calculate $a$.} } 
\\

\begin{tabular}{cc|cc}
\multicolumn{4}{c}{\textbf{Other results}} \\
\hline \hline 
\multicolumn{2}{c|}{Asteroseismology}  & \multicolumn{2}{c}{Model atmosphere} \\
\hline
\text{ } & \multicolumn{1}{c|}{KQ Pup A}  & \text{ } & \multicolumn{1}{c}{KQ Pup B} 

\\
\hline

$T_{\rm eff} \: \rm [K] $ & $3660_{\pm 170}$\tablefootmark{*} & $T_{\rm eff} \: \rm [K] $ & 19\,900   \\ 
$\log(L / \rm L_{\odot} )$ & 4.55-4.77\tablefootmark{*} & $\log(L / \rm L_{\odot} )$ &  3.95$_{\pm 0.4}$  \\ 
$M \: \rm [M_{\odot}]$ & 7.1-11.8  &  $T_{(\tau=20)} \: \rm [K]$ & 20\,000 \\  \cline{1-2}
\multicolumn{2}{c|}{Interferometry} &  $v \sin i \: \rm [km\, s^{-1}]$ & 190$^{+70}_{-70}$ \\ \cline{1-2}
$R_{\rm A} \: \rm [R_\odot]$ & $ 509^{+19}_{-17} $ & $v_\infty \: \rm [km\,s^{-1}]$ & 880\tablefootmark{**}  \\ 
$\log(L_{\rm A}/ \rm L_{\odot})$ & $4.62^{+0.08}_{-0.06}$ &  $v_\mathrm{turb} \: \rm [km\,s^{-1}]$ &  7  \\ \cline{1-2}
\multicolumn{2}{c|}{Evolutionary models}   & $\log (\dot{M} / (\rm\mathrm{M_\odot}\,\mathrm{yr}^{-1}))$ & -8.5  \\ \cline{1-2}
age [Myr]& 18.1-38.0  & $X_\mathrm{H}$ & 0.73  \\ 
$M_\mathrm{Ba}$ [$\rm M_\odot$] & $\sim$7.5-11.5 &  $\xi_\mathrm{phot}, \xi_\mathrm{max} \: \rm [km \,s^{-1}]$ & 7, 70 \\
$M_\mathrm{Bb}$ [$\rm M_\odot$] & $\sim$1.2-8.1 & $\log(g) \: \rm [cgs] $ & $3.61$\tablefootmark{***}\\ 
        \hline

        \end{tabular}

\tablefoot{  \tablefoottext{*}{From \citet{rossi92} and \citet{healy24}.}
\tablefoottext{**}{Cut at 200\,km\,s$^{-1}$.}
\tablefoottext{***}{Assumed; see Sect. \ref{chapter:powr}} }
\end{table}

We also investigated the $O-C$ values of the eclipse timings in the TESS data to investigate whether we could measure a light travel time due to the orbital motion of KQ Pup B. The $O-C$ residuals suggest that during periastron, the eclipses occur later than in the previous two TESS sectors (up to a difference of $\sim 1 \: \rm hr$). This is consistent with our orbital solution presented in Fig. \ref{fig:pmoired} and with the parameters listed in Table \ref{table:table_results}, because for roughly half of the orbit, closer to the periastron, Ba+Bb are accelerating away from us and 
increasing their heliocentric distance. The last two sectors also have more activity compared to the first two sectors (see Fig. \ref{fig:kq_pup_tess_raw}), which would agree with Ba+Bb getting closer to KQ Pup A. But further TESS observations along the orbit would be needed to build an $O-C$ curve which would fully constrain the light-time effect in the system.

\subsection{Spectro-interferometric signature of the Ba+Bb orbit}
\label{sect:spectro-interf}
We also investigated short-term variations of the recovered positions of KQ Pup Ba+Bb relative to A from VLTI. As we discussed, Br$\gamma$ emission is either associated with (one of) the hot components (Ba, Bb) or the disk. Even in the case of the disk, the positional variations could still correspond to KQ Pup Bb, as demonstrated in simulations by \citet{rubio25}, where the strongest Br$\gamma$ emission can arise from a part of the disk disturbed or accreted by the companion. The differences between the predicted orbit and the measured Br$\gamma$ position lie at the edge of the error bars, i.e., between $\sim 0.1-0.3 \: \rm mas$ in different VLTI observations. Assuming $M_{\rm Ba+Bb} \sim 14 \: \rm M_{\odot}$ as found in the previous subsection, the resulting semi-major axis would be $a_{\rm Ba+Bb}$ $\sim 0.3 \: \rm au$, i.e., $\sim 0.4 \: \rm mas$. 
The recovered positional variations of Br$\gamma$ could therefore be compatible with the existence of the Ba+Bb system, but it remains within the errors.
More VLTI-GRAVITY observations will be necessary to precisely measure the astrometric orbit of Ba+Bb, especially those at longer baselines, i.e., higher angular resolution. Such observations could also help to resolve the source of the Br$\gamma$ emission. We also emphasize that non-interacting RSG+B binaries in the VLTI-GRAVITY sample normally do not show the Br$\gamma$ emission feature, implying that in this case, we detect it because of an unusual increased flux contribution, likely from the accretion disk emission. 

\begin{figure}
    \centering
    \includegraphics[width= 1.0\columnwidth]{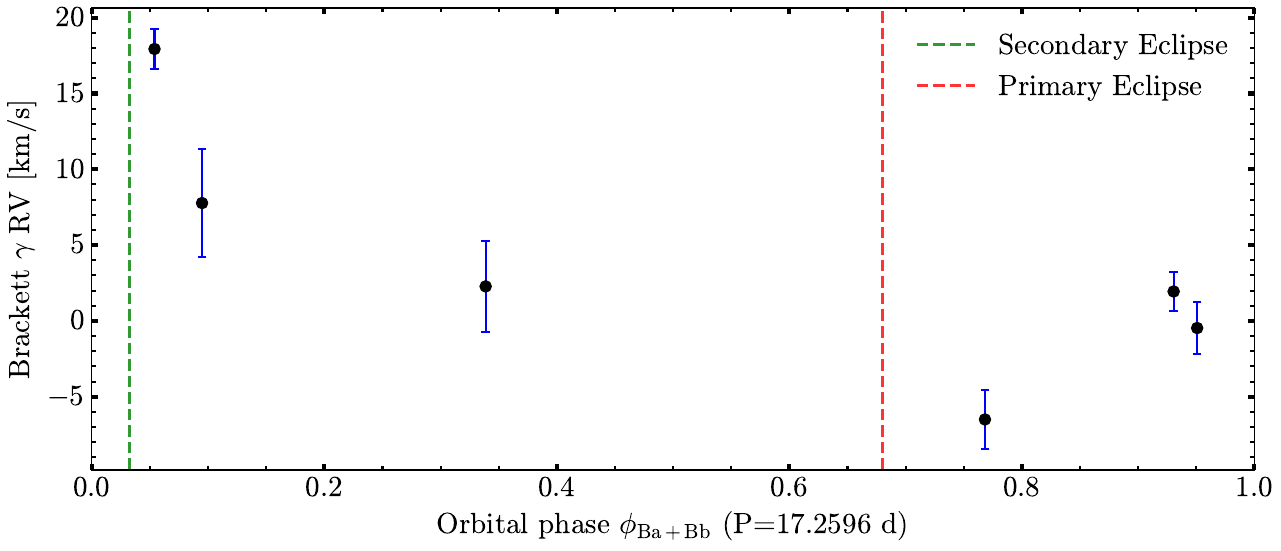}
    \caption{Brackett $\gamma$ RV phased with $P_{\rm Ba+Bb}$. The RV shifts due to the $A+B$ orbit were removed. Apart from five VLTI epochs taken with the SMALL configuration, which were used for the orbital fit, we also used an additional observation taken at the MEDIUM configuration (see Table \ref{table:table_vlti}).}
    \label{fig:br_gamma_rv}
\end{figure}

Due to the spectro-interferometric capabilities of VLTI-GRAVITY, we can also study the wavelength shift of Br$\gamma$ as determined from the fit to the interferometric observables, namely T3PHI, DPHI, and $|V|$ (Fig. \ref{fig:pmoired_obs}). This has an advantage over using the Balmer lines, as we are not affected by complicated spectral profiles, yielding more reliable RVs. Indeed, the RVs of Br$\gamma$ show a strong correlation with the period of $P_{\rm Ba+Bb}=17.2596 \: \rm d$, see Fig. \ref{fig:br_gamma_rv}. Brackett $\gamma$ forms over a more compact region of the disk than H$\alpha$ \citep[e.g.,][]{klement17}, and therefore it likely traces the inner region of the disk, which is more sensitive to the Ba+Bb orbit. This is in agreement with the variability of H$\beta$, which also traces a more compact region of the disk than H$\alpha$ and shows stronger V/R variability close to $\sim 17 \: \rm d$ (see Fig. \ref{fig:kq_pup_spectra_rv_2}).

\subsection{Orbital modeling of Ba+Bb}\label{chapter:BaBb}

We calculated some representative binary models with the PHOEBE code to simulate the Ba+Bb pair \citep{Prsa-2016,Conroy-2020}. We used the physical constraints presented in this work, plus the phase differences between the primary and secondary eclipses in the TESS light curve.   

Based on the luminosity estimates, we first calculated the brightnesses of the stars in the TESS passband to deblend the light curve from the contribution of KQ Pup A, and to estimate the true eclipse depths. Since the TESS passband is very similar to the \textit{Gaia} $G_{\rm RP}$ passband, we used the bolometric correction (BC) tables published for \textit{Gaia} by \citet{Jordi-2010}. 
The BC$_{\rm RP}$ values change quite rapidly around the physical parameters of both A and Ba, but we estimate that for A, BC$_{\rm RP}$ is between $-0.2$ and 0.5\,mag, while for Ba it is between $-2.3$ and $-2.9$\,mag. From these, we estimated the absolute magnitudes and the flux ratios, and found that the dilution factor (the ratio of the combined flux of the system relative to the flux level of Ba+Bb) is between 12--55 or 22-70, for $\log (L_{\rm A}/L_\odot) = 4.55$ or $ 4.77$, respectively. With these dilution factors, we estimate that the true eclipse depth of Ba+Bb is in the range of 0.02 to 0.20\,mag. 

\begin{figure*}
    \centering
    \includegraphics[width=0.495\linewidth]{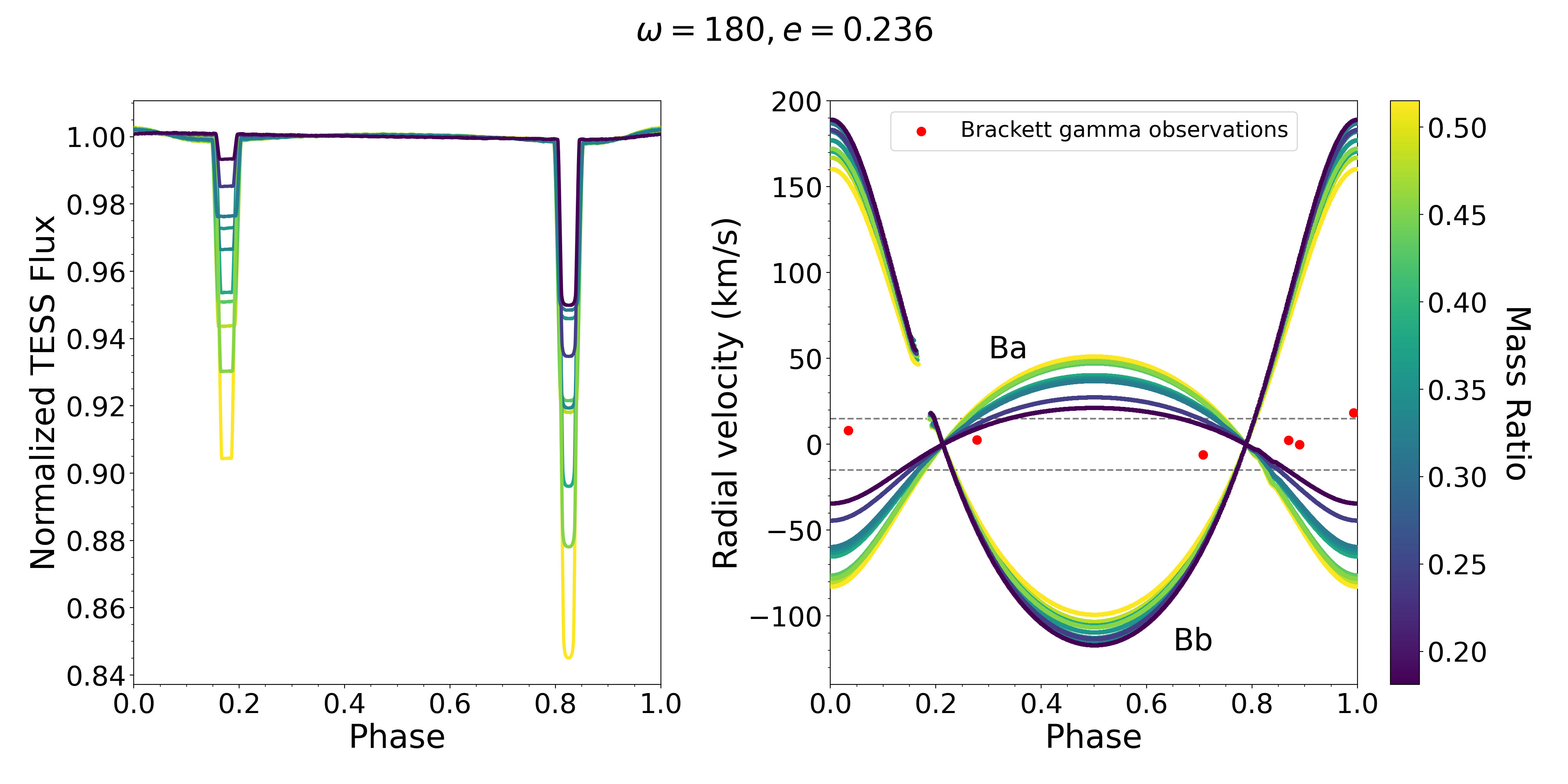}
    \includegraphics[width=0.495\linewidth]{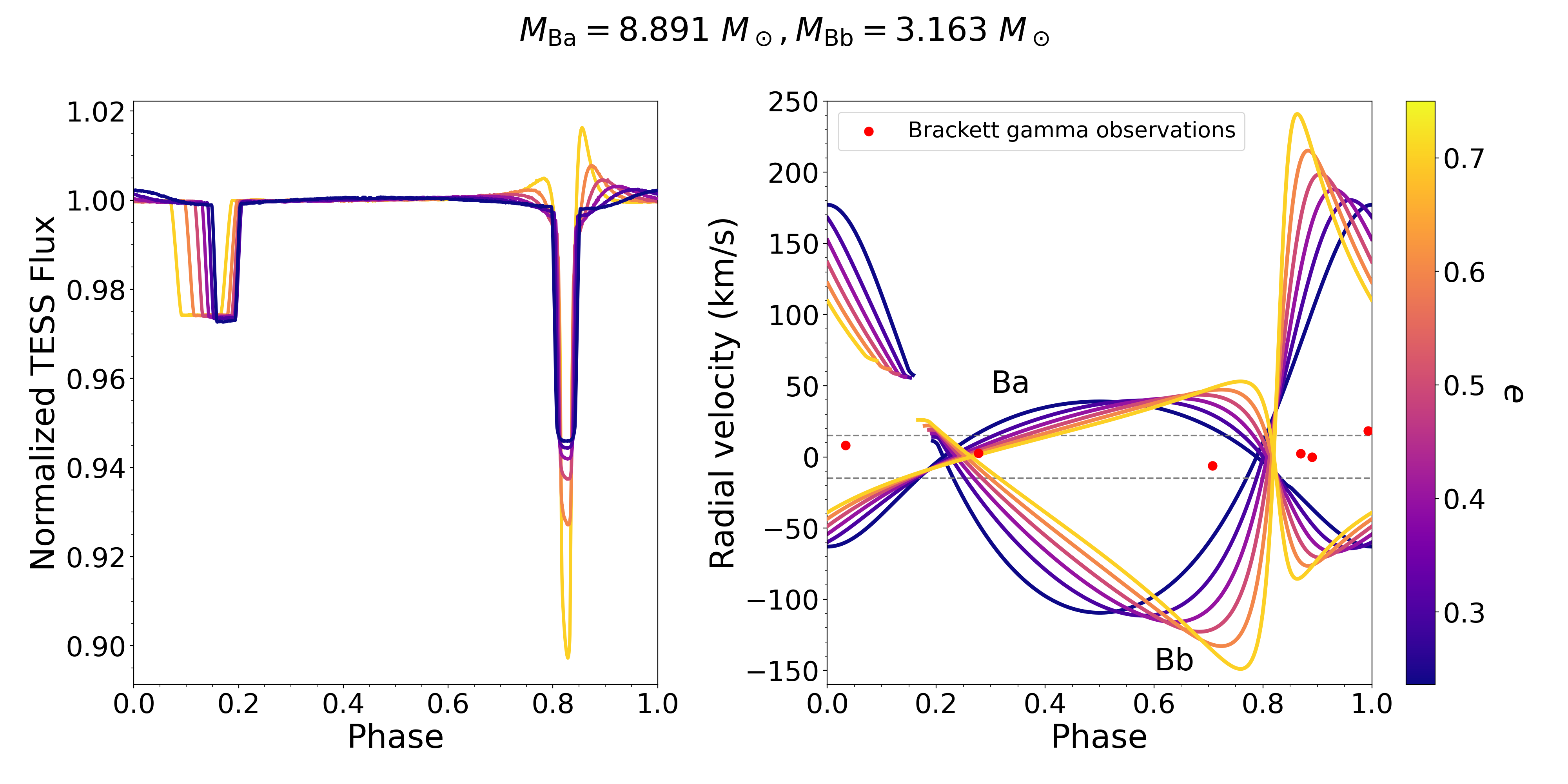}
    \caption{Preliminary PHOEBE models for the Ba+Bb pair. {\em Left panel:} Light curves in the TESS passband and RV curves of both components, calculated with $\omega=180^\circ$ and $e=0.236$. The color bar indicates the mass ratio of the two stars. The Br$\gamma$ RV data are shown in red, while the dashed gray lines indicate the observed $\pm 10\, \rm km\:s^{-1}$ velocity scatter.
    We note that since the Ba+Bb pair is embedded, the measured RVs (Fig. \ref{fig:pmoired}) do not show an RV amplitude similar to PHOEBE.
    {\em Right panels:} 
    Effect of using the different values of eccentricity ($e=0.236-0.7$) and argument of pericenter ($\omega=109.7-180^\circ$) displayed on the light curves in the TESS passband as well as the RVs, respectively. The color bar indicates the eccentricity of the Ba+Bb pair.}
    \label{fig:kq_pup_phoebemodel}
\end{figure*}

The offset of the secondary eclipse from the 0.5 phase between primary eclipses clearly indicates a significantly eccentric orbit for Ba+Bb. Assuming an orbit where the semimajor axis is perpendicular to the line of sight (defined as $\omega = 0^\circ$ or $180^\circ$ as the argument of pericenter in PHOEBE, depending on the direction of the orbit), eclipses happen $90^\circ$ apart from pericenter. This is the shortest and fastest arc a star can take around another star between eclipses, so the differences between the eclipse phases can constrain the minimum eccentricity. 

We therefore modeled the orbit in PHOEBE with $\omega=180^\circ$ first, and the eclipses being central ($i=90^\circ$). We chose four mass values for Ba based on the isochrone points near the location of Ba in Sect. \ref{chapter:evol}: 8.3, 8.9, 9.4, and 9.9\,$M_\odot$. We then selected a few test points for Bb, at luminosities $\log(L/L_\odot) = 2$, 2.5, and 3.0, as well as two low-mass test points at 1.5 and 2.0\,$M_\odot$.
We calculate the minimum eccentricity of the orbit to be $e_{\rm Ba+Bb,min} \ge 0.236$. If we increase the eccentricity in the model, the direction of the semimajor axis needs to be adjusted to match the eclipse times, by keeping the $e\cos\omega$ parameter constant. We thus mapped the allowed ranges for $\omega$ where $e\cos\omega = -0.236$. The critical value of $e=1.0$ is reached at $\omega = 103.65^\circ$ or $\omega = 256.35^\circ$. 

The preliminary PHOEBE models with the semimajor axis fixed perpendicular to the line of sight are presented on the left of Fig.~\ref{fig:kq_pup_phoebemodel}, where we colored the selected model combinations according to the Bb/Ba mass ratio.
The light curves in the first panel of Fig.~\ref{fig:kq_pup_phoebemodel} show flat-bottomed eclipses, which are clearly different from the observed, V-shaped eclipses in Fig.~\ref{fig:kq_pup_tess}. Furthermore, central eclipses, or eclipses with the same impact parameter, would make the differences between eclipse depths more pronounced than what we see in the observations. The TESS light curve therefore suggests that the impact parameters for the two eclipses may be different, and that both eclipses are partial or grazing, suggesting an inclination slightly offset from 90$^\circ$. We also observe that for higher-eccentricity models (the right panels of Fig.~\ref{fig:kq_pup_phoebemodel}), mutual heating at periastron would increase the brightness of the pair at the primary eclipse. Although this is difficult to test with the dipper-like variations present in the TESS data, the relatively quiescent Sector 7 does not show strong signs of such increases. Sector 61 does show a possible brightening, but that is more likely due to an accretion burst, as mutual heating would repeat in every orbit. Therefore, we rule out $e\sim0.7$ models, and limit the range of likely eccentricities to between $0.236 \le e \lesssim 0.5$. The growing phase difference between the secondary and primary eclipses (see the right panels of Fig.~\ref{fig:kq_pup_phoebemodel}) also suggests the exclusion of higher $e$. However, this upper limit for $e$ is only a qualitative limit, not a precise quantitative one.

We tested whether the RVs of Br$\gamma$ could be attributed to the Ba component, which we intended to fit while modeling the system. However, it is clear from the right panels of Fig.~\ref{fig:kq_pup_phoebemodel} that the models do not match the expected RV range, nor follow the trend of the observations. Likewise, older RV data for Ba+Bb (Fig.~\ref{fig:pmoired}) show a scatter of only $\lesssim 10$ km\,s$^{-1}$ and are taken over a long period of time and are thus affected by the light-time effect and potential dynamical effects from the A+B orbit as well as by the accumulation of uncertainties over hundreds of Ba+Bb orbits. This confirms that Ba+Bb are embedded as discussed in Sects. \ref{chapter:rv}, and therefore, we cannot directly measure the velocity amplitude of Ba or Bb. This also shows that the Br$\gamma$ emission is not coming from the photosphere, but is formed over the inner region of the disk instead, and cannot be used to constrain the orbital parameters of the Ba+Bb components. 

Without RVs for Ba and Bb, the orbital parameters cannot be constrained accurately enough for attempting to model the orbital inclination. Without that, we are unable to constrain the impact parameters of the eclipses, and we have to restrict the models to central eclipses. With this caveat, we find that the radius ratio of the two components should be between $0.22 < R_{\rm Bb}/R_{\rm Ba} < 0.40$. However, it is also possible that the eclipses are partial or grazing, allowing for radius ratios closer to unity.

\section{Mass transfer and evolutionary scenarios}
\label{chapter:triple}

\subsection{Nature of the disk and the Ba+Bb pair}
\label{chapter:disk}

The obtained mass ratio $q_{\rm AB} \sim 1.44$ suggests that either the primary hot component had a lower initial mass than the RSG ($M_{\rm A} \gtrsim M_{\rm Ba}$, with the remaining mass in Bb), as otherwise Ba would have already evolved to an RSG, or, there has been significant mass transfer in the system.
Based on the results described in Sect. \ref{chapter:kq_pup_syst}, the hot components Ba+Bb host a disk. We further explore the properties of the inner binary Ba+Bb and their disk to determine whether it could be fueled by accretion from the RSG, or an interaction between the Ba+Bb pair (e.g., Algol-type binary, \citealt{rensbergen21}). It could also be a decretion disk, as in the case of Be stars. Often, it is assumed that past mass transfer plays a key role in the formation of Be stars, as evidenced by Be+sdOB binaries, which host a fast-rotating Be star and a stripped helium star \citep[e.g.,][]{wang21, lechien25}. The last scenario appears less likely, as based on the determined $v \sin(i)$ of KQ Pup Ba (with $i\sim 90^\circ$), Ba is rotating at $v_{\rm rot}/v_{\rm crit}\sim 0.4_{\pm 0.15}$, i.e., likely below the typical $v_{\rm rot}$ range of Be stars \citep{nardini25} and not enough to form a decretion disk, while no spectral signatures of stripped helium stars were found (no \ion{He}{ii} lines and only weak \ion{He}{i} lines). 
The high eccentricity of the Ba+Bb pair ($0.236 \le e \lesssim 0.5$) would also make a past or ongoing mass transfer between Ba+Bb less likely, as the mass transfer typically results in a circular orbit, which is also the case for the majority of Be+sdOB binaries \citep[e.g.,][]{lechien25}. However, the eccentricity of the inner binary could be caused by the von Zeipel-Lidov-Kozai effect \citep{vonzeipel10, kozai62,lidov62,naoz13,ito19} as well as the angular-momentum exchanges with the wind accreted from KQ Pup A \citep[e.g.,][]{krynski25}. 

To study the properties of the disk, we investigated V/R ratios of Balmer lines in our high-cadence STELLA spectra (2024-2026). A companion passing through or near the disk can inflict density waves seen as smaller-scale V/R variations of a similar timescale as the orbital period \citep[e.g.,][]{okazaki91, rubio25}.
In Fig. \ref{fig:kq_pup_spectra_rv_2}, we can see that the V/R variations appear to be phased with the period of the TESS eclipses of $P_{\rm Ba+Bb}=17.2596 \: \rm d$, especially for the H$\beta$ variations. Indeed, a period analysis of the V/R ratios and peak separation of H$\alpha$ and H$\beta$ revealed several periods close to $\sim 17\ \rm \, d$. This shows the Balmer lines are sensitive to the Ba+Bb orbit, especially at the inner regions of the disk. However, there is also a very strong trend of about $\sim 90 \: \rm d$ (apart from a longer trend due to the 26-yr orbital period), which is quite similar in timescale to the RSG pulsation at the beginning of 2025. 

\begin{figure}
    \centering
    \includegraphics[width=1\columnwidth]{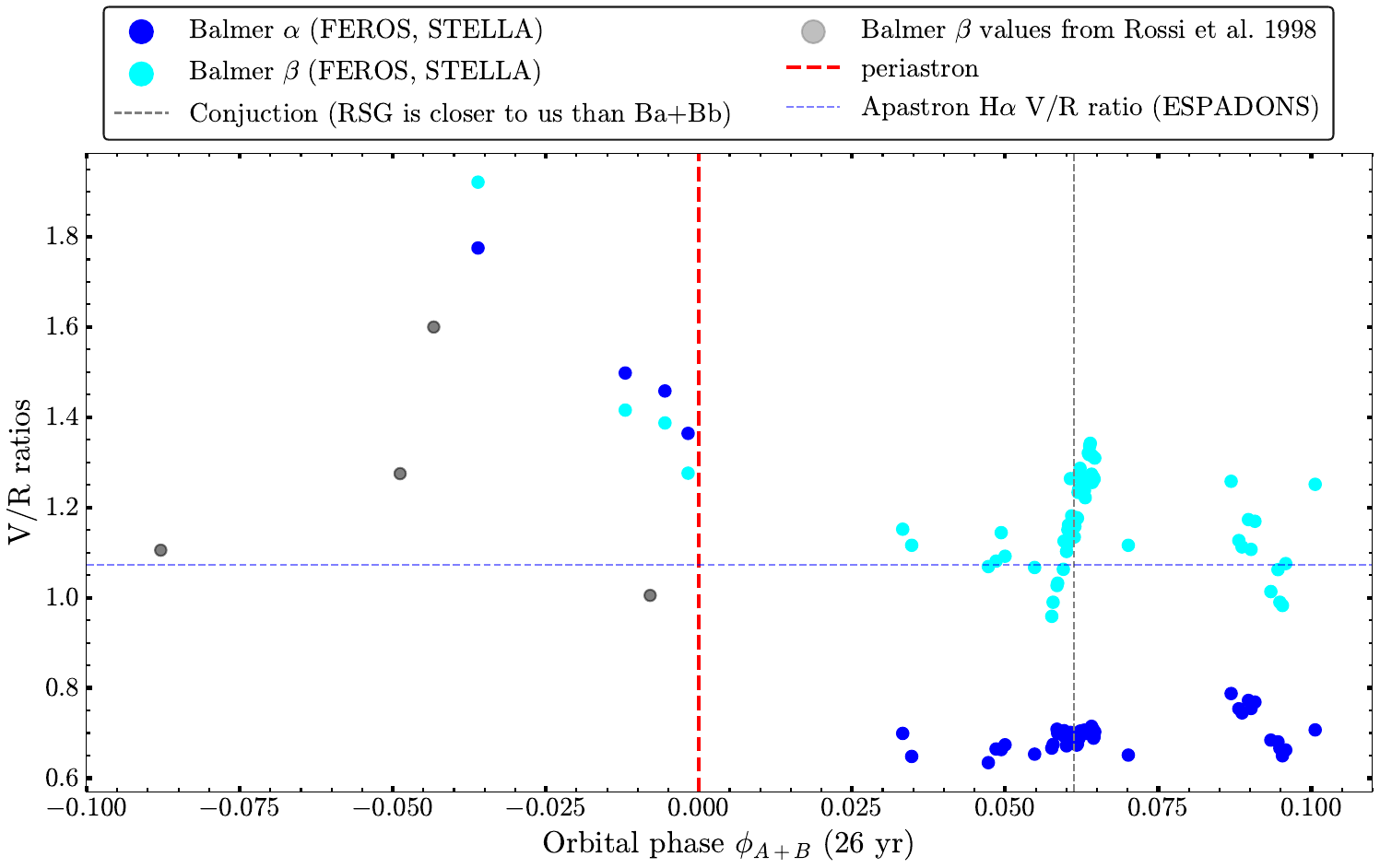}
    \includegraphics[width=1\columnwidth]{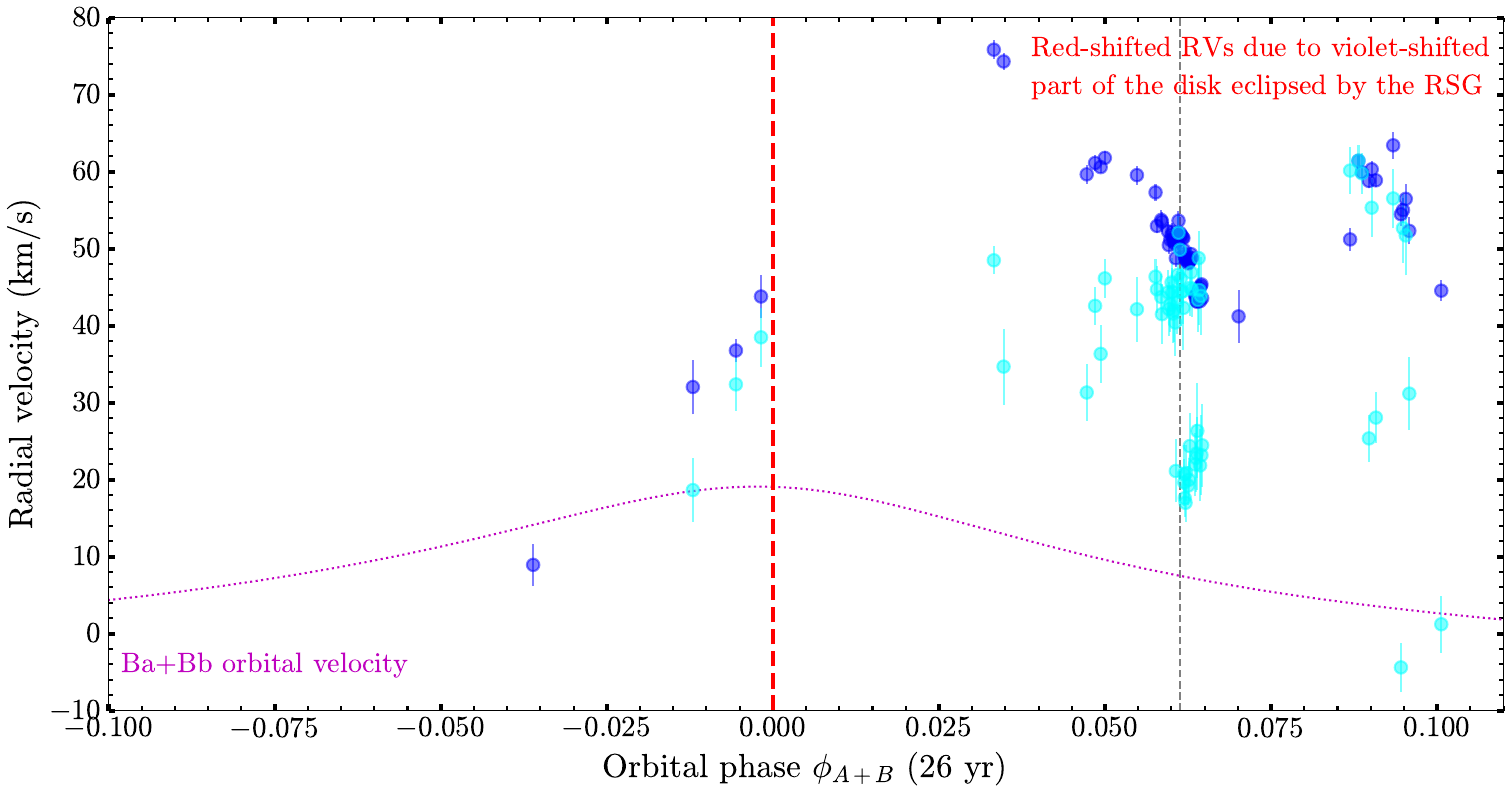}    
    \caption{Variability of H$\alpha$ and H$\beta$ based on the full data (STELLA, PLATOspec, FEROS, and ESPaDOnS), including archival data. 
    {\em Upper panel):}
    V/R ratios. A large increase before periastron can be seen, followed by a strong decrease, likely due to interaction but also geometrical effects.
    {\em Lower panel:}
    Estimated RVs. Strong red-shifted RVs 
    result from the partial eclipse of the disk by the RSG.}
    \label{fig:kq_pup_vr_rv}
\end{figure}

To characterize the properties of the disk during the full A+B orbital cycle, we also analyzed Balmer profiles in the FEROS (1996-1997) and ESPaDOnS (2010) spectra, as well as profiles from \citet{rossi98}, see Fig. \ref{fig:kq_pup_vr_rv}. As discussed before, by apastron ($\phi_{\rm A+B} \sim 0.5$), all Balmer emission lines significantly weaken, and only emission in H$\alpha$ remains significant, showing a weak, narrow profile (Fig. \ref{fig:kq_pup_spectra_opt}). This is consistent with measurements from \cite{rossi98}, which also suggest the weakest Balmer emission lines between $\phi_{\rm A+B} \sim 0.5-0.9$. From then, the emission starts increasing in strength quickly, reaching the strongest emission by $\phi_{\rm A+B} \sim 0.95$. This is consistent with the emergence of a possible disk occultation in the last two sectors of TESS data (starting between $\phi_{\rm A+B} \sim 0.9-1.0$, see Fig. \ref{fig:kq_pup_tess_raw}) and the strengthening of the shell spectrum after $\phi_{\rm A+B} \sim 0.9$. According to the spectra from \citet{rossi98}, the emissions remain strong until $\phi_{\rm A+B} \sim 0.4$. Thus, the variability of Balmer lines clearly shows that the disk properties strongly depend on the A+B orbital phase, which rules out an ongoing mass transfer between Ba+Bb or a decretion disk, as in those scenarios, the disk signatures would be visible at all A+B orbital phases. Instead, the disk appears to form shortly before the periastron, at $\phi_{\rm A+B} \sim 0.95$, and thus it must be fueled by accretion from the RSG at periastron. According to \citet{rivinius24}, smaller, newly formed disks tend to have larger V-R peak separations than larger, decaying disks. This is strongly supported by the variability of the peak separation of H$\alpha$, which showed $\sim 140-150 \: \rm km \: s^{-1}$ shortly before the periastron ($\phi_{\rm A+B} \sim 0.95-1.0$, FEROS), $\sim 130-140 \: \rm km \: s^{-1}$ following the periastron ($\phi_{\rm A+B} \sim 0.03-0.11$, STELLA), and $\sim 70\: \rm km \: s^{-1}$ at apastron ($\phi_{\rm A+B} \sim 0.5$, ESPaDOnS).

However, this does not fully explain the behavior of the Balmer lines following the periastron. According to all previous works \citep[e.g.,][]{rossi98}, the Balmer lines usually showed V/R $\sim 1$ or $>1$, while different Balmer lines showed similar V/R to each other. Before the periastron ($\phi_{\rm A+B} \sim 0.95$), H$\alpha$ and H$\beta$ showed V/R $\sim 2$. But following the periastron ($\phi_{\rm A+B} \sim 0.03-0.11$), H$\beta$ (as well as higher Balmer lines) showed V/R $\sim 1$, while H$\alpha$ has remained at V/R $\sim 0.7$. The RVs, especially those of H$\alpha$, were significantly red-shifted with respect to the expected orbital velocity of Ba+Bb, as well as with respect to the FEROS RVs of H$\alpha$ and H$\beta$ just before the periastron. Furthermore, based on the determined positions of B with VLTI (see Fig. \ref{fig:pmoired_2}), most of the STELLA observations were taken during the conjunction of the system (the beginning of 2025), with the hot components very close to the RSG, at a projected distance of only two angular radii of the RSG, while based on the determined orbit, the RSG was closer to us than the hot components. From this, we conclude that the outer regions of the disk (probed by H$\alpha$) were eclipsed by the RSG. Namely, the violet-shifted emission component of the disk was eclipsed, resulting in the apparent red-shifted RVs. This can be also supported by the case of VV Cep, as velocity shifts of H$\alpha$ up to $30-40 \: \rm km \: s^{-1}$ were also reported when the hot component was coming in and out of the eclipse \citep{wright77}, while a similar behavior was also shown for the shell absorption lines of VV Cep in the UV \citep{bauer00}. For KQ Pup, the inner regions of the disk (e.g., H$\beta$) were less affected, although still showing similarly strong trends as H$\alpha$, such as a sudden change in V/R and RVs at the smallest angular separation from the RSG, at $\phi_{\rm A+B}\sim 0.06$ (see Fig. \ref{fig:kq_pup_vr_rv}). These trends could potentially also be explained by the RSG pulsation of $\sim 90 \: \rm d$, resulting in different fractions of the disk being eclipsed. However, it is also possible that a larger part of the disk, including the Ba+Bb pair, would show a dimming due to the dense extended RSG atmosphere. Indeed, this is confirmed by the flux contribution of B to the total $K$-band flux measured with VLTI-GRAVITY (see Table \ref{table:table_vlti_results}). At the time of the conjunction (observations between December 2024 and March 2025), the determined flux contribution of B to the $K$-band continuum was negligible, about $\sim 0.0-0.2 \%$, but for the recent observations (between November 2025 and January 2026) at a larger angular separation, the flux contribution has increased to $\sim 0.3-0.6 \%$. This demonstrates that near the conjunction following the periastron, Ba+Bb were partially eclipsed by the extended atmosphere of the RSG. This could potentially also be supported by the reported drop of far-UV flux after $\phi_{\rm A+B} \sim 0.9$ \citep{gonzalez02}. 

Based on the above, we can estimate the sizes of the emitting regions of the disk, see the sketch of the system in Fig. \ref{fig:pmoired_2b}. Because H$\alpha$ showed the strongest geometrical effect (shift of RVs) due to the eclipse, the angular radius of the H$\alpha$ emitting region must be of the same order as the angular separation between the RSG surface and Ba+Bb at $\phi_{\rm A+B}\sim 0.06$, i.e., $\sim 2.95 \: \rm mas $, but it may be even larger, perhaps comparable to the angular radius of the RSG of $\sim 5.9 \: \rm mas$, since even a year after the conjuction (around the beginning of 2026), V/R of H$\alpha$ still remains at $\sim 0.7$. That could once again be supported by the case of VV Cep, as the size of the emitting region of H$\alpha$ was also estimated to be of a similar scale as the RSG \citep{pollmann20}. Thus, converting to a physical size, the resulting radius of the outermost region of the disk probed by H$\alpha$ would be $\gtrsim 500 \: \rm R_\odot$ or $\gtrsim 2.3 \: \rm au$ (this is much larger than the decretion disks of Be stars; e.g., \citealt{klement17}). The radius of the inner disk regions probed by higher Balmer lines, such as H$\beta$, would be smaller, i.e., $<2.3 \: \rm au$. Considering the orbital period, mass, and the resulting semi major axis of Ba+Bb of $a_{\rm Ba+Bb}$ $\sim 0.3 \: \rm au$, it is clear that the accretion disk is circumbinary, with the inner disk radius of $\gtrsim 0.5 \: \rm au$ (assuming the max $e_{\rm Ba+Bb}$). As shown before, Br$\gamma$ traces the innermost accreting region of the disk.

\subsection{Wind Roche-lobe overflow}
\label{chapter:wrlof}
As discussed in the previous subsection, the disk is very likely fueled by accretion from the RSG. Nonetheless, the orbit of KQ Pup A and the inner binary Ba+Bb are wide enough to prevent a $\sim 10 \: \rm M_\odot$ star from filling its Roche-lobe. This is true even when considering that the Roche-lobe radius at periastron is smaller, at around $700 \: \rm R_\odot$ (using the formula from \citealt{eggleton83} with the orbital separation replaced by the periastron distance), while RSG models of $M_{\rm i}=8-12 \: \rm M_\odot$ do not typically extend above $\sim 500 \: \rm R_\odot$ \citep[e.g.,][]{MIST1-2016}. Indeed, this is in agreement with our measured photospheric radius for KQ Pup $R_{\rm A} = 509^{+19}_{-17} \: \rm R_\odot $ determined from our VLTI observations (see Table \ref{table:table_results}). Therefore, KQ Pup fills its Roche Lobe only by $\sim 70 \% $ at periastron. Furthermore, previous studies showed that mass-transferring tertiaries are rare and in many cases drive the inner binary to undergo mass transfer first, or even merge \citep{de_Vries14, tonen20}. In the absence of other evidence of past mass transfer between the outer tertiary and the inner binary, as well as between the inner binary components (Sect. \ref{chapter:disk}), we worked under the assumption that we are observing the system before having undergone any significant mass transfer, apart from wind accretion.

We conclude that the disk is fueled by wind Roche-lobe overflow \citep[wind-RLOF;][]{mohamed07}, i.e., by accretion from the dense RSG wind and extended atmosphere, and not by regular photospheric Roche-lobe overflow (RLOF) from the RSG. Balmer emissions strongly depend on the orbital configuration of A+B, likely due to a circumbinary accretion disk forming around the hot companions shortly before periastron, and dissipating by apastron. For eccentric accreting systems, significant changes along the orbit would be expected \citep{okazaki07, lajoie11, saladino19, krynski25, mukhija26}. The wind-RLOF scenario could also be supported by V/R variations showing a variability close in timescale to the RSG pulsations, $\sim 90 \: \rm d$ (see Fig. \ref{fig:kq_pup_spectra_rv_2}). If the disk is fueled directly from the RSG wind, the accretion rate (and hence also disk properties) would show modulation by the pulsations, because the pulsations could aid in filling the Roche lobe. 
For many symbiotic systems, the cool giant component does not fill its Roche lobe as well; therefore, the wind-RLOF mechanism is the expected main mode of mass transfer \citep{merc25, boffin25}, possibly also aided by pulsations or tidal deformation \citep{merc24}. Recently, mid-IR VLTI imaging of molecular and dust layers of the AGB star $\pi^1$ Gru by \citet{drevon26} demonstrated the wind-RLOF mechanism.

The wind-RLOF scenario can also be aided by the presence of the extended envelope on top of KQ Pup A. Based on recent RSG models, the hydrostatic radius of RSGs may be much larger than their photospheric radius \citep{ercolino24, gonzalez24}. Indeed, RSGs are known to show extended molecular layers \citep[e.g.,][]{arroyo15, gonzalez24} that may reach up to several stellar radii. Consistently, our VLTI data of KQ Pup A indicate a size of CO molecular layers of $\sim 1.13 \:  R_{\rm A}$ in the near-IR (see Appendix \ref{chapter:vlti}). Compared to other sources with similar CO extension by \citet{gonzalez24} and a similar IR excess \citep[e.g.,][]{cruzalebes19}, this indicates a possible atmospheric extension at mid-IR wavelengths up to about $\sim 5-10 \: R_{\rm A}$. Thus, considering the periastron separation between the RSG and the hot components of $\sim 2100 \: \rm R_\odot$ ($\sim 4 \: R_{\rm A}$ or $\sim 9.9 \: \rm au$), the extended atmosphere of the RSG would easily fill the Roche Lobe at periastron. Using the previous results, we can also estimate when the wind-RLOF began and thus constrain the size of the extended atmosphere and the outermost hydrostatic radius of the RSG. Based on the emergence of disk accretion signatures (emergence of Balmer emission and disk occultations in TESS) appearing around $\phi_{\rm A+B} \sim 0.95$, the wind-RLOF would begin at a separation of $\sim 13.6 \: \rm au$ ($\sim 6\:  R_{\rm A}$), with a corresponding Roche lobe radius of the RSG of $\sim 4.7 \: \rm au$ ($\sim 50 \%$ filled by the RSG). A weaker wind accretion to Ba+Bb would be expected to occur at all orbital phases, but clearly, the accretion rate is not enough to balance the dissipation rate of the circumbinary disk around Ba+Bb. Assuming wind-RLOF occurring only between $\phi_{\rm A+B} \sim 0.95-0.05$ and the disk decaying by apastron ($\phi_{\rm A+B} \sim 0.5$), we estimate the dissipation timescale of the disk to be $\sim 10 \: \rm yr$, which is comparable to dissipation timescales of Be disks \citep{ghorey21,quigley25}.

With simple theoretical calculations regarding the density structure in the RSG wind and accretion rate on the inner binary (see Appendix\,\ref{appendix:mesa:atmosphere}), we also conclude that the accretion rate of material on the inner binary is about $<10\%$ of the mass-loss rate from KQ Pup A, and is driven by wind-RLOF. Given the short lifespan of KQ Pup A as an RSG, this mass accretion of its wind has negligible consequences in terms of the mass budget of the hot components. If KQ Pup A does reach core-collapse, which is likely given the mass estimate, this extended envelope may provide flash-ionization features in the supernova at early times \citep{ercolino24} and modify the light curves \citep{kurfurst20}. 

\subsection{Constraints from stellar evolutionary models}
\label{chapter:evol} \label{chapter:isochrones}

Considering the previously determined masses, the lack of evidence of significant mass transfer, and the comparable depths and shapes of the primary and secondary eclipses in TESS, it is likely that Bb is also a main-sequence (MS) star. To test whether Ba and Bb could still be a pair of MS stars by the time KQ Pup A developed into an RSG, we employed the publicly available MESA evolutionary models \citep{Paxton2011, Paxton2013, Paxton2015} from the MIST database \citep[MESA Isochrones and Stellar Tracks,][see Appendix\,\ref{appendix:mesa:timescales} for details]{dotter16,MIST1-2016}. Using the observed mass of KQ Pup A (Table\,\ref{table:table_results}) and the fact that it appears as an RSG, we obtain an age constraint of $18.1-38.0 \: \rm Myr$ (see Fig.\,\ref{fig:evol1}). Assuming that all three stars are coeval, and that Ba and Bb have not yet exchanged mass via RLOF, with $M_\mathrm{Ba}>M_\mathrm{Bb}$, we obtain the mass-constraints of $5.8\,M_\odot\lesssim M_\mathrm{Ba}\lesssim11.8\,M_\odot$ and $  1.2\,M_\odot<M_\mathrm{Bb}\lesssim8.5\,M_\odot$ (see Appendix\,\ref{appendix:mesa:timescales}), with the lower limit of $1.2 \:  \rm M_\odot$ based on the minimum mass of Bb necessary to reach MS within the age constraint. 

\begin{figure}
    \centering
    \includegraphics[width= 1.0\columnwidth]{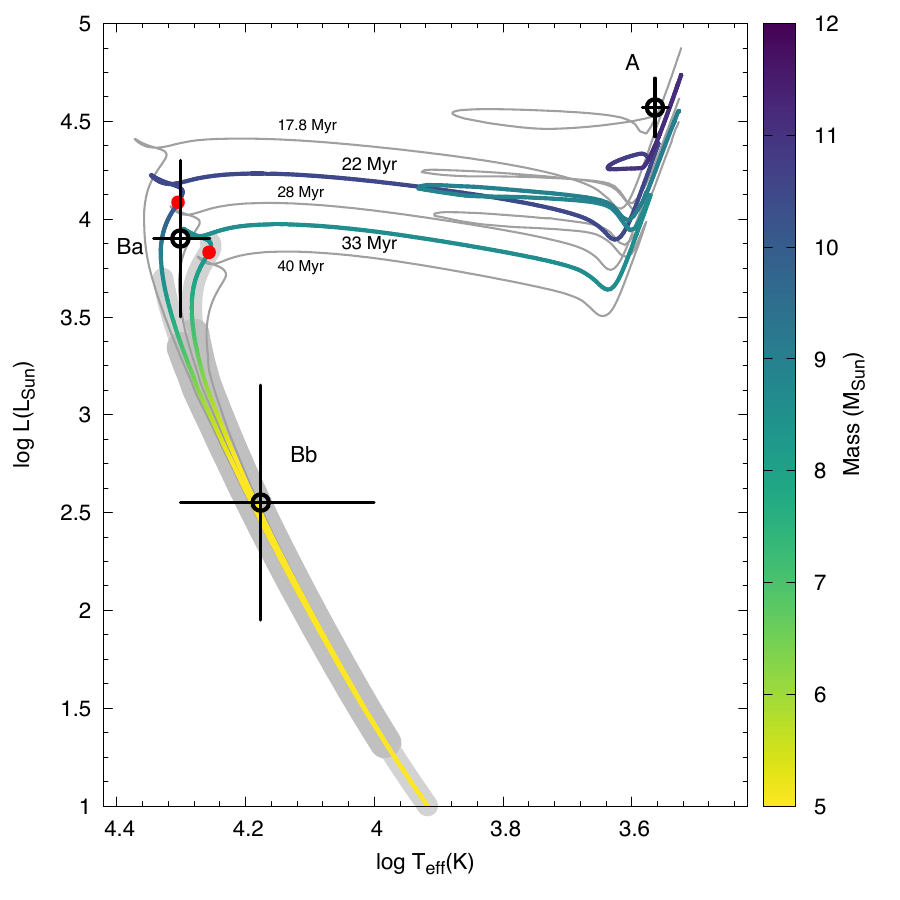}
    \caption{Positions of KQ Pup A and Ba on the HRD compared to isochrones interpolated from the MIST database. Two isochrones are color-coded by mass. Red dots indicate the TAMS. Additionally, the position corresponding to the most likely mass for Bb of $4.3\,M_\odot$ is indicated, while the area shaded in gray shows the range of models that fit the mass constraints for Bb. Gray lines indicate the age limits, and the middle value  is from evolutionary models.}
    \label{fig:kq_pup_isochrone}
\end{figure}

Based on the numerous determined physical parameters for KQ Pup A and Ba, we further tested whether they are indeed coeval stars and whether we could estimate additional properties from evolutionary tracks. We therefore compared their Hertzsprung–Russell diagram (HRD) positions to isochrones from MIST \citep{dotter16}, selecting models with solar composition ([Fe/H] = 0), and rotation enabled ($v_{\rm init}/v_{\rm crit}=0.4$, based on our $v \sin i $ value). We used the online interpolator\footnote{\url{https://waps.cfa.harvard.edu/MIST/interp_isos.html}} to calculate isochrones near the age limits $18.1-38.0 \: \rm Myr$ and at a few intermediate values (see Fig.~\ref{fig:kq_pup_isochrone}). 

As shown in Fig.~\ref{fig:kq_pup_isochrone}, star A is at the RSG branch, at, but more likely above the blue loops that indicate the He-core burning phase. We note that we are not overly concerned by the offset of star A from the models: as \citet{joyce20} showed, changes to various model parameters, especially the mixing length and mass loss, can induce shifts up to $\pm200$\,K in high-mass RSG models. Initial masses for models around the position of A are between 9--12 $\rm M_\odot$, while the present-day mass of the RSG could be lower by up to $ \rm \lesssim0.5M_\odot$ (see Appendix \ref{appendix:mesa:timescales}). 

The other well-constrained star, Ba, appears to lie close to the MS. Here we plot the star at values $\log{L} = 3.9\pm0.4$ (accounting for a contribution from Bb up to $\log{L} = 3.0$), and at $T_{\rm eff} = 20000\pm2000$\,K. We highlight the terminal-age main sequence (TAMS) model points with red dots: the star is very close to them, indicating that it is nearing, or just past the end of its MS lifetime. Based on the luminosity uncertainties for Ba, we estimate an age range of $27\pm7$\,Myr for the system.
Indeed, the position of Ba (Fig.\,\ref{fig:kq_pup_isochrone}) fits well between the isochrones at 22 and 33\,Myr, and the mass constraints that arise (10.5--11.0 $\rm M_\odot$ and 7.3--8.5 $\rm M_\odot$ from each isochrone, respectively) are also well within the values inferred just from the constraints in age and masses. The mass ranges for A and Ba along the isochrones are therefore in good agreement with other mass inferences, and their positions are compatible with coeval evolution.  This agrees with RSG+B systems studied by \citet{patrick25}, where the majority of systems could also be explained by co-eval evolution.

It is much harder to constrain Bb on the isochrones, since at the moment we can only infer its mass from the difference between the combined mass of the B components and the inferred mass for Ba. Given the large uncertainties, the possible values for Bb span masses around 1.2--8.1 $\rm M_\odot$ (based on the evolutionary tracks) or 2.1--7.0 $\rm M_\odot$ (based on the astrometric mass range for Ba+Bb). Therefore, we highlight the ranges defined by these two limits for Bb with thinner grey (wide range) and thicker grey (narrow range) areas in Fig.~\ref{fig:kq_pup_isochrone}. We note that the plot only extends to $\log{(L/L_{\odot})}=1.0$ in order to highlight the differences between the post-MS branches, but the models only reach the lowest mass and the ZAMS point at $\log{(L/L_{\odot})}\approx0.23$. We also include a fiducial point with large error bars around $4.3\,M_\odot$, the most likely mass range for Bb based on the mean dynamical masses and the PHOEBE models.

\section{Discussion}
\label{chapter:discussion}

\subsection{Classification of RSG binaries}

In the literature, there is confusion about the classification of binary RSGs. \citet{pant20} compiled a sample of about 100 Galactic RSG binaries based on spectral type, several of which may actually not be RSGs from an evolutionary point of view. Indeed, the more recent list by \citet{healy24} gives a smaller number. Meanwhile, \citet{munari21} selected M supergiants from the sample by \citet{pant20}, considering them all as VV Cephei-type binaries. However, in the new large samples of RSG+B binaries in the Local Group \citep[e.g.,][]{neugent20,patrick22, patrick25}, the majority of their binaries were not reported as interacting, except for 4 companions embedded in the RSG wind by \citet{patrick25}. \citet{neugent19} also reported that one of the systems in the sample is the first known RSG+Be pair, based on Balmer emission lines, and \citet{neugent20} classified several other binary candidates as RSG+Be. However, we note that all the previously known VV Cephei binaries \citep{cowley69} also show Balmer emission lines, as previously discussed, and were considered to host a Be companion, including for KQ Pup \citep{altamore92} and VV Cep \citep{hutch71}. RSG binaries were also sometimes confused with the $\zeta$ Aurigae group, which consists mostly of earlier and less massive red giants.
Historically, some of the VV Cephei binaries were also misclassified as classical symbiotic stars, i.e., systems consisting of an evolved red giant and a white dwarf \citep[e.g.,][]{merc19}. Nonetheless, VV Cephei-type binaries could likely also be classified as symbiotic B[e]-type (symB[e]) stars \citep{lamers98, munoz24}, interacting symbiotic systems consisting of a cool (super)giant and a hot component, enshrouded in a nebula producing forbidden emission lines. 

We reiterate that VV Cephei binaries were classified by \citet{cowley69} as K/M supergiants showing hydrogen emission lines associated with their hot companions, while low-ionization forbidden emission lines forming in the ionized nebula are also present. Clearly, based on the discussion above, VV Cephei binaries are a special case of interacting RSG binaries. The newly discovered RSG+Be \citep{neugent20} and embedded systems \citep{patrick25} belong to this special class. The stars in the class could be further categorized, depending on whether the photospheric lines of hot companions (rotationally broadened absorptions of \ion{H}{i} and \ion{He}{i}) remain visible, or whether hot companions are embedded (e.g., KQ Pup and VV Cep).

Furthermore, the VV Cephei phenomenon could be a transitory evolutionary phase, possibly occurring only near periastron for wider eccentric orbits. \citet{ercolino24} showed that binary interaction of (initially) wide RSG+B pairs can cause large blue loops in the HR diagram before exploding as various types of SN. Indeed, several post-RSG yellow hypergiants (YHG) showing signs of binary interaction were identified, most famously the recent possible transition of WOH G64 to a YHG \citep{munoz24}, while suddenly showing VV Cephei properties. Likewise, some other YHGs are thought to be post-RSG objects and show signs of past binary interaction, such as V366 Nor, which has a disk-hosting a companion \citep{kourniotis25}, also showing similar properties to VV Cephei binaries. Meanwhile, V772 Cen, one of the original VV Cephei binaries, no longer shows the VV Cephei properties\footnote{Archival spectra available at \url{https://archive.eso.org/scienceportal/home}}.

\subsection{Classification of hot components in VV Cephei-type binaries}

As shown in this work, but also for example for the VV Cep system \citep{pollmann20}, the \ion{H}{i} emission likely arises in an accretion disk rather than a decretion disk. Therefore, the Be classification of some RSG companions may be incorrect. This is also evident by the likely too low $v_{\rm rot}$ of the hot companion in the KQ Pup system, while a low value was also found for the assumed RSG+Be pair in \citet{neugent19}.
To find out whether the KQ Pup Ba+Bb pair could correspond to some of the known classes of short-period massive binaries, we compared the available spectra. 

Natural candidates for comparison would be Be+sdOB binaries, which are also routinely studied using the Br$\gamma$ hydrogen line \citep{klement24,klement25}. However, as discussed in Sect. \ref{chapter:disk}, we have not found significant evidence of a Be or a stripped component, while the orbital periods of such systems are typically of the order of months \citep[e.g.,][]{lechien25}.
In the UV region, there is similarity to the spectra of the episodic Be shell star Pleione and the B[e] star FS CMa, as noted by \citet{bauer00}. We also found similarities with B[e] supergiants. For example, GG Car shows a similar IUE spectrum, while it shares other properties - an eclipsing companion of about $\sim 30 \: \rm d$, and a dense CSM and disk \citep{marchiano12, kraus13, porter21, kashi23}. However, the evolutionary status of B[e]sg stars is uncertain. In Fig. \ref{fig:kq_pup_comp}, we show a comparison of UV spectra of KQ Pup to $\phi$ Per (Be+sdOB) and GG Car (B[e]sg).

\subsection{Implications for the multiplicity of massive stars and evolution}
\label{chapter:evolution}

Based on the eclipses detected in TESS light curves (Table \ref{table:table_candidates}) and a list of Galactic RSGs by \citet{healy24}, it appears that 5 out of 44 Galactic binary RSGs have a third component, or, if including RSG candidates (thus also including FR Sct), 6 out of 79. This implies that many other RSG binaries must be hierarchical triple systems as well, just without eclipses. 

Therefore, this could be a common initial configuration for massive systems \citep{frost25, bordier26}, suggesting that many present short-period massive binaries or supposedly single (possibly runaway) massive stars were influenced by RSGs. For instance, even though the VV Cephei binaries were likely incorrectly thought to host Be companions, such RSG systems may still play a role in the formation of Be stars. Indeed, \citet{li26} modeled wide massive systems undergoing wind-RLOF (orbital periods of > 1000 d, peaking around 10\,000 d), including from RSG donors. In their models, for an RSG+B system, the B star accreted about $\sim 0.15 \: \rm M_\odot$, which allowed it to acquire enough angular momentum to be spun up to the critical rotation, potentially becoming a Be star. This could also play a role in the formation of Be+sdOB systems, since, as shown above, short-period binary hot companions of RSGs appear to be common (and two of our eclipsing candidates show ongoing interaction, i.e., semi-detached eclipses, see Appendix \ref{appendix:tess}). Likewise, several B[e]sg stars with short-period companions (tens to hundreds of days) are known \citep[e.g., ][]{kraus14, maravelias18, porter22, cidale26}, while showing UV spectra and circumbinary structures similar to VV Cephei binaries. Such structures could thus be leftover material from the RSG phase. Furthermore, once the RSG component explodes as a SN, it may leave behind a neutron star or black hole. Therefore, binary or triple RSG systems could very well be progenitors of high-mass Be/sg X-ray binaries \citep[e.g.,][]{pablo11, vali25}. This highlights the importance of understanding the evolution of RSG systems, as it could help unveil the origin of several important groups of stars with unclear evolutionary pathways, for example,  the ones above, allowing us to uniquely study such systems during the first $\lesssim 40 \: \rm Myr$ of their evolution.

Specifically for the KQ Pup system, once the Ba leaves the MS (assuming it has not yet), it will initiate mass transfer with Bb (see Appendix \ref{appendix:mesa}), possibly evolving into an Algol-type system \citep{mkrtichian22}. Such systems are considered one of the likely immediate progenitors of Be+sdOB systems and, more generally, Be stars \citep[e.g.,][]{rivinius24}. This evolution may be significantly influenced by the wind-RLOF from A (see Sect. \ref{chapter:wrlof} and \citealt{li26}), and ultimately by the upcoming SN explosion of A, which could either alter the orbit or even unbind the system \citep[e.g.,][]{lu19}. Alternatively, \citet{tomassini26} recently studied a likely post-RSG system, AFGL 4106, that exhibits a mass distribution similar to that of the KQ Pup system but with the primary being the more massive component, likely a post-RSG YSG, and the secondary an RSG. This unlikely evolutionary configuration could indeed be explained by a system with a similar initial configuration as KQ Pup, but where the hot companions interacted or merged \citep[e.g.,][]{tonen20} before the outer wide component became an RSG.

\section{Conclusions}
\label{chapter:conclusions}
We report the discovery of the first hierarchical triple RSG system, KQ Pup, consisting of an outer RSG (A) and two hot components (Ba+Bb) in a tight inner binary. This discovery was accomplished with an innovative combination of several instruments and methods, with some of them being used in this way for an RSG system for the first time. We combined high-precision VLTI-GRAVITY astrometric measurements, enabled by the detection of the hydrogen Br$\gamma$ line in the near-IR, with RV measurements to determine the orbital parameters and dynamical masses of the system. The derived masses agree with those estimated for KQ Pup A from asteroseismology and for KQ Pup Ba from evolutionary models. These results give an unexpected mass ratio for the system, with the Ba+Bb components being more massive than the RSG component, i.e., $M_{\rm Ba+Bb} \sim 14 \: \rm M_{\odot} $ and $M_{\rm A} \sim 10  \: \rm M_{\odot} $, making the RSG a less-massive outer tertiary in the system despite it dominating the optical light. Such a configuration is compatible with MESA-MIST evolutionary models, yielding an approximate age of the system of $ 18.1-38.0 \: \rm Myr$. 

The astrometric and RV constraints on the wide A+B orbit also allowed us to obtain an orbital parallax of the system independent from the \textit{Gaia} trigonometric parallax. We found the parallax to be $\pi =1.24^{+0.05}_{-0.04}\: \rm mas$, translating to a distance of $d = 805_{-28}^{+30} \: \rm pc$, which is in agreement with the \textit{Gaia} DR3 value ($1.36 \pm 0.14 \: \rm mas$, \citealt{gaia21}). This is the first time that an orbital parallax has been determined for a Galactic RSG, and it is also significant for this system, as the \textit{Gaia} astrometric parallax of KQ~Pup may be affected by the orbital motion.  

KQ Pup offers an opportunity to constrain the hydrostatic radius of its RSG component through hydrodynamic modeling of its well-constrained phase-dependent mass transfer and disk formation. Rather than being due to direct mass transfer between the stellar components, the disk in the KQ Pup system (and likely also other VV Cephei binaries) is fueled by wind-RLOF from the dense RSG atmosphere to the inner binary, as the strengths of the Balmer lines strongly depend on the orbital phase, while the RSG fills its Roche lobe at periastron only by $\sim 70\%$. Strong Balmer emissions appear after $\phi_{\rm A+B}\sim0.95$ (with the Roche Lobe filled by $\sim 50\%$) and are accompanied by the emergence of possible disk occultations in the TESS data and the strengthening of the shell spectrum in UV. Overall, this demonstrates the Ba+Bb circumbinary accretion disk forming shortly before periastron and decaying by apastron. This is compatible with the detection of the Br$\gamma$ feature, and it is the first such detection for an RSG system.
The Br$\gamma$ line is routinely used to study other types of interacting massive stars with VLTI-GRAVITY, such as Be+sdOB systems or accreting young stellar objects \citep[e.g.,][]{gravity24, klement25}.

KQ Pup does not show full eclipses by the RSG component as in the case of its more famous counterpart, VV Cep. However, after the periastron, the outer parts of the accretion disk are eclipsed by the RSG, as evidenced by the behavior of H$\alpha$. The outer regions of the circumbinary disk likely have a similar radius as the RSG, i.e., $\sim 2.3 \: \rm au$. Meanwhile, the VLTI-GRAVITY data show that the flux contribution of B to the total $K$-band flux dropped during this partial outer eclipse, suggesting that Ba+Bb are dimmed by the extended atmosphere of the RSG.

The discovery of KQ Pup Bb was primarily enabled by the detection of eclipses in the TESS light curves, and we also list other independent indicators of its existence:
\begin{itemize}

    \item TESS eclipses: We detected primary and secondary eclipses for KQ Pup Ba+Bb in all four sectors (2019-2025). The eclipses have a period of $P_{\rm Ba+Bb}=17.2596 \: \rm d$.

    \item Orbital parameters: We tested representative Ba+Bb models in PHOEBE to reproduce the observed light curve. The phase offset between the eclipses shows that the orbit is eccentric, which we estimate to fall between $0.0236 \leq e \lesssim 0.5$ based on the timing of the eclipses and lack of clear signatures of mutual heating at periastron. This also defines an allowed range for the argument of periastron, but more detailed modeling will require RV data that unambiguously trace Ba+Bb.

    \item Balmer and Brackett lines: Our high-cadence STELLA spectra show variations of Balmer emission profiles on a similar timescale as the Ba+Bb orbit, i.e., $\sim17 \: \rm d$, especially for H$\beta$, while the RVs of Br$\gamma$ also show similar variations.

    \item Mass ratio of A to Ba+Bb: The mere fact that the mass of B is higher than that of A can be explained only by a Bb component for a coeval evolution. If Ba on a wide orbit were a single component more massive than the A component, it would already have to be more evolved than the RSG component, unless significant mass transfer occurred. 
\end{itemize}

Determining the nature of the population of binary hot components in RSG systems will require further study and modeling. For KQ Pup Ba, we found the best-matching PoWR model with $T_{\rm eff} = 19\,900 \: \rm K$, $v \sin i  \sim 190  \: \rm km \, s^{-1}$ ($v_{\rm rot}/v_{\rm crit}\sim0.4$), and $\log(L/L_{\odot}) \sim 3.95 $, making it a B2 MS star near the TAMS. Most likely, KQ Pup Bb is also an MS star with a minimum mass of $ \gtrsim 1.2 \: \rm M_{\odot} $ rather than a stripped star, as no typical lines of stripped He-rich subdwarfs were detected. The Ba+Bb pair is embedded in the dense CSM and disk, preventing us from fully characterizing the stars. This is a similar situation as some massive young stellar objects \citep[e.g.,][]{frost21}.

We have also listed several other candidates for eclipsing companions in RSG systems that will require further observations. Our findings have important implications for the multiplicity of massive stars, as we have found evidence of a third component for about $10 \%$ of the known Galactic RSG binaries \citep[e.g.,][]{healy24}. Such a large fraction of eclipsing systems suggests that many other RSG binaries are also, in fact, hierarchical triple systems. And in some of them, if similar to KQ Pup, more massive components could be hiding in the light of the RSG component. Binary and triple RSG systems could be progenitors of some types of interacting massive stars (e.g., Be+sdOB, B[e]sg systems) and high-mass Be/sg X-ray binaries.

Our results demonstrate that several methods can be used to study such systems and detect new companions. New high-cadence spectroscopy would be sufficient to find more non-eclipsing candidates based on short-term V/R variations in VV Cephei binaries. To determine more precise properties of hot components, including those of KQ Pup Ba+Bb, high-resolution, high-cadence observations in the UV are required, such as with HST. For such systems, using VLTI-GRAVITY to study the Br$\gamma$ line allowed us to determine precise dynamical masses and study spatial properties of the companion at high-angular resolution. Meanwhile, future VLTI observations will allow us to constrain the dynamical masses of the system further, especially when combined with the forthcoming epoch astrometry from Gaia DR4.

\section*{Data availability}

Table \ref{table:table_rvs} is only available in electronic form at the CDS via anonymous ftp to cdsarc.u-strasbg.fr (130.79.128.5) or via https://cdsarc.cds. unistra.fr/viz-bin/cat/J/A+A/vol/page.

\begin{acknowledgements}
     We thank the anonymous referee for useful comments that significantly improved the quality of the paper.

     We thank E. Paunzen for the first look at SMEI data for KQ Pup, and to Gregory Henry for providing the APT light curves. Fruitful discussions with P. Krynski, H. Boffin, J. Merc, K. Ohnaka, A.K. Dupree, and W.-R. Hamann are also acknowledged. 

     DJ acknowledges support from the ESO Studentship. DJ and JK were partially supported by grant GA \v{C}R 25-15910S.

     GGT is supported by the German Deutsche Forschungsgemeinschaft (DFG) under Project-ID 496854903 (SA4064/2-1, PI Sander), and acknowledges financial support by the Federal Ministry for Economic Affairs and Climate Action (BMWK) via the Deutsches Zentrum f\"ur Luft- und Raumfahrt (DLR) grant 50 OR 2503 (PI Sander). 

     VR and AACS are supported by the German \textit{Deut\-sche For\-schungs\-ge\-mein\-schaft, DFG\/} in the form of an Emmy Noether Research Group -- Project-ID 445674056 (SA4064/1-1, PI Sander). VR and AACS further acknowledges financial support by the Federal Ministry for Economic Affairs and Climate Action (BMWK) via the Deutsches Zentrum f\"ur Luft- und Raumfahrt (DLR) via the DLR grant 50 OR 2306 (PI Ramachandran/Sander).
     This project was co-funded by the European Union (Project 101183150 - OCEANS).

     ACR acknowledges funding from the Netherlands Organisation for Scientific Research (NWO), as part of the Vidi and Aspasia research program BinWaves (project number 639.042.728, PI: de Mink).

     This research was supported by the `SeismoLab' KKP-137523 \'Elvonal grant of the Hungarian Research, Development and Innovation Office (NKFIH) and by the LP2025-14/2025 Lendület grant of the Hungarian Academy of Sciences.

     Based on observations made with the Very Large Telescope Interferometer (VLTI) at the Paranal Observatory of European Southern Observatory (ESO).

     Based on data obtained with the STELLA robotic telescopes in Tenerife, an AIP facility jointly operated by AIP and IAC.

     Based on observations obtained at the Canada-France-Hawaii Telescope (CFHT) which is operated by the National Research Council of Canada, the Institut National des Sciences de l'Univers of the Centre National de la Recherche Scientique of France, and the University of Hawaii.

     We acknowledge the use of TESS High Level Science Products (HLSP) produced by the Quick-Look Pipeline (QLP) at the TESS Science Office at MIT, which are publicly available from the Mikulski Archive for Space Telescopes (MAST). Funding for the TESS mission is provided by NASA’s Science Mission directorate.

    Based on INES data from the IUE satellite.

PLATOSpec was built and is operated by a
consortium consisting of the Astronomical Institute ASCR in Ondrejov, Czech
Republic (ASU), the Thüringer Landessternwarte (Thuringian State Observatory - Germany), the Universidad Catholica in Chile (PUC - Chile), and minor partners include Masaryk University (Czechia), Universidad Adolfo Ibanez (Chile) and Institute for PLasma Physics of the Czech Academy of Sciences (Czechia).  DP acknowledges financial support from the FWO in the form of a junior postdoctoral fellowship No. 1256225N. Financing for the modernisation and front end of the 1.52-m telescope was provided by AsU and personal costs were partly financed from grant LTT-20015. Financing for the construction of PLATOSpec was provided by the Free State of Thuringia, under the "Directive for the Promotion of Research PUC is acknowledging the support from ANID Fondecyt n. 1211162 and n. 1251299, and ANID QUIMAL ASTRO20-0025. Use of the 1.52-m telescope was made possible
through an agreement between ESO and the PLATOSpec consortium.

This research made use of NASA’s Astrophysics Data System Bibliographic Services, as well as of the SIMBAD and VizieR databases operated at CDS, Strasbourg, France.
\end{acknowledgements}

%
%
\bibliographystyle{aa} 
\bibliography{bibliography} 

%

\begin{appendix}
\onecolumn

\section{Observing logs for KQ Pup} 
\label{chapter:vlti}
In Table \ref{table:table_vlti}, we list our VLTI-GRAVITY observations. In Table \ref{table:calibrators}, we list properties of the VLTI calibrators, including parameters of MARCS spectra used to calibrate the flux of calibrators \citep{gustaf08}. The obtained spectral transfer function was used to calibrate the flux of our science target. Lastly, in Table \ref{table:table_vlti_results}, we list the results obtained from fitting the Br$\gamma$ in our VLTI dataset when fitted per epoch.

Additionally, we also analyzed the atmospheric extension of KQ Pup A, following a similar workflow as in \citet{wittkowski18} and \citep{jadlovsky26}. We fitted a uniform disk (UD) model to the CO molecular bands $^{12}$C$^{16}$O (2-0) at $2.29 \: \rm \mu m $ and $^{12}$C$^{16}$O (3-1) at $2.32 \: \rm \mu m $ and compared it to the photospheric angular diameter $\theta_{\rm A, UD}$. In all VLTI epochs, the obtained angular diameters are similar and show an average CO atmospheric extension of $\sim 6.63 \: \rm mas$, i.e., $\sim 1.13 \: R_{\rm A}$ when divided by $\theta_{\rm A, UD}$.

\begin{table}[htbp]
        \centering
        \caption{Lists of observations of KQ Pup taken with VLTI-GRAVITY.} 
        \label{table:table_vlti} 
        \setlength{\extrarowheight}{3pt}
        \begin{tabular}{cccccc}
\hline \hline 
\text{Date} & \text{Time [UT]} & \text{Exposure [s]}  & \text{Seeing ['']} & \text{ Coherence time [ms]} & \text{Configuration}  
\\
\hline
2024-12-04 & 04:35:02 & 3.0 & 0.41 & 4.54 & \text{SMALL}    \\ 
2025-01-28 & 02:52:58 & 3.0 & 0.82 & 7.93 & \text{SMALL}    \\ 
2025-03-06 & 03:00:11 & 3.0 & 0.80 & 6.68  & \text{SMALL}   \\ 
2025-11-19 & 07:31:36 & 3.0 & 0.86 & 6.58 & \text{SMALL}   \\ 
2025-11-24 & 05:30:14 & 3.0 & 0.62 & 4.32 & \text{MEDIUM}   \\ 
2026-01-25 & 05:20:12 & 3.0 & 0.73 & 5.22 & \text{SMALL}   \\ 
        \hline
        \end{tabular}
\tablefoot{
Only the observations taken with the SMALL VLTI-AT configuration (A0 B2 D0 C1) were used for the orbital solution.}
\end{table}

\begin{table}[htbp]
        \centering
        \caption{Properties of the calibrators used for VLTI-GRAVITY observations of KQ Pup.}
        \label{table:calibrators} 
        \setlength{\extrarowheight}{3pt}
        \begin{tabular}{cccccc}
\hline \hline 
\text{Calibrator} & \text{Sp. Type}  & \text{LDD [mas]} & \text{UDK [mas]}  & $T_{\rm eff} \: \rm [K] $ & $\log g$    
\\
\hline
HD 58215 & K4III & 2.54 $\pm$ 0.25 & 2.48 & 4000 & 1.5\\ 
d Vel & G6III & 1.70 $\pm$ 0.15 & 1.66 & 4500 & 2.0  \\ 
        \hline
        \end{tabular}
\tablefoot{Listed are the limb-darkened disk (LDD) and K band diameters (UDK) from \citet{bourges17}, while the right part of the table shows properties of MARCs models \citep{gustaf08} used for flux calibration (solar metallicity).}
\end{table}

\begin{table}[htbp]
        \centering
        \caption{Results obtained from fitting the Br$\gamma$ line in the VLTI-GRAVITY data using PMOIRED.} 
        \label{table:table_vlti_results} 
        \setlength{\extrarowheight}{3pt}
        \begin{tabular}{ccccccc}
\hline \hline 
\text{Date} & \text{$\theta_{\rm A, UD} \: [\rm mas]$} & \text{ $f_{\rm B}$}  & \text{ $\rm f_{\rm Br\gamma}$ } & \text{ $\rm FWHM_{\rm Br\gamma} \: \rm [nm]$ }   & \text{ $\rm \Delta RA  \:\rm [mas] $ }   & \text{ $\rm \Delta DEC \:\rm [mas]$ }  
\\
\hline 
2024-12-04 & 5.86 & $\sim 0$ & 0.032 & 1.22 & $-2.76_{\pm0.15}$ & $-4.35_{\pm0.12}$ \\ 
2025-01-28 & 5.84 & 0.002 & 0.034 & 0.97 &  $-3.88_{\pm0.11}$ & $-3.32_{\pm0.10}$\\ 
2025-03-06 & 5.93 & $\sim 0$ & 0.027 & 1.32 & $-4.95_{\pm0.47}$ & $-2.27_{\pm0.37}$ \\ 
2025-11-19 & 5.89 & 0.003 & 0.032 & 0.99  & $-9.07_{\pm0.04}$ & $+2.23_{\pm0.04}$\\ 
2026-01-25 & 5.94 & 0.006  & 0.032 & 1.04 & $-9.91_{\pm0.15}$ & $+3.69_{\pm0.29}$ \\ 
        \hline
        \end{tabular}
\tablefoot{Listed are the angular diameter $\theta_{\rm A, UD}$ of KQ Pup A (KQ Pup B was fitted as a point source), flux contribution $f_{\rm B}$ of KQ Pup B, full width at half maximum (FWHM) of the Br$\gamma$ line, and positions of Br$\gamma$ recovered from relative astrometry.
The positions $\Delta RA$ and  $\Delta DEC$ are relative positions of B with respect to A. Positions of A and B with respect to their center of mass can be recovered using $ \vec{pos}_A = - q/(1+q) \cdot \vec{sep}$ and $ \vec{pos}_B = 1/(1+q)\cdot\vec{sep}$, where $q_{\rm AB}=M_{\rm B} / M_{\rm A}$ is the mass ratio and $\vec{sep} = \vec{pos}_B - \vec{pos}_A =  \rm (\Delta RA, \Delta DEC)$. Flux $f_{\rm B}$ is the fraction of the contribution of B to the total flux of the system, i.e., B contributes about $0.6 \%$ of the total $K$-band flux during the last observation.} 
\end{table}

\begin{figure*}[htbp]
    \includegraphics[width=1\textwidth, keepaspectratio]{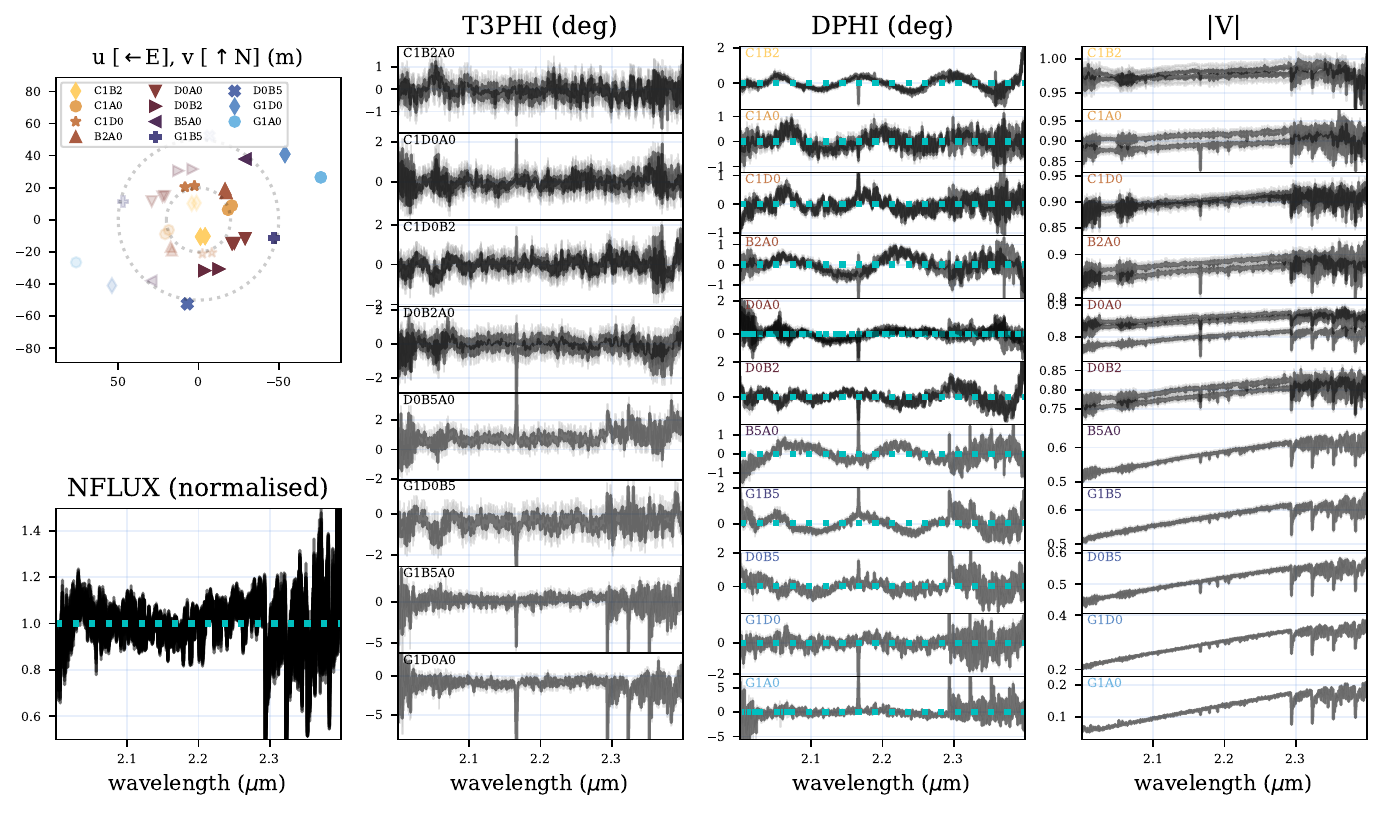}
    \caption{Same as Fig. \ref{fig:pmoired_obs} but for the full wavelength range of the VLTI-GRAVITY dataset for KQ Pup. Apart from Br$\gamma$ at $ 2.167 \: \rm \mu m $, the data include features typical of RSGs, such as \ion{Ti}{i} lines between $ \sim 2.18-2.22 \: \rm \mu m $ and strong CO molecular bands at $ \sim 2.29-2.40 \: \rm \mu m $.} 
    \label{fig:pmoired_obs_full}
\end{figure*}

\begin{table*}[htbp]
        \centering
        \caption{List of our determined RVs used for the orbital fit in this work. The full table is available in machine-readable format.} 
        \label{table:table_rvs} 
        \setlength{\extrarowheight}{3pt}
        \begin{tabular}{ccccccc}
\hline \hline 
\text{Source} & \text{Date} & \text{MJD [d]} & \text{RV$_{\rm A}$}  & \text{Error} & \text{RV$_{\rm B}$} & \text{Error}  
\\
\hline
\citetalias{cowley65} & 1918-02-06 & 21630 & 9.7 & - & - & -   \\ 
\citetalias{cowley65} & 1918-02-19 & 21643 & 10.4 & - & - & -   \\ 
\citetalias{cowley65} & 1918-02-21 & 21645 & 9.8 & - & - & -   \\ 
.. & .. & .. & .. & .. & .. & ..   \\ 
.. & .. & .. & .. & .. & .. & ..   \\ 
STELLA & 2026-01-31 & 61071.6 & 29.295 & 0.013 & - & -   \\ 
STELLA & 2026-02-03 & 61074.5 & 29.324 & 0.015  & - & -   \\ 
STELLA & 2026-02-06 & 61078.5 & 29.472 & 0.016 & - & -   \\ 
        \hline
        \end{tabular}
\tablefoot{The table includes archival RVs from \citet{cowley65}.
MJD was calculated as HJD - 2400000 d. Methods for determining the RV of the A and B components in various spectra are described in Sect. \ref{chapter:rv}. Data for B include the determined RV from H$\alpha$  based on the emission centroid (ESPaDOnS), from the interferometric fit to Br$\gamma$ (VLTI-GRAVITY), and from cross-correlation of IUE spectra. }
\end{table*}

\FloatBarrier 
\onecolumn

\section{Raw TESS data for KQ Pup and other promising candidates \label{appendix:tess}}

\subsection{Light curves}
\begin{figure*}[htbp]

    \includegraphics[width=1.0\textwidth]{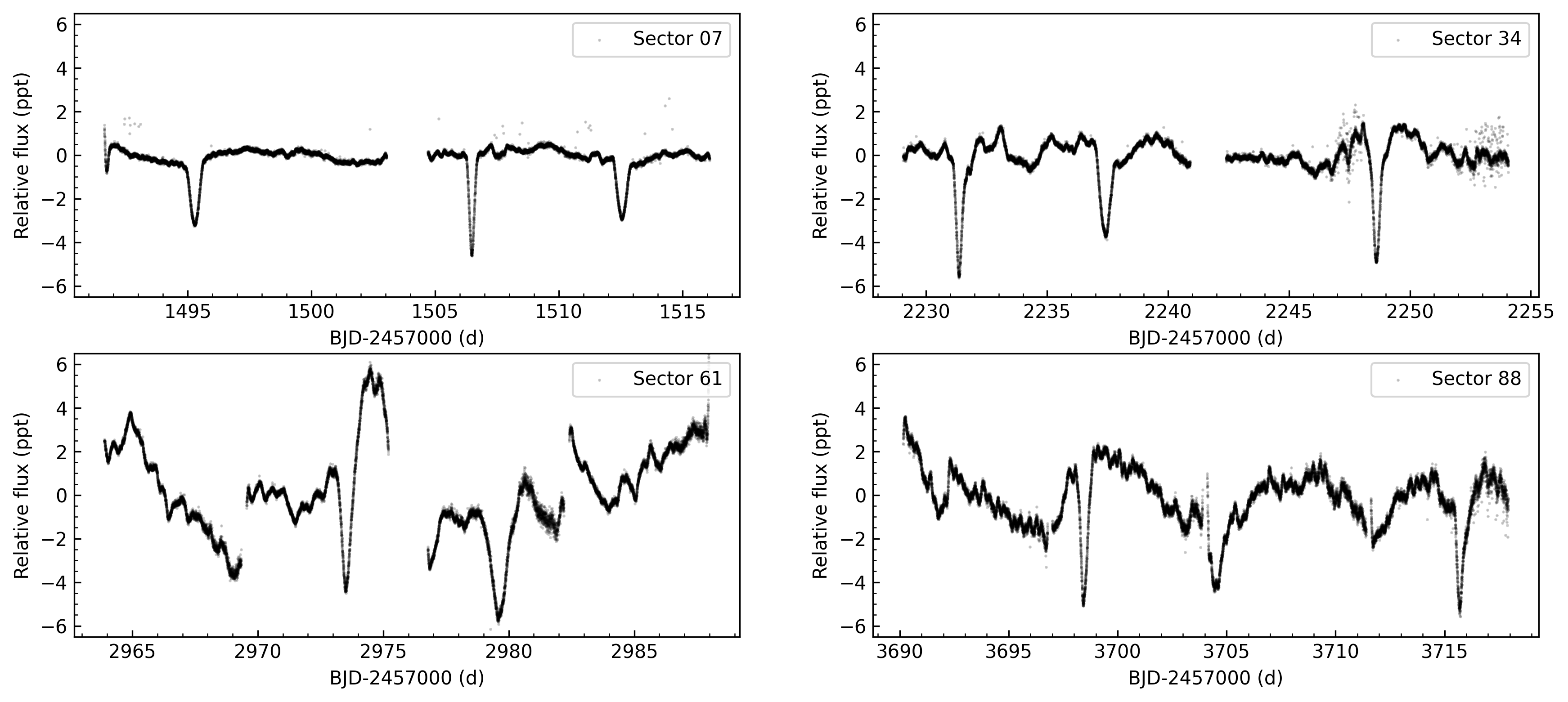}
    \caption{
    SPOC data (120 s exposures) for the four available sectors in the TESS dataset for KQ Pup - sectors 7 (January 2019), 34 (January 2021), 61 (January 2023), and 88 (January 2025).}

    \label{fig:kq_pup_tess_raw}
\end{figure*}

\begin{figure*}[htbp]

    \includegraphics[width=1.0\textwidth, keepaspectratio]{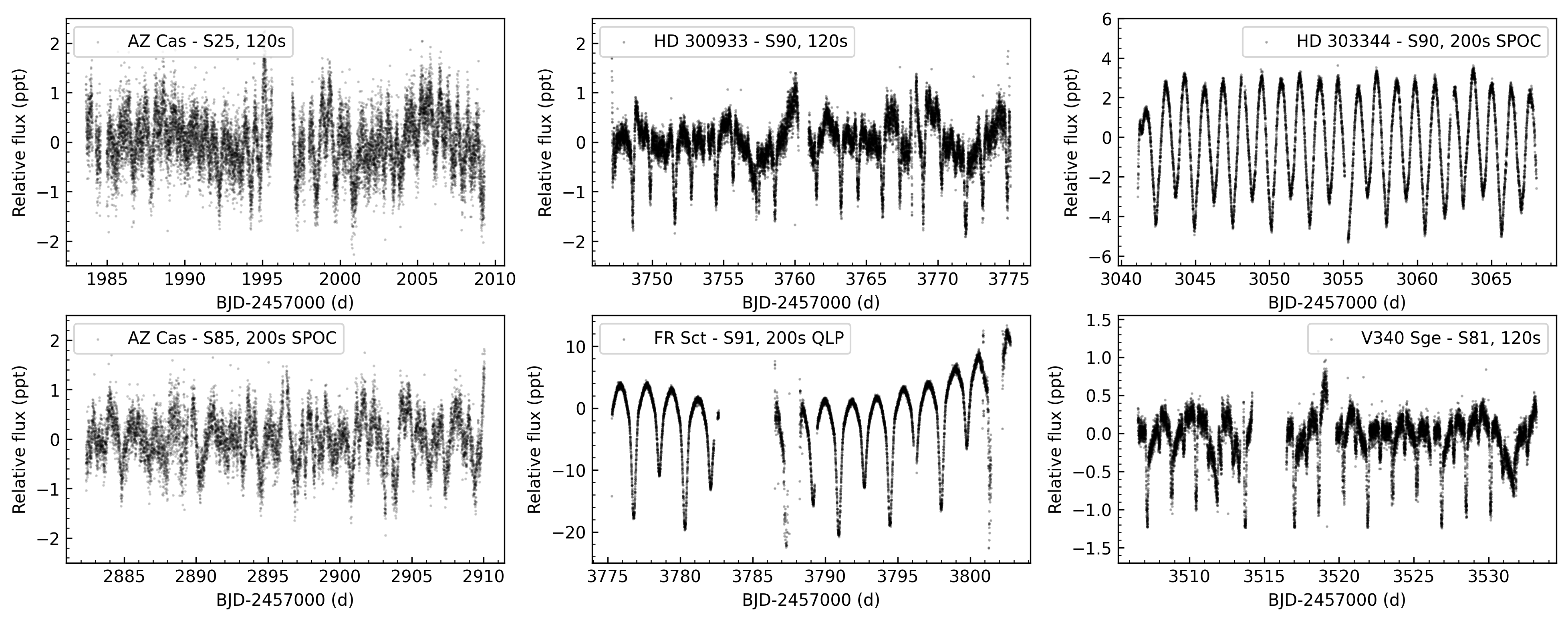}
    \caption{
    Examples of TESS data (for selected sectors, see the legend in each plot) for the other promising candidates. 
    We used 200\,s or 120\,s (where available) cadences. We show two sectors for AZ Cas, whose classification as an EB is uncertain.}

    \label{fig:tess_other_candidates}

\end{figure*}

 \subsection{Notes on individual targets}
Below, we report some additional notes on individual targets and we also list discarded candidates in Table \ref{table:table_candidates_discarded}.

\begin{table*}[htbp]
        \centering
        \caption{List of hierarchical triple RSG candidates from TESS discarded due to a likely blend with nearby EBs of similar periods or due to signal offset.} 
    \label{table:table_candidates_discarded} 
        \setlength{\extrarowheight}{3pt}
        \begin{tabular}{ccccccc}
\hline \hline 
\text{Star} & \text{Spectral type}  & \text{Period [d]} & \text{H em.}  & \text{[Fe II] em.} & \text{Multi-core Na I}     & \text{Nearby eclipsing source}  \\ 
\hline
CD-32 4371 & M3I:  & 0.94 &  & & & \text{Gaia DR3 5594921650170191744} \\ 
BD+53 2693 & K2Ia  & 1.83 &  & & & \text{V2263 Cyg} \\ 
HD 115336 & K3.5Iab  & 2.18 &  & & & \text{Gaia DR3 5859054064419031040} \\ 
HD 142696 & M0Ib  & 1.25 &  & & & \text{Gaia DR3 5884691789303621504} \\ 
V349 Car & M3Iab  & 1.22 & X & weak & \checkmark   & \text{EB signal offset from photocenter} \\ 
V1092 Cen & M2.5Iab  & 10.26 & X & X & X & \text{Gaia DR3 5335719343117052544} \\ 
TIC 44430172 & M1Iab & 2.65 & X &  &  & \text{EB signal offset from photocenter} \\ 

\hline

        \end{tabular}
\tablefoot{The spectral binary signatures are listed based on archival spectra, when available.}

\end{table*}

\begin{figure}[htbp]

    \includegraphics[width=0.45\textwidth, keepaspectratio]{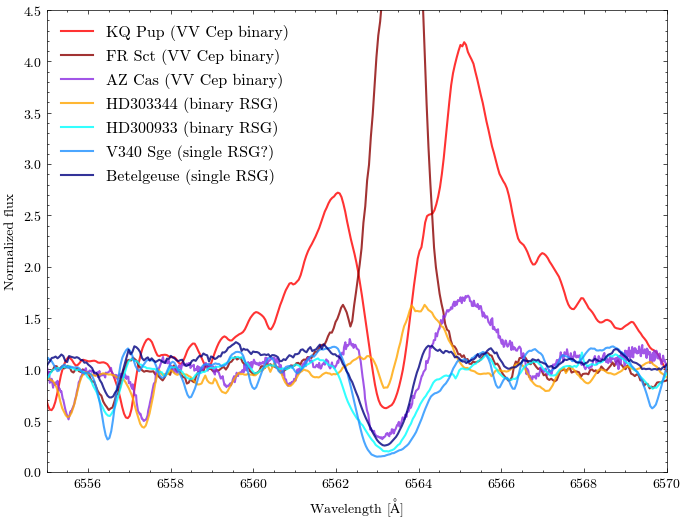}  
    \includegraphics[width=0.45\textwidth, keepaspectratio]{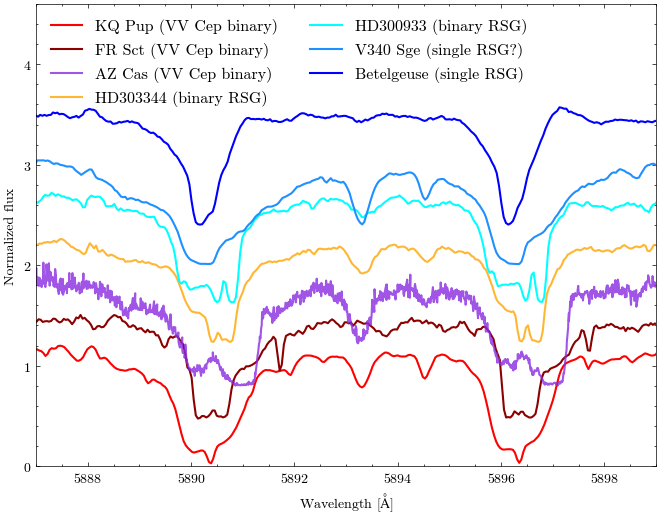}
    \caption{
    {\em Upper panel:}
    Comparison of the H$\alpha$ line for our promising candidates, using PLATOSpec spectra. We can see that only the VV Cephei binaries show the Balmer emission. Single RSGs only show the deep absorption component.
    {\em Lower panel:}
    Same but for the \ion{Na}{i} doublet, which shows multiple cores for binary RSGs. We compare the profiles to Betelgeuse, a well-known RSG without a hot massive B-type companion.}

    \label{fig:platospec}
\end{figure}

\paragraph{FR Sct and HD 303344:} These candidates show semi-detached eclipses (possibly similar to Algol-type binaries; see, e.g., \citealt{mkrtichian22}), making them interesting targets for future observations. We also found that HD 303344 shows Balmer emission in the spectra (Fig. \ref{fig:platospec}), possibly adding it to the VV Cephei group. FR Sct was part of the initial list of 13 VV Cephei binaries by \citet{cowley69}. \citet{pigulski07} reported likely photometric eclipses of $P=3.53 \: \rm d$ but this target was never followed up or confirmed. Considering that eclipses of FR Sct were detected with ASAS, that would also make it the most accessible target for ground-based photometry. 

\paragraph{AZ Cas:} This star is also a well-known VV Cephei binary; it was part of the initial sample of 13 stars \citep{cowley69}. The orbital period is $\sim 9.3 \: \rm yr$ and it undergoes full eclipses, with a totality lasting about $\sim 100 \: \rm d$ (AAVSO). Based on our reported TESS eclipses, AZ Cas would become the first known RSG system showing eclipses both within the inner pair and by the outer RSG, if confirmed.

\paragraph{Wide systems:} There are some candidate systems with large angular separation between the RSG and their companions, i.e., similar to the Antares system ($\sim$ 5 arcsec between Antares A and B) or even larger. Eclipses detected for such companions are difficult to separate from blends with nearby unrelated EBs. In the case of the promising candidate, V340 Sge, it has a wide companion at $\sim$30 arcsec \citep{ccdm}, showing the same eclipses, while a discarded candidate, BD+532693, is surrounded by 4 B-type stars (2 of them Be-type) within $\sim$35 arcsec, one of them showing the same eclipses. To judge whether such wide systems are physically associated and thus to confirm whether such systems are wide hierarchical triple systems, would be beyond the scope of this paper, and it will become clearer with the release of upcoming epoch astrometry in Gaia DR4.

\paragraph{VV Cep:} We also hypothesize that VV Cep, the most famous star of the VV Cephei sample, could be a triple system, based on its reported V/R variations. This star experiences full eclipses (e.g., AAVSO), with the totality lasting about $\sim 450 \: \rm d$. \citet{pollmann20} reported strong V/R variations of about $\sim 42 \: \rm d$, including during the totality, while there are also V/R variations related to its $20.36 \: \rm yr$ orbital period. Therefore, as in our case, where V/R variations were sensitive to the orbit of KQ Pup Ba+Bb, the shorter V/R variations could correspond to a hypothetical VV Cep Bb. We note that we have not found a clear signature of repeating eclipses in the TESS data for VV Cep, although the $\sim 42 \: \rm d$ period is longer than the typical length of TESS sectors of $\sim 27 \: \rm d$. The TESS light curves also appear to show a dipper-like variation in some sectors.

\FloatBarrier
\twocolumn

\begin{figure*}[htbp]
    \section{Auxiliary plots for KQ Pup and other candidates}
    {\includegraphics[width=1\textwidth, keepaspectratio]{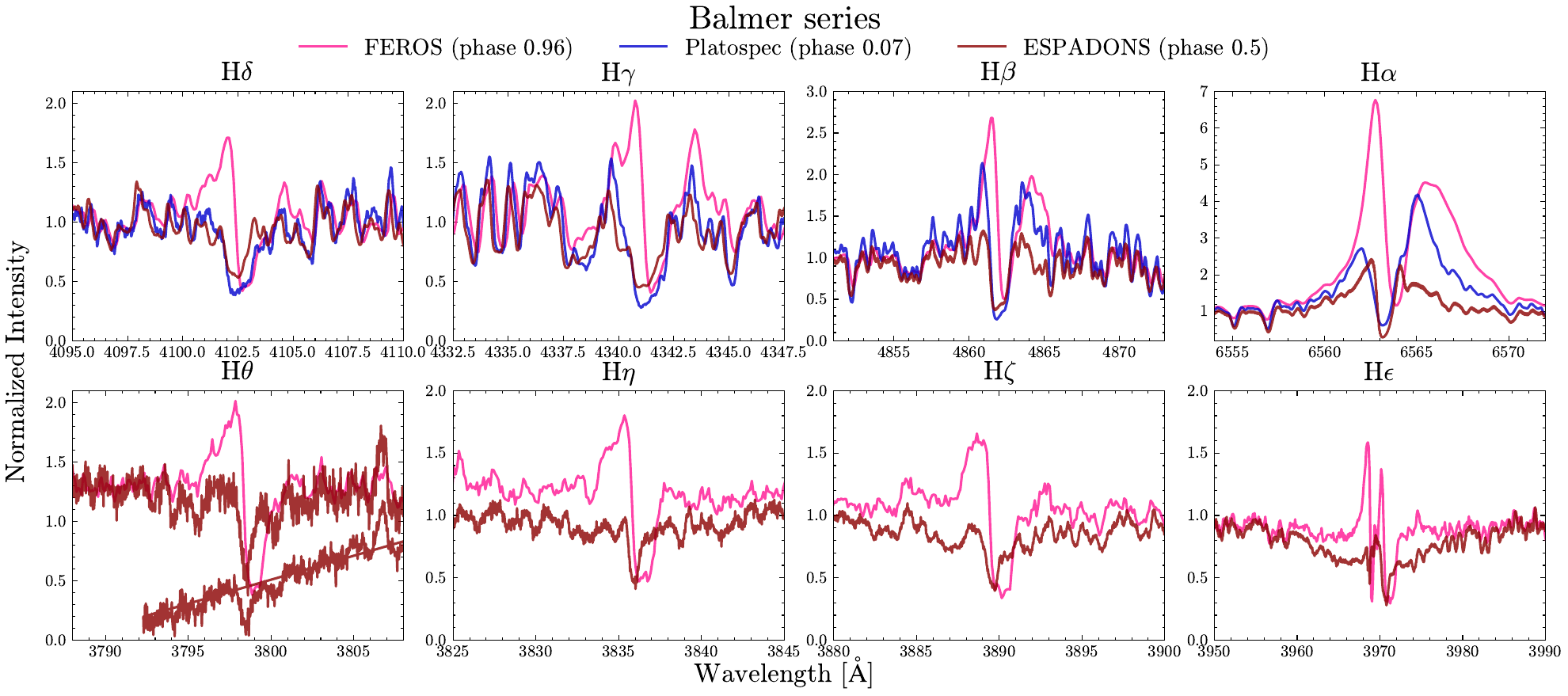}}
    {\includegraphics[width=1\textwidth, keepaspectratio]{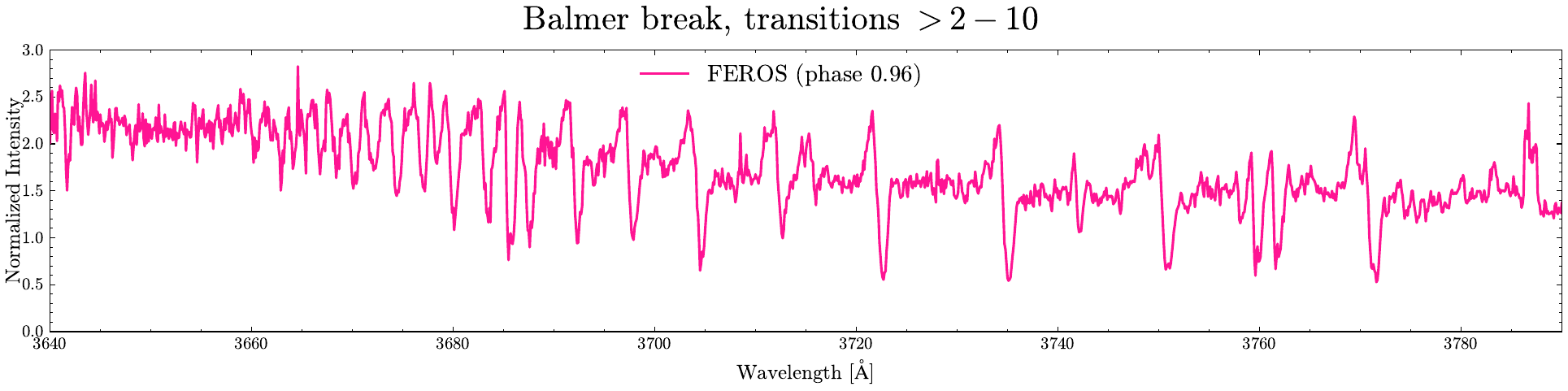}}
    \caption{Balmer series, from H$\alpha$ (2-3) to H$\theta$ (2-10), based on ESPaDOnS, FEROS, and PLATOSpec for different orbital phases $\phi_{\rm A+B}$ of KQ Pup. While the majority of the optical spectrum is dominated by the RSG, the Balmer lines form near the hot component. Single RSGs usually do not show the highest Balmer transitions. The lower plot shows all Balmer lines up to the Balmer break near periastron. Archival optical spectra also show that the V/R ratio is dependent on the phase $\phi_{\rm A+B}$ of the 26-yr orbital period \citep[e.g.,][]{rossi98}. We note that the central sharp feature in H$\epsilon$ is caused by \ion{Ca}{ii} H line forming in the ISM, and the wide depression is likely an instrumental effect in ESPaDOnS spectra. }

    \label{fig:kq_pup_spectra_opt}
\end{figure*}

\begin{figure*}[htbp]
    {\includegraphics[width=0.5\textwidth, keepaspectratio]{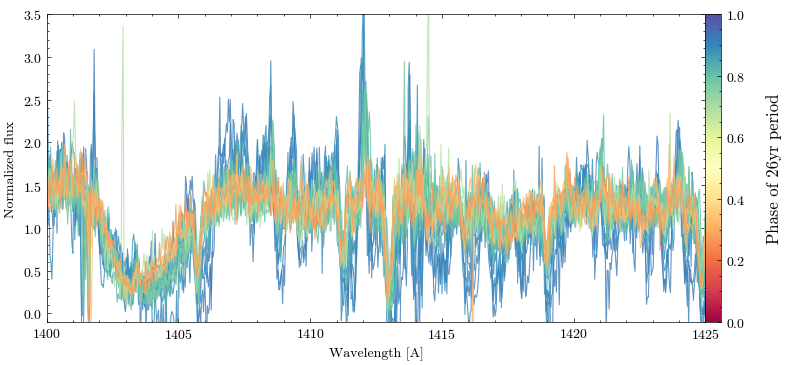}}
    {\includegraphics[width=0.5\textwidth, keepaspectratio]{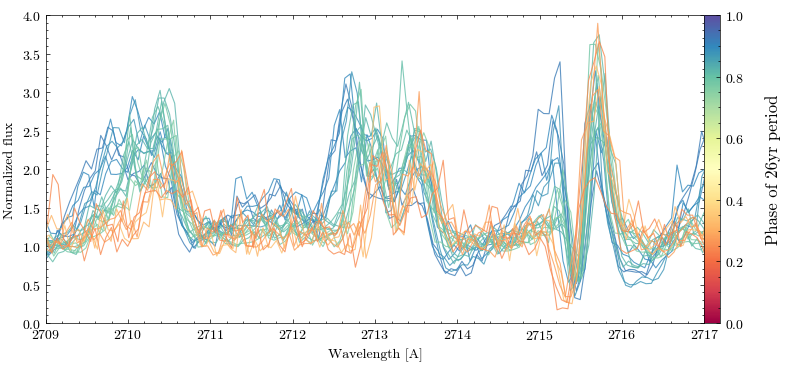}}

    \caption{Important features from the IUE spectra of KQ Pup.
    {\em Left panel:}
    Resonance and narrow lines in the SWP region, colored based on $\phi_{\rm A+B}$. We observe the narrow lines broadening and deepening near periastron. 
    {\em Right panel:}
    Emission and P Cygni profiles in the LWP/R region. We see that for most of the 26-year orbit, the spectra show regular P Cygni profiles, and their absorption cores move toward longer wavelengths, while the emission lines move in the opposite direction. Near periastron, some of the P Cygni profiles appear inverted. }

    \label{fig:kq_pup_spectra_uv}
\end{figure*}

\begin{figure*}[htbp]
    {\includegraphics[width=1\textwidth, keepaspectratio]{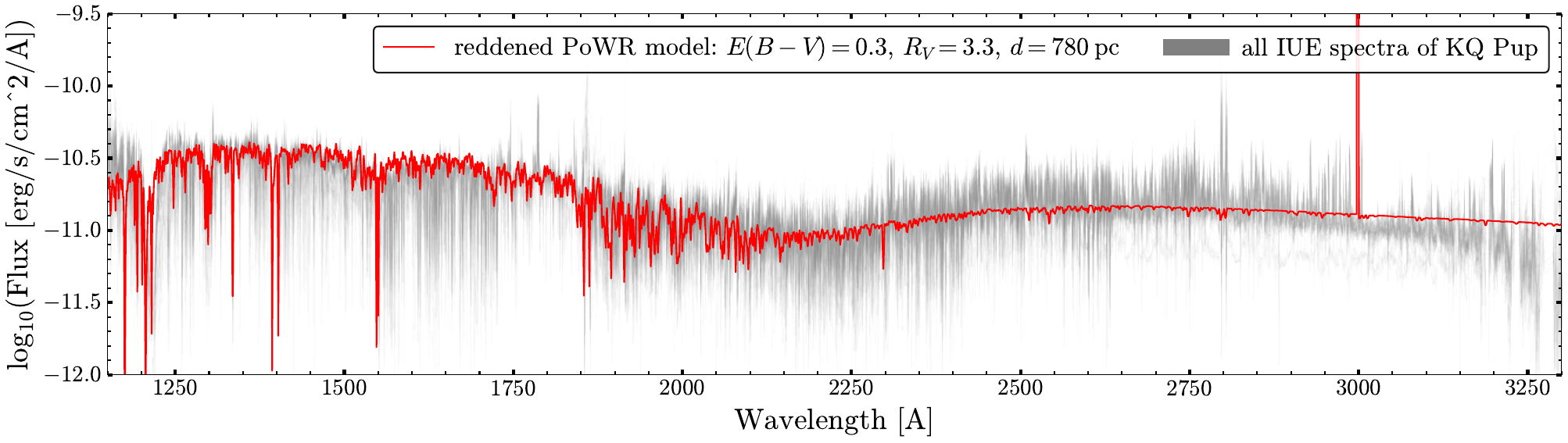}}
    \caption{Reddened PoWR model compared to all IUE SWP and LWP/R binned spectra of KQ Pup. The model agrees with the observations for a reddening of $E(B-V) = 0.30$ with $R_\mathrm{V} = 3.3$ and assuming the distance of 780\,pc \citepalias{bailer21}.  }

    \label{fig:KQPUP_UV_SED}
\end{figure*}

\begin{figure*}[htbp]
    {\includegraphics[width=0.48\textwidth, keepaspectratio]{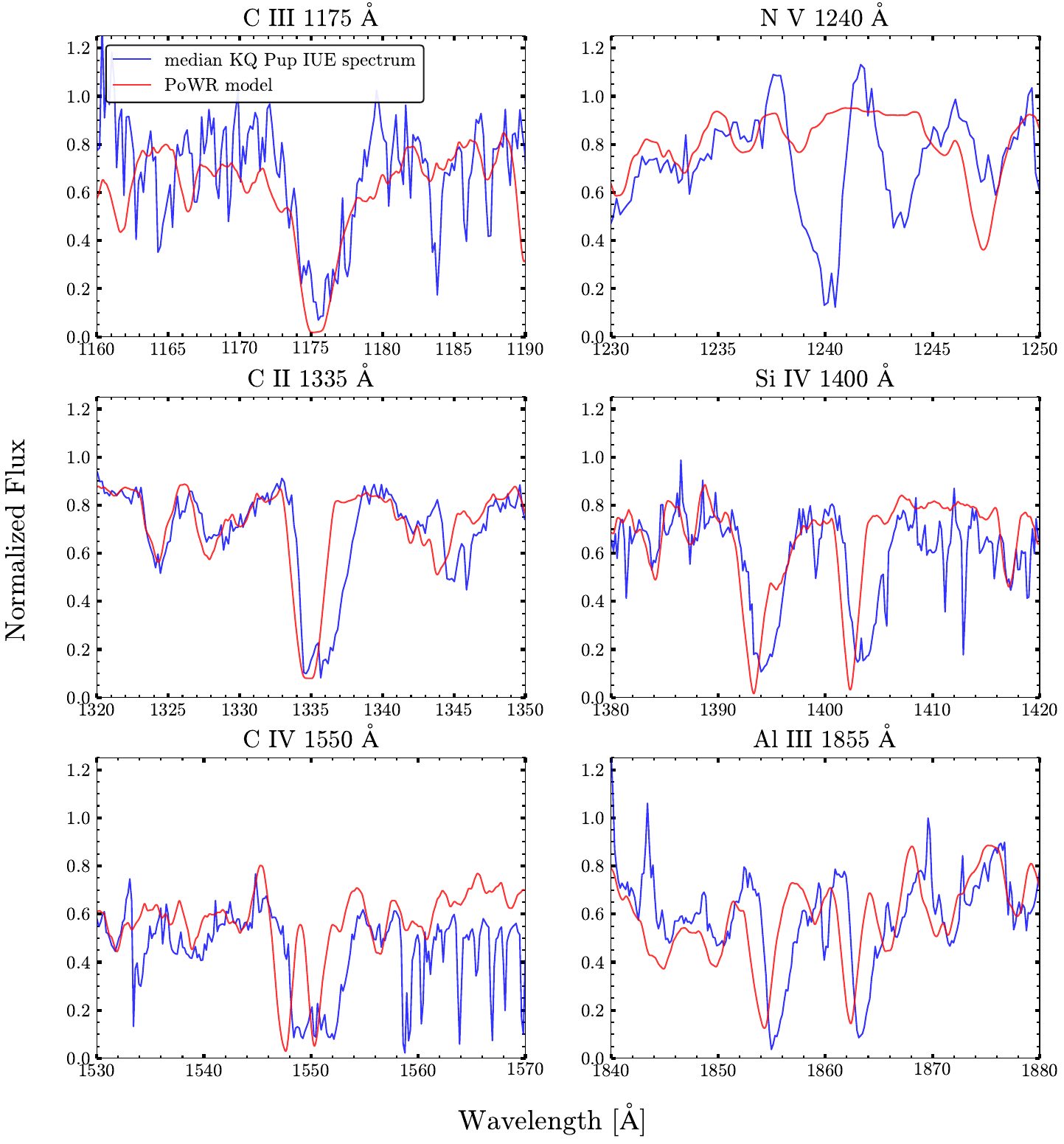}}
    {\includegraphics[width=0.52\textwidth, keepaspectratio]{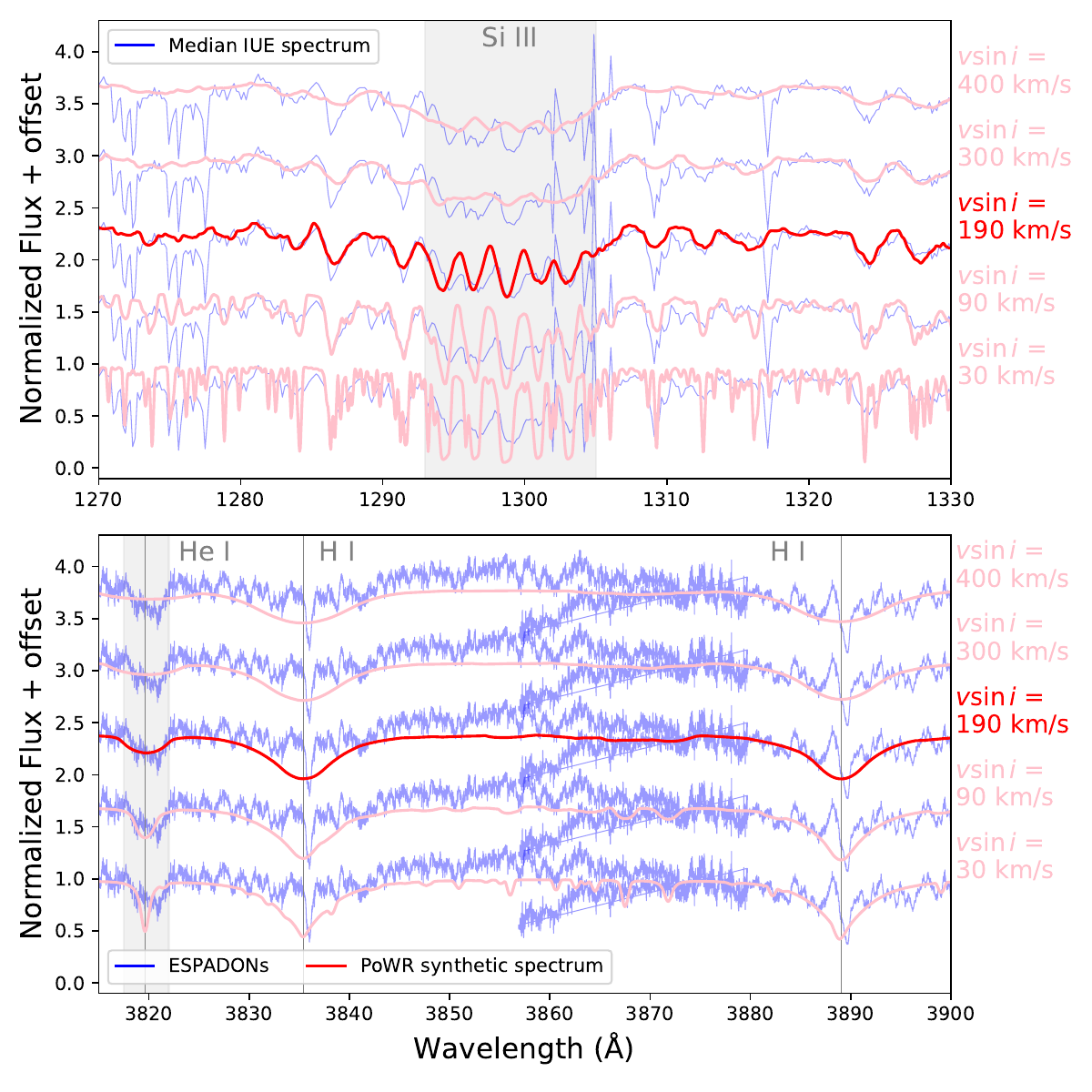}}

    \caption{
    {\em Left panel:}
    Best-fit of the PoWR model atmosphere (red) to KQ Pup IUE data (blue), showing regions of the SWP spectra.
    {\em Right panel:}
    Example of determining $v \sin i$ from the IUE SWP and optical spectra. The rotationally broadened \ion{He}{i} line at 3819$\AA$ is also reproduced by the fit; other \ion{He}{i} lines below $4000 \: \rm \AA$ do not appear to show such broadening, likely due to their proximity to H$\eta$ and H$\zeta$.
    }

    \label{fig:kq_pup_spectra_uv_fit}
\end{figure*}

\begin{figure}[htbp]
    {\includegraphics[width=0.25\textwidth, keepaspectratio]{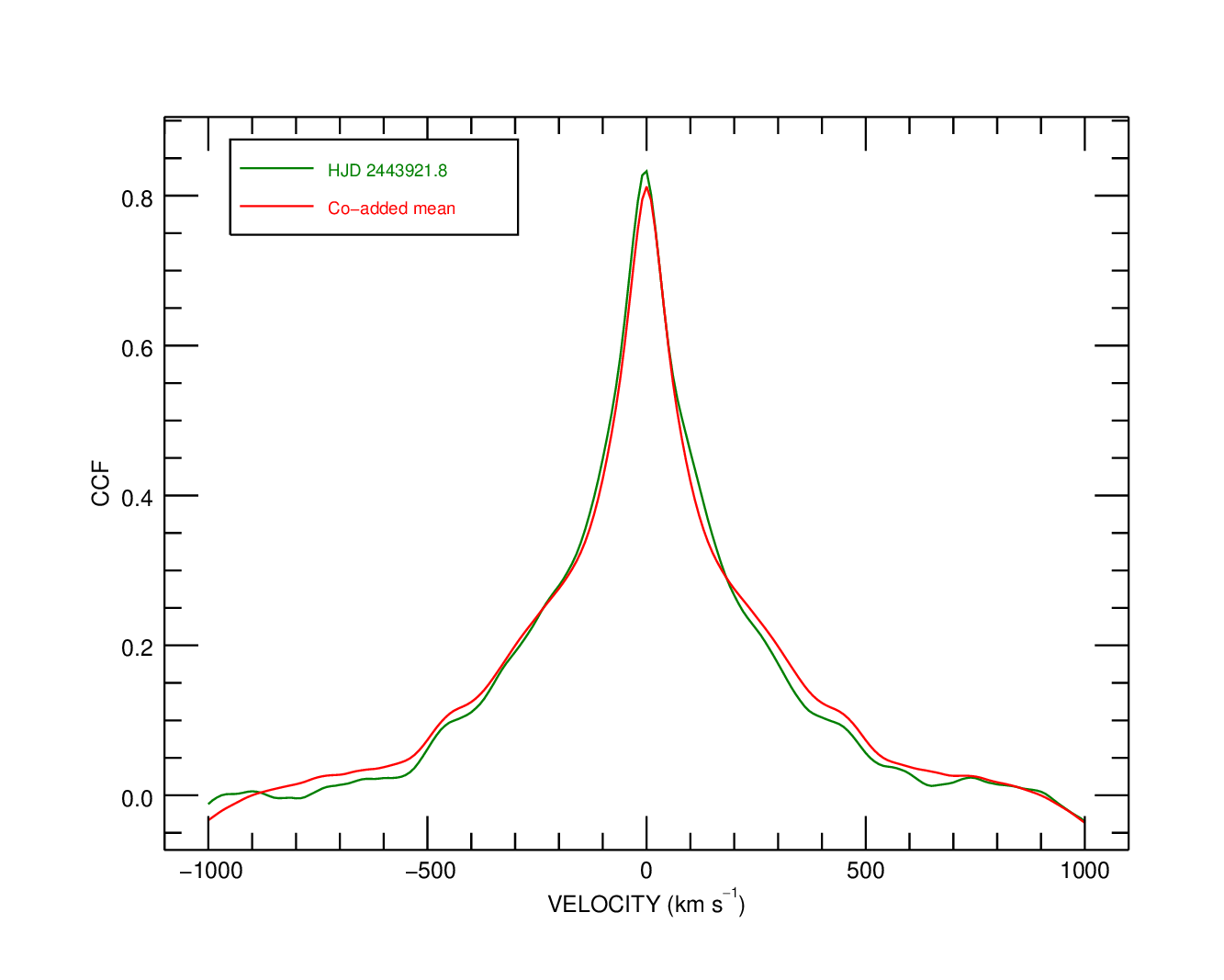}}{\includegraphics[width=0.25\textwidth, keepaspectratio]{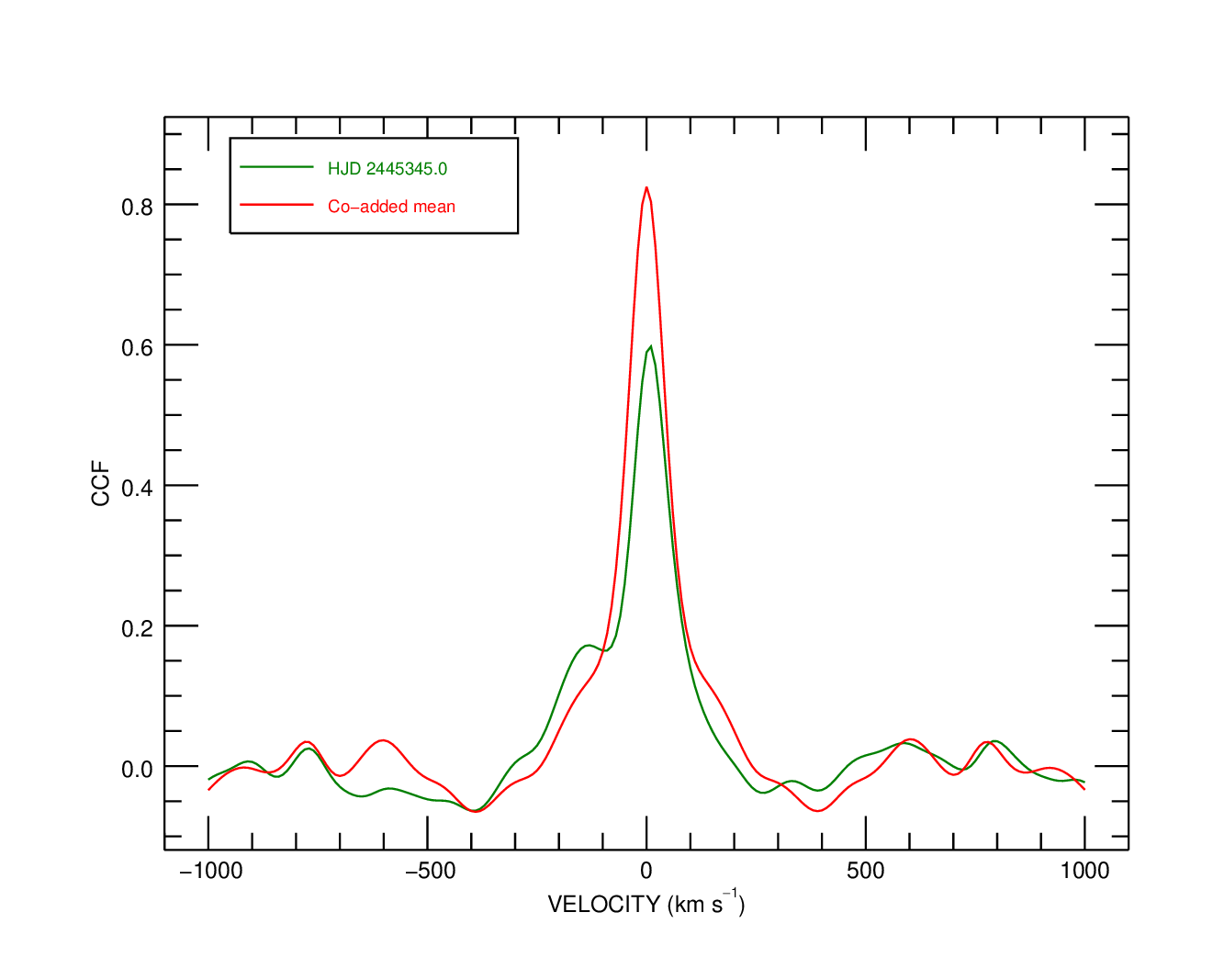}}
    \caption{Examples of CCFs for IUE SWP (left) and LWP (right) spectra of KQ Pup. Red lines are the co-added average spectra, and green lines are selected epochs, shown for comparison. No significant secondary peaks were found in any epoch.}
    \label{fig:ccfs}
\end{figure}

\begin{figure}[htbp]
    {\includegraphics[width=0.45\textwidth, keepaspectratio]{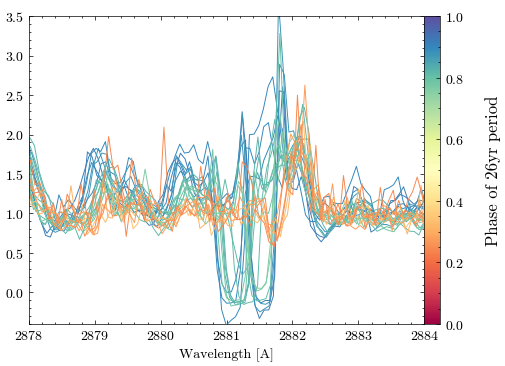}}
    
    \caption{
    Example of variable profiles in the UV spectra of KQ Pup. The line shows a deep absorption profile near periastron ($\phi_{\rm A+B} \sim 0.7-0.95$), while before that, no such deep absorption is present. Colored based on $\phi_{\rm A+B}$.}
    \label{fig:kq_pup_spectra_uv_2}
\end{figure}

\begin{figure}[htbp]
    \includegraphics[width=0.45\textwidth, keepaspectratio]{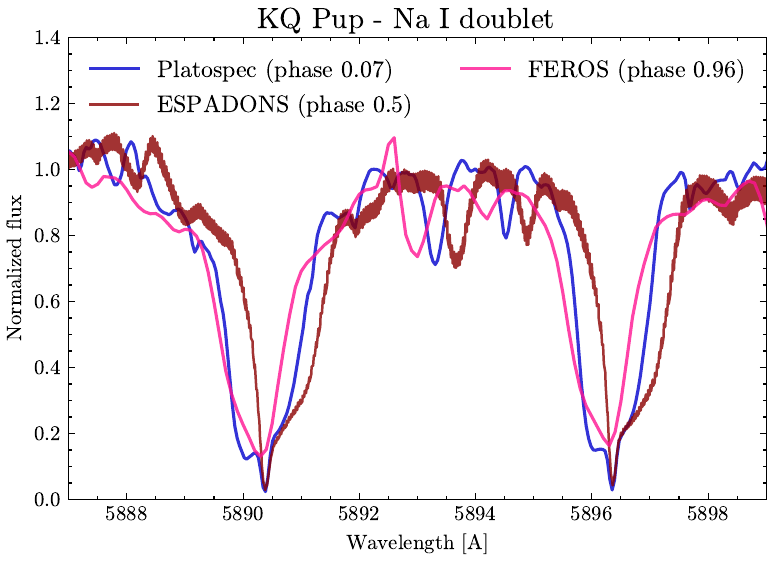}  
    \caption{\ion{Na}{i} doublet for KQ Pup near periastron and apastron. We see a clear relation to the 26-year orbital period of A+B, with the sharp absorption component staying in place, but disappearing before periastron (FEROS).}

    \label{fig:kq_pup_spectra_nai}
\end{figure}

\begin{figure}[htbp]
    {\includegraphics[width=0.5\textwidth, keepaspectratio]{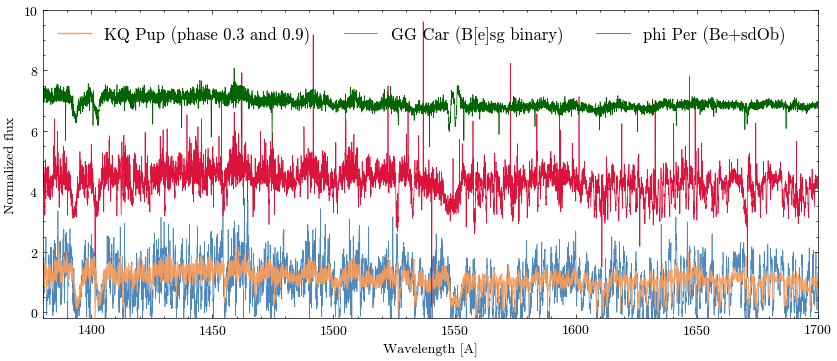}}
    \caption{
    Comparison of KQ Pup (for phase 0.3, orange, and 0.9, blue) to a typical Be+sdOB ($\phi \: \rm Per$) and a known B[e]sg binary GG Car, using IUE SWP spectra.}

    \label{fig:kq_pup_comp}
\end{figure}

\FloatBarrier 
\twocolumn

\section{Stellar evolutionary models}\label{appendix:mesa}
\subsection{Preliminary constraints on the timescales and masses}\label{appendix:mesa:timescales}
\begin{figure}[]
    
    \centering
   \includegraphics[width=1\linewidth, keepaspectratio]{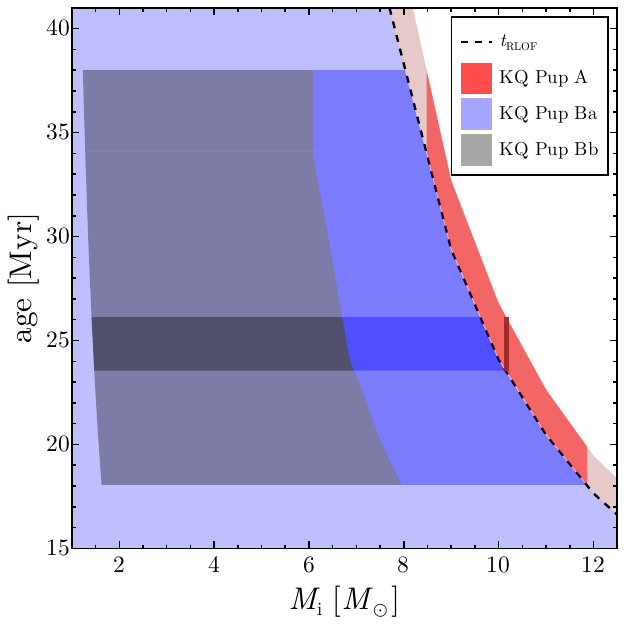}  
    \caption{Age-mass diagram from the MIST stellar models \citep{dotter16}. Different regions mark how a single star of given initial mass and age would appear, either as a main-sequence star (light blue) or an RSG (light red). The region highlighted in red shows the parameter space in age and mass for KQ Pup A. Similarly, the region highlighted in blue and gray shows the parameter space for KQ Pup Ba and Bb, within the constraints of the mass and age of KQ Pup A and the assumption that Ba is more massive than Bb. The regions with darker highlighting show the mass constraints for KQ Pup Ba and Bb assuming the mean value for the mass of KQ Pup A. The dashed line reports the age range for which higher masses would have left the main-sequence and trigger Case\,B RLOF, assuming an orbital period of $17.26\,\mathrm{d}$.}

    \label{fig:evol1}
\end{figure}

    Using stellar evolution models from MIST \citep{dotter16}, assuming solar metallicity and initial rotation of $40\%$ critical, we can infer the age of the system as well as constrain the mass of KQ Pup Ba. To do this, we use the data available (VLTI global+RV fit) in Table\,\ref{table:table_results}, which gives us $M_\mathrm{A}\sim8.5-11.1 \: \rm M_\odot$. Accounting for wind mass loss until halfway through core-He burning, the stellar evolution models suggest an initial mass $M_\mathrm{A,i}\sim8.5-11.9 \: \rm M_\odot$. For KQ Pup A to be observed as an RSG, it must have an age $\sim18.1-38.0\,\mathrm{Myr}$ (Fig.\,\ref{fig:evol1}). 

    Within this age constraint, we can infer the possible initial mass ranges that KQ Pup Ba and KQ Pup Bb must have, assuming both Ba and Bb are main-sequence stars, with Ba being the more massive one. The observed values are likely to be smaller by $\lesssim0.5 \: \rm  M_\odot$ due to wind-mass loss between ZAMS and the moment at which the system is observed. We also assume that the triple system has evolved coevally, and that KQ Pup Ba and Bb have not experienced any form of mass transfer. 
    
    Under these assumptions, we can constrain the minimum mass of Bb for it to have successfully reached the main-sequence by the time of the observation. For our range of ages, Bb must have initial masses above $\sim1.2 \: \rm M_\odot$ \citep[MIST, ][]{dotter16}. As Bb cannot hold more than half the mass of the Ba+Bb system, we derive that $M_\mathrm{Bb}\leq \frac 1 2 M_\mathrm{Ba+Bb}\sim \frac 1 2 qM_\mathrm{A,i}$. For the given range of uncertainties in $M_\mathrm{A}$, this yields $1.2 M_\odot\lesssim M_\mathrm{Bb}\lesssim 8.5 \: \rm  M_\odot$  (see Fig.\,\ref{fig:evol1}). 

    For the mass of Ba, we note that being the more massive star in the binary, we can constrain its minimum mass to $M_\mathrm{Ba}\geq \frac 1 2 M_\mathrm{Ba+Bb}\sim \frac12qM_\mathrm{A}$. At the same time, it cannot be more massive than KQ Pup A at the ZAMS; otherwise it would have evolved to an RSG, and therefore $M_\mathrm{Ba}\lesssim M_\mathrm{A,i}$. Within the current uncertainty in $M_\mathrm{A}$, the mass constraint of Ba is $5.8\: \rm \mathrm{M_\odot}\lesssim M_\mathrm{Ba}\lesssim 11.8 \: \rm M_\odot$ (see Fig.\,\ref{fig:evol1}). These values are compatible with those derived from astroseismic measurements (see Sect. \ref{chapter:asteroseismology})

    The orbital period of $17.26\,\mathrm{d}$ is such that RLOF only occurs after one of the two stars leaves the main-sequence (see Fig.\,\ref{fig:evol1}). It is worth noting that this does not necessarily prove that we are observing the system before such an event. Indeed, if mass transfer has already occurred, KQ Pup Ba may actually be the initially less massive companion that accreted a lot of mass, while KQ Pup Bb is the initially more massive companion, which is now partially stripped. The observations support that Ba and Bb have not interacted yet (see Sect.\,\ref{chapter:triple}), but it cannot be fully ruled out. For example, if future observations classify Ba as a Be star, which would suggest that the star accreted material, causing it to spin rapidly and form a decretion disk, this may have been driven by the accretion of material from the winds of KQ Pup A \citep{li26}, instead of mass transfer from the initially more massive but now dim Bb. The nonzero eccentricity of Ba+Bb is also noteworthy, as mass transfer typically circularizes orbits \citep[e.g.,][]{lechien25}, but it may also have been driven by eccentricity pumping \citep{krynski25} from the interaction with the RSG winds. 

\subsection{Extended atmosphere}\label{appendix:mesa:atmosphere}
We can estimate the local density that the extended structure of KQ Pup A would have close to the Ba+Bb binary. We follow \cite{ercolino24} and construct an extended envelope assuming a quasi-hydrostatic, isothermal stratification with turbulent pressure using a single-star of $M\sim11.6  \: \rm M_\odot$ halfway through core-helium burning, with $\log L/L_\odot = 4.5$, $T_\mathrm{eff}=3\,420\ \mathrm{K}$ and $R=526  \: \rm R_\odot$ (cf. the interferometry estimates in Table\,\ref{table:table_results}). While the photospheric radius is about $\sim 4$ times smaller than the orbital separation at periastron ($\sim 2\,100  \: \rm R_\odot$), the extended hydrostatic layers may be much larger \citep{ercolino24}.

Assuming that the inner binary is outside the low-density hydrostatic layers of KQ Pup A, we stitch a $\beta$-law wind structure (with $\beta=5$ and $v_\infty=25\,\mathrm{km}\,\mathrm{s}^{-1}$) above the sonic point ($v_{\rm sound} = 5-15 \: \rm km s^{-1}$) of the extended atmosphere \citep[see][]{ercolino24}. With this, we note that the density contrast of the material swept up by the inner binary at periastron passage with KQ Pup A is one to two orders of magnitude higher than at apastron. The Bondi-Hoyle accretion \citep{bondi1944} during periastron passage from KQ Pup A to the inner binary is about $10\%$ the mass-loss rate from KQ Pup A, while it can be significantly higher at apastron (due to the smaller velocities). However, we note that the low velocity and relatively high mass of the binary result in an accretion radius that is within a factor of two of the orbital separation, meaning that the Bondi-Hoyle model is not suitable to address the accretion of material by the Ba+Bb binary. Indeed, if the strength of the Balmer emission traces the accreted material from KQ Pup A, the Bondi-Hoyle mechanism would imply the opposite behavior. This therefore strengthens the scenario that KQ Pup A is undergoing wind-RLOF to the inner Ba+Bb binary \citep{mohamed07}. 

\end{appendix}

\end{document}